\documentclass[preprint]{aastex}

\newcommand{\<}{$<$}

\newcommand{\nh}{\mbox{$N_{\rm H}$}}

\newcommand{\HII}{\ion{H}{2}}
\newcommand{\skipthis}[1]{}

\newcommand{\psqcm}{{\rm cm}^{-2}}
\newcommand{\persqcm}{\rm \,cm^{-2}}

\newcommand{\ps}{{\rm s}^{-1}}

\newcommand{\erg}{{\rm ergs}}

\def\micron{\hbox{$\mu$m}}
\newcommand{\be}{\begin{equation}}
\newcommand{\ee}{\end{equation}}
\newcommand{\e}{et al.\ }

\newcommand{\asec}{$^{\prime\prime}$}
\newcommand{\amin}{$^{\prime}$}

\slugcomment{\today}

\shorttitle{RCW 108}
\shortauthors{Wolk et al.}

\begin{document}


\title{X-ray and IR Point Source Identification and Characteristics \\
    in the Embedded, Massive Star-Forming Region RCW 108}


\author{Scott J. Wolk,
Bradley D. Spitzbart,
Tyler L. Bourke, Robert A. Gutermuth}
\affil{Harvard--Smithsonian Center for Astrophysics,
       60 Garden Street, Cambridge, MA 02138}
\author{Miquela Vigil}
\affil{Lincoln Laboratory, Massachusetts Institute of Technology, 
Lexington, MA 02420}  
\author{Fernando Comer\'on}
\affil{European Southern Observatory, Karl-Schwarzschild Strasse 2,
       D-85748 Garching bei M\"unchen, Germany}




\begin{abstract}
We report on the results of an approximately 90~ks {\it Chandra}
observation of 
a complex region that hosts multiple sites of recent and active
star formation in ARA OB1a. The field is centered on the embedded
cluster RCW~108--IR and
includes and a large portion of the open
cluster NGC 6193. We detect over 420 X--ray 
sources in the field and combined these data with deep near-IR,
$Spitzer/IRAC$ and MSX
mid--IR data. We find about 360 of the X-ray sources have near--IR counterparts.
We divide the region into 5 parts based on the X-ray point source
characteristics and extended 8\micron\ emission.  The most clearly
defined regions are the central region --  identified by embedded sources
with high luminosities in the both the near--IR and X-ray as well as high X-ray
temperatures ($\sim$ 3~keV) and the eastern region -- 
identified by low extinction
and  $\sim$ 1~keV X-ray temperatures. Other regions, identified by their
directional relationship to RCW~108--IR are less uniform -- 
representing combinations of the first two regions, independent star
formation epochs, or both.   
The cluster members range in X-ray 
luminosity from $~10^{29}$ to $ 10^{33}$ ergs s$^{-1}$. 
Over 18\% percent of the cluster members
with over 100 counts exhibit flares.  
All sources with over 350 counts are variable.
Overall about 10\% (16\% in RCW~108--IR)  appear to have optically thick 
disks as derived from their position in the (J$-$H), (H$-$K) diagram.
The disk fraction becomes much higher when IRAC data are employed.
The largest fraction of X-ray sources are best
described as possessing some disk material via a more detailed extinction fitting.  
We fit the bulk of the X-ray spectra as absorbed Raymond-Smith 
type plasmas and 
find the column to the RCW~108-IR members varies from $10^{21}$ to 
$10^{23}\persqcm$.  
We find that the field contains 
41 candidate O or B~stars and 
estimate that the total number of pre--main sequence stars in the 
field is about 1600
$\pm$ 200.  Approximately 800 are confined to
the 3\amin\ ($\sim$ 1.1~pc) central region.
\end{abstract}


\keywords{\ion{H}{2} regions -- ISM; individual (RCW~108) -- stars; formation -- 
X-rays; point sources -- X-rays; stars}


\section{Introduction}

\begin{figure}[t]
\plotone{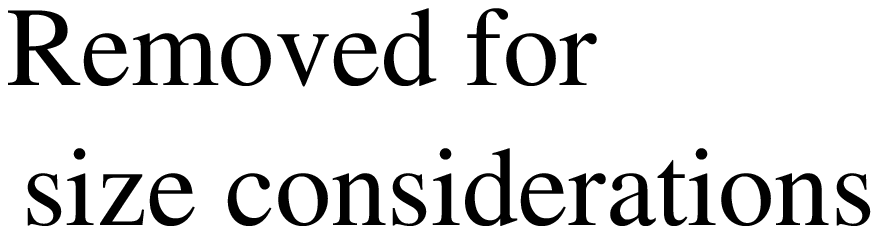}
\caption{An optical image of the RCW~108 region (adapted from image by 
Robert Gendler and Martin Pugh 2006;  
http://www.robgendlerastropics.com/NGC6188MP.html). The ACIS-I field of view is
outlined as a square about 17\arcmin\ on a side.  The bright stars to the
east are the HD 150135/6 complex of O stars.  These stars are the
center of NGC 6193 (circle), NGC 6188 is identified as the ridge of
emission.  The inset shows an IRAC image of  
RCW~108--IR  showing the detail of the ionization front 
(BGOR=3.6\micron, 4.5\micron, 5.8\micron\ and 8.0\micron\
respectively).  The edge of the diffuse emission is fairly sharp.}
\label{introfield}
\end{figure}

Ara OB1a may be one of the best examples of
triggered  star formation in the local galaxy.  The triggering in the
most active portion is easily imagined from images of the
NGC~6193/RCW~108-IR complex such as Figure~\ref{introfield}.  
Ara OB1 was first studied in detail by Whiteoak (1963) using
photographic photometry and objective prism spectroscopy to identify
about 35 O and B star members.  He identified OB1a with a relatively
nearby cluster associated with open clusters NGC~6193 and NGC~6167
while Ara OB1b was found to be about twice as distant. 
Whiteoak's work was a follow up to an
H$\alpha$ survey by Rodgers, Campbell and Whiteoak (RCW; 1960) who cataloged
181 H$\alpha$ emission regions in the southern sky. 
NGC~6193 is an open cluster dominated by a pair of O stars, HD 150135
(O6.5V; Walborn 1972) and HD 150136 (O3+O6V; Niemela \& Gamen 2005).   These are the brightest optically revealed O stars in
the association and are thought to be responsible for ionizing the
bright rim of emission to the west (NGC 6188), which separates NGC 6193
from RCW 108. The youth of the region is clear in sky survey plates
and Figure~\ref{introfield} which show concurrent regions of ionized gas and
dust lanes. 

Ara OB1a is a compact association covering about 1 sq. degree ($\sim$
25~pc on a side) around a
central cluster -- NGC 6193. 
The distance to NGC 6193 is 1.3$\pm$0.2 kpc 
and the cluster is associated with the HII region RCW 108
(Herbst \& Havlen 1977).
There is some confusion in the literature as to what is actually
meant by RCW~108. The original
definition of Rodgers, Campbell and Whiteoak (1960) refers to all the
region where H$\alpha$ nebulosity is detected, for which they give a size
of 210\arcmin\ $\times$ 120\arcmin\ centered at: l, b= 336.49, --1.48. Straw et
al. (1987) used RCW~108--IR to refer to the embedded IR cluster about
15\arcmin\ to the west of the O stars in NGC~6193 that has been
identified with IRAS~16362-4845.   The confusion arises when
RCW~108--IR is abbreviated by dropping the ``--IR''.  For the remainder of
this paper we will refer to the embedded cluster as RCW~108--IR.

\subsection{NGC~6193}
NGC~6193 is the open cluster to the east of the bright emission rim NGC~6188.
In a region about 40\amin\ ($\sim$ 15~pc) on a side, 
Herbst \& Havlen (1977) measured over 700 stars photometrically. Of these, 
59 had photometric distances and reddening consistent with cluster
membership.  Their survey appears complete to spectral types earlier than A0. 
Their best fit to the photometric distance was 1320$\pm 120$pc.
Several other groups have estimated the distance to the cluster 
using photometric parallax techniques and obtain results 
ranging from $\approx$1100 to
$\approx$ 1400pc (Moffat \& Vogt 1973, Fitzgerald 1987,
Kalcheva \& Georgiev 1992).
Vazquez \& Feinstein (1992) obtained an age of about 3 Myr by fitting
the upper main sequence, they also determine a distance of about 1410
$\pm 120$pc.  We will use 1.3 kpc for this paper to be consistent with
recent studies. For linear distances we will usually use parsecs,
 1\amin = 0.38~pc at 1.3 kpc.  The mean extinction is low towards
 NGC~6193 (A$_V \sim 0.5$).

From the X-ray perspective, there is an image from
the zero--order of a 100~ks $Chandra$ observation  (Skinner \e 2005).  
Since the effective
area of the HETG zero-order is only about 15\% that of the unobscured
ACIS-I, this observation is relatively shallow.
Still, the observation should be fairly
complete to 1 solar mass. It reveals 43 X-ray sources in the central square
0.76~pc on a side.
Only 11 of these had previous optical identifications, however all 43
were detected at H band by 2MASS and are likely cluster members. In addition to
those sources cataloged by Skinner et al.~there are about 30 X-ray
point sources visible outside of the central 0.8~pc on a side 
region in this data set. Most of these have I-band magnitudes
consistent with cluster membership and are between 0.8~pc and
2.58~pc from HD 150136. 

\subsection{RCW~108--IR}

A bright rim (NGC~6188) marks the border between the \HII\ region and a 
 dark nebula containing RCW~108--IR.
The bright rim is produced by the ionization 
of the molecular cloud hosting  RCW~108--IR by the two central stars
of NGC 6193.
 Herbst and Havlen (1977) proposed this was a site of active star formation.

RCW~108--IR is a young, compact cluster partially embedded in its parent molecular cloud.
Straw et al. (1987) used photometry at IR wavelengths (1.2-100 $\mu$m)
to perform the first survey of the cluster.
They reported on 55 objects; of which $<$ 20 have optical counterparts.
In addition to the point sources, there is diffuse IR emission, 
The total luminosity of the cluster is about 1.8$\times 10^5
$L$_\odot$, dominated by two early O stars.
The full aggregate of O and B stars appears responsible for the
ionization of the diffuse emission (Comer\'on et al. 2005). The
primary exciting source star, identified by Straw \e as IRS~29,  is
resolved into multiple sources by Comer\'on et al. The brightest of these 
is a probable O9 star.
Straw et al. also identify at least 1 protostar based on high
luminosity and low temperature. They put forward a
physical model of the region in which photoionization of the older
stars in the Ara OB1a association ionized the bright rim (NGC 6188)
which lies at the edge of a finger of molecular material which
thickens from east to west.
IRS~29 is the second strongest
water maser seen by SWAS (second to the BN object on Orion;
Melnick-private communication) with a peak flux of over 7000 Jy in the
557 GHz line.  A follow up study suggests that gaseous H$_2$O is
largely restricted to a thin layer of gas near the cloud surface.


There have been two recent near-IR through mm studies of RCW~108--IR.
Urquhart \e (2004) compared the 2MASS and MSX observations of the
cloud with radio recombination lines and radio continuum using the
Australia Telescope Compact Array (ATCA).  With synthesized beamwidths
of 12\asec\--20\asec, they find
RCW~108--IR is an ultracompact \HII\ region, 
less than 0.1pc in scale with a core--halo morphology.  
They discuss 8 new sources within the UCHII region and three
25~$\mu$m\ sources set back about 0.95~pc from the bright
rim.  A more detailed study has been recently published by Comer\'on
et al. (2005).  From their high resolution ($\sim$ 25\asec) CO map they estimate a
total mass of the cloud at 8000 M$_{\odot}$ with $< 200 $M$_{\odot}$
in the molecular
concentration harboring the compact \HII\ region. They produced a
high resolution JHK$_s$ map of the region and identify 25 stars whose luminosities suggest
spectral types earlier than A0 under the assumption that there is no significant
circumstellar contribution to the K-band flux within the diffuse 
IR nebula.  Comer\'on \e conclude that the ionization of the UCHII region is provided
by this aggregate. Strangely, they detect no stars later than A0 in
the core.  They suggest that low--mass star formation has yet to
commence here. Deeper adaptive optics 
imaging does reveal fainter members of the cluster, but not in the large 
numbers that would have been expected from an extrapolation of the K luminosity 
function of the brightest members toward fainter magnitudes
(Comer\'on \& Schneider 2007).
 IRS 29 is found to be an O9 star from the visible
spectrum of the compact \HII\ nebula, 
which is in agreement with its infrared photometry. Urquhart \e (2004)
confirm the O9 spectral type via radio flux measurements. 
Out of 4365 stars brighter than K$_s$=14.5 in the whole 13\arcmin
$\times$ 13\arcmin\ field ($\sim$5~pc on a side), 87 are found to have strong disk signatures, most being
located in the molecular cloud that contains RCW108--IR and along the
inside of the bright rim -- NGC~6188.

We recently observed RCW~108--IR, the southwestern portion of NGC 6193
and the dividing rim -- NGC~6188  using the $Chandra$ X-ray
Observatory. The goals of the observation were to:
i) detect the effects of triggered star formation
ii) identify deeply embedded PMS stars via their X--ray emission and
  derive an initial mass function and an X-ray luminosity function
  and iii) investigate the X--ray properties of the embedded sources.
In \S 2, we outline the X-ray observations made specifically
for this paper and supporting IR
and mm wave observations. In \S 3, we combine these data to understand the
basic morphology of the region, dividing the field into five regions
based on a core region and the  
cardinal directions which are well aligned to NGC~6188. The X-ray
characteristics of the X-ray sources and their regional dependencies
are discussed in the next two section. In \S 6, near-IR and mid-IR data are
brought in to study the gas to dust ratio, K-band luminosity functions
and OB stars among the X-ray sources. The more extreme X-ray sources
are discussed in \S 7.  We compare our work with other recent results
in \S 8 and summarize our results in \S 9.

\section{Observation and Data Reduction} 

This paper will focus on X-ray observations of RCW 108-IR and NGC 6193
made by the $Chandra$ X-ray Observatory (CXO).  But we will also
incorporate several other observations, many of which have not been
discussed in the published literature.  The observations include 
near-IR photometry,  Mid-IR observations by $Spitzer/IRAC$ and MSX, 
as well as mm--wave observations performed at the Swedish-ESO Submillimetre Telescope
(SEST).  After detailing the reduction of the primary X-ray data, the
other data reduction is summarized in this section.

\subsection{X-ray Data}
The field was observed by $Chandra$ on October 25 2004 starting at 02:37
UT  for 92.2 ks of total time and 88.8 ks of so called ``good-time'' 
(ObsId 4503). The ACIS was
used in the nominal imaging array (chips I0-I3) which provides  
a field of view of approximately 17\arcmin $\times$ 17\arcmin\ ($\sim$
6.5~pc on a side).
The aimpoint was at 16:39:58.7, -48:51:54.4  (J2000.0).
In addition, the S2 and S3 chips were on and located over 
IRAS~16379--4856.  About 20 point sources were detected associated
with this object, however the analysis of these data are not presented
here because they are far off-axis. 

The background was slightly variable during this observation.  
For the first 5 ks the background
was about 5\% higher than the median, and a similar high period 
which lasted about 2ks occurred about 20 ks into the observation.
The background was about 3\% lower than the median for the last 10 ks. 
The data used in these analysis were reprocessed using the default 
 ``repro 3'' provided by the Chandra X-ray Center (CXC) data systems. 
As such, they were processed through the standard CIAO pipeline at the
CXC, using their version DS7.6.  This version of the
pipeline automatically incorporated a noise correction for low energy
events. It has been noted that this filter can remove good events 
from the cores of bright point sources, resulting in an
underestimation of the X-ray flux (L. Townsley private communication).
This proved not to be an issue in this case since the count rate did
not exceed 0.04 counts per second for any source.

To identify point sources, photons
with energies below 300 eV and above 8.0 keV were filtered out.  
The excluded energies
generally lack a stellar contribution due to the low instrumental
throughput and the relatively cool thermal spectra of coronal
emission.  By filtering the data as
described, contributions from hard, non-stellar sources and
monochromatic detector background are attenuated.  
A monochromatic exposure map was generated in the standard way using an
energy of 1.49 keV which is a reasonable match to the expected 
peak energy of the sources and the $Chandra$ mirror transmission.  
WavDetect was then run on a series of flux
corrected images binned by 1, 2 and 4 pixels. The resulting
source lists were combined and this resulted in the detection of 429 sources. 
For the bulk of the X-ray analysis we will restrict ourselves to the
337 source detections with significance greater than  3.5\footnote{The
  src\_significance given by the CIAO tool WavDetect is not in
  units of $\sigma$ so no false alarm probability is associated. While the
  relative value of src\_significance is meaningful, 
  it is not clearly defined in a
  statistical sense (V.~Kashyap Private communication).}.
  These are tabulated in Table~\ref{XSources}.  However, all X-ray 
sources returned by WavDetect are statistically significant and
we present follow-up information on the faint sources as well.  The
fainter sources are listed in Table~\ref{XSources_faint}.
In each of these tables, The first column contains a running src
number, the second an official IAU style designation.  Columns 3 and 4
are the R.A. and Declination of the centroid of the X-ray source
followed by the off axis angle in column 5.
At each source position an extraction ellipse was calculated following
Wolk \e (2006) updated for the appropriate satellite roll
angle \footnote{$Chandra$ roll angle is the angle between the detector
  vertical and north. The point spread function is strongly dependent
  on {\em both} the off-axis angle ($\Theta$) and the rotation angle 
($\Psi$) of the incident point source}.  This provides an
extraction ellipse containing 95\% of the source flux. The number of
counts in this region are listed in column 6.  
For each of the 429 sources, a background ellipse is identified. 
The background is an annular ellipse with the same center, eccentricity, and 
rotation as the source.  The outer radius is 6 times the radius of the source 
and the inner radius is 3 times larger than the source.  From this region any 
nearby sources are subtracted with ellipses 3 times the size of the source 
ellipse.  The net counts are calculated by subtracting the background
counts  (corrected for area) and multiplying the result by 1.053 to
correct for the use of a 95\% encircled energy radius. This is the
value given in column 7.  Table~\ref{XREF} contains a cross-referencing
for all X-ray sources with a counterpart found in the SIMBAD database
with an offset of $<$ 2\arcsec.

\subsection{MSX observations}
The Midcourse Space Experiment (MSX) obtained observations of
 RCW~108--IR and its surroundings  
at 8.3, 12.1, 14.7 and 21.3 $\mu$m as part of its Galactic
Plane survey (Price \e 2001).  The four bands
are labeled A (6.8-10.8$\mu$m), C (11.1-13.2 $\mu$m), D (13.5-15.9
$\mu$m), and E (18.2-25.1 $\mu$m).  The A and C band emission is a
combination of thermal dust emission and emission from polycyclic
aromatic hydrocarbon (PAH) bands at 7.7, 8.6 and 11.2 $\mu$m.  The D
and E bands trace thermal emission of cool dusty objects. The details
of the reduction of these data are discussed by Price \e (2001). 

\subsection{Near--IR observations}
In the near-IR we use J, H and K$_s$ photometry obtain from 2MASS
project and from a deep pointed observation of RCW~108--IR.  The latter
data are discussed by Comer\'on \e (2005) while the 2MASS point source
catalog reductions are summarized by Cutri \e (2003).
JHK$_s$ photometry of sources in the central 
region of RCW 108-IR is taken from deep adaptive optics observations carried 
out with the VLT and described in Comer\'on and Schneider (2007).
We will discuss the details of the matching of the near-IR
point sources and X-ray sources in \S6.


\subsection{Spitzer/IRAC observations}
RCW~108--IR was observed with the Infrared Array Camera (IRAC;
Fazio \e 2004) on--board the $Spitzer$ Space Telescope (Werner \e 2004).
A very shallow IRAC map of RCW 108-IR and the surrounding region 
was made as part of the $Spitzer$
``Early Release Observations''.  These were very limited programs
primarily with aesthetic, not scientific, goals. 
The mean exposure was about 3.2
seconds per band and total observing time of about 30 minutes.
Observations were made in September 2005 (PID~112). The total area
covered with 4 band coverage was roughly 23\arcmin\ ($\sim$8.75~pc) 
on a side, 
aligned roughly north-south. Two band coverage extends 5\arcmin\ to the
north and south - band~2/4 ([4.5\micron] and [8.0\micron] respectively) 
in the north and
band~1/3  ([3.6\micron] and [5.8\micron] respectively) in the south. 
Mosaics and related data were downloaded
from the $Spitzer$ archive in software version S12.4.0.  Additional
processing to remove artifacts and perform point source photometry  
was performed using customized
software described by Gutermuth \e (2007; PhotVis see \S 2.1).  
Sources were then classified into infrared disk class (c.f. Lada \&
Adams 1992, Allen \e 2004, Megeath \e 2004) following the prescription
in  Gutermuth \e (2007; \S 4.1) which uses cuts in a multidimensional
color space including J, H, K$_s$, and all 4 IRAC bands. About 2600
sources were detected in all four band within 
the full IRAC field. About 20,000 sources were detected in at least one IRAC
band.  Diffuse emission is
clear in the 5.8 and 8.0 \micron\ bands and traces the
lower resolution MSX data.

\subsection{Millimeter--wave Observations}

Observations at 1.2 mm were obtained in September 2002 and November 2002
with the 37-channel hexagonal bolometer array SIMBA at the Swedish-ESO
Submillimetre Telescope (SEST) located in La Silla, Chile at an altitude of
2370 meters.  SEST, now defunct, was a 15 m diameter telescope and operated in
the frequency range 70 - 365 GHz with a typical pointing accuracy of
3$\arcsec$ in azimuth and elevation.  The HPBW of a single SIMBA element was
about 24\arcsec, and the separation between elements in the sky is
44\arcsec.  In the final maps, pixels were 8\asec\ on a side with a beamwidth of 
24\arcsec\ and a sensitivity of 235 mJy/beam rms.
SIMBA was set to fast scanning mode and the size for each map
was 20\arcmin$\times$ 20\arcmin\ with a scanning speed of 160\arcsec/second.
Four maps
were made of RCW 108 with different scanning angles. The data reduction and analysis was carried out using the
MOPSI\footnote{Mapping On-Off Pointing Skydip is a software package
  for infrared, millimeter and radio data reduction developed and
  constantly upgraded by R. Zylka.} software package (Chini \e 2003).

\section{Morphology} 
\begin{figure}[t]
\plotone{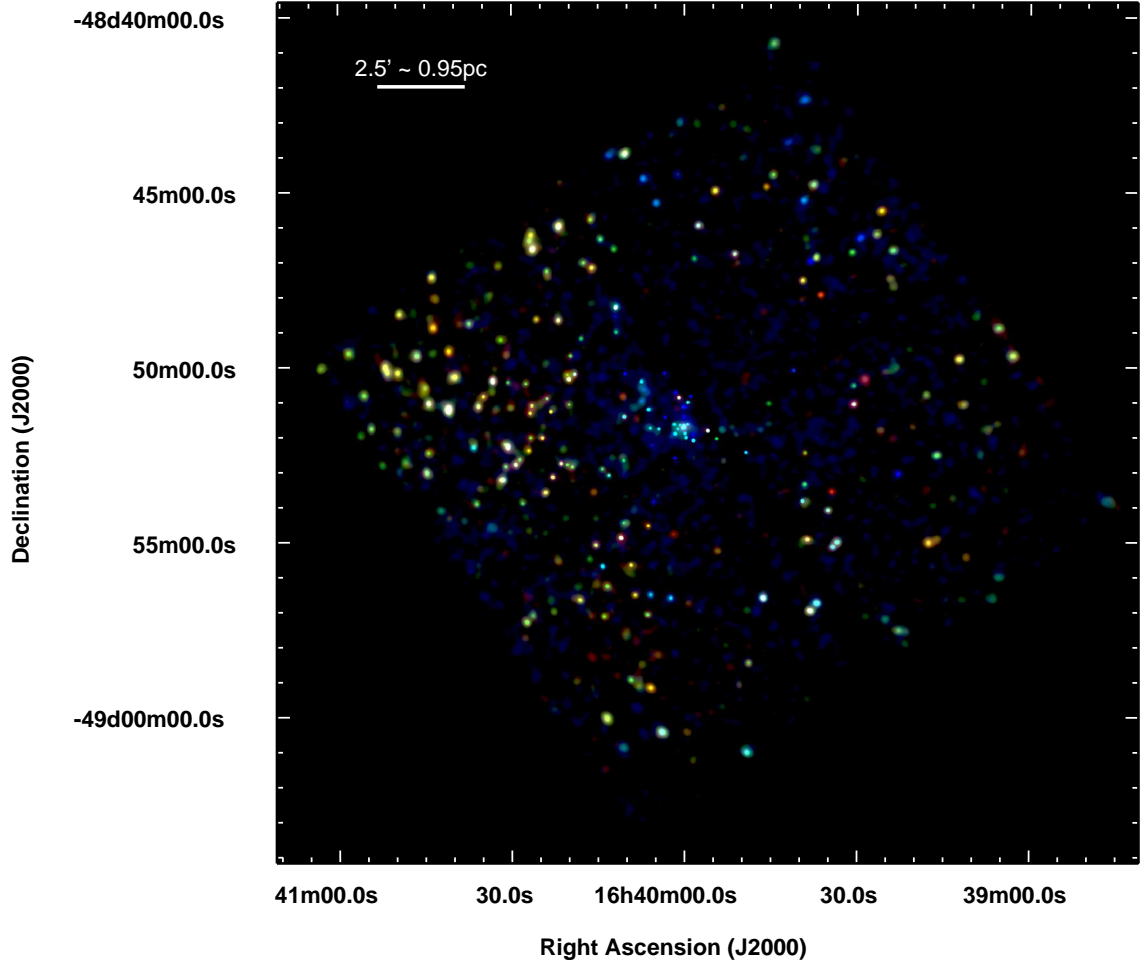}
\caption{An adaptively smoothed version of the X-ray data. 
Photon energies between 300eV and 1.1 keV are plotted in red,
0.9 -- 2.4 keV in green and 2.1 -- 8.0 keV in blue.  Note the
concentration of blue sources in the middle and a marked absence of
source to the north and the south of this core.  The eastern region
has a multitude of sources at a variety of colors.}  
\label{fullfield}
\end{figure}

It is clear from Figure~\ref{fullfield} that the X--ray sources toward
the center are harder than those towards the outside. Further, there
is a remarkable hourglass shaped region, $\sim$3~pc in extent, virtually
devoid of sources except for those sources in the core. This is
indicative of tremendous extinction.  In this section, we investigate
the relationship of this extinction to mid-infrared emission as
observed by MSX and mm emission seen in the SEST observations.  We also study
whether the X-ray sources to the south and west are similar to those to the
north and east.  This would be expected if RCW~108--IR sits in a dark
cloud in front of the larger NGC 6193 cluster and if sources from the
older, more
revealed cluster are seen around all sides of RCW~108--IR.

\subsection{Long Wavelength Morphology}

\begin{figure}[t]
\plotfiddle{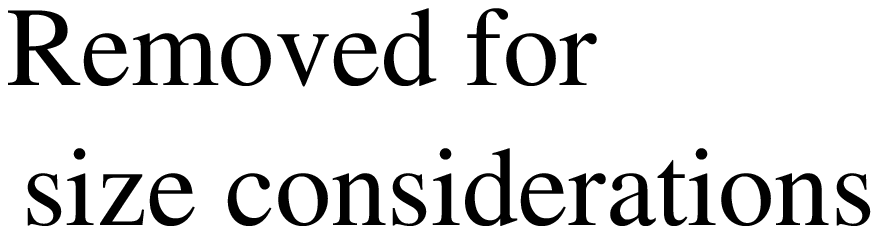}{0.0in}{0.}{270.}{270.}{122.}{0.}
\plotfiddle{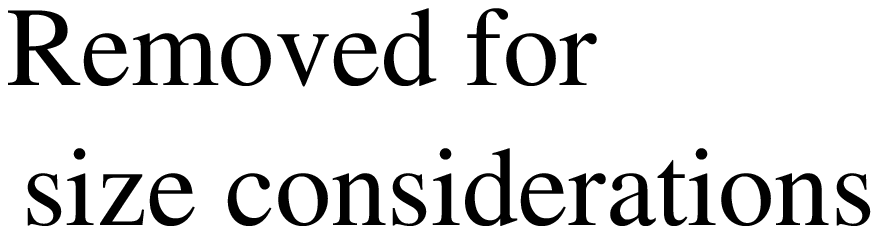}{0.0in}{0.}{315}{270.}{100.}{0.}
\caption{Top: An MSX image of approximately the same area as
  Figure~1. (BGR= A Band (8.28\micron),
     C Band  (12.13\micron ). and 
     E Band (21.3 \micron) respectively.  Two IRAS sources in the
  region are identified.  The smaller box coincides with RCW~108--IR, shown in the lower panel.
 Bottom: A smoothed X-ray image of the central 
3~pc $\times$ 3~pc region of RCW~108--IR (RGB=, 0.5-1.1 keV, 
     0.9-2.4 keV and 
     2.1-8.0 keV respectively). 
The bright source near the center is IRS~29, an O9 star.
The contours represent MSX A band data at 
4.7$\times 10^{-6}$,
8.9$\times 10^{-6}$,
1.3$\times 10^{-5}$,
1.7$\times 10^{-5}$,
5$\times 10^{-5}$,
1$\times 10^{-4}$ and 
2$\times 10^{-4}$ W m$^{-2}$ sr$^{-1}$.}
\label{smoothzoom}
\end{figure}

The most extensive mid-IR emission is seen in the MSX A band data 
(Figure~\ref{smoothzoom}).
These data show linear emission running south to north through the
ACIS field, coincident with NGC~6188.   
The emission has a very sharp edge on the east side
where the emission rises by a factor of about three (to about 4$\times 10^{-6}$ 
W m$^{-2}$ sr$^{-1}$) over a single resolution element.  Emission
levels remain elevated throughout the western two-thirds of the ACIS
field.  The A band emission peak is nearly coincident with the O star within the IRS~29 complex 
at 4.5$\times 10^{-4}$W m$^{-2}$ sr$^{-1}$.  
MSX C and D band data trace the A band data at about 2/3 and 1/5 the irradiance
level respectively.  Since the D band excludes PAH emission, 
this indicates that the emission is dominated by thermal dust.

The morphology of MSX E-Band emission roughly follows the A--band emission.
However, the 21 $\mu$m emission is concentrated in one fairly circular clump
coincident with a concentration of emission at 1.2 mm.  
The extent of the 21 $\mu$m
emission is about the same as the A--band emission in the east--west
direction, and about 1~pc less in the north--south direction (in
radius).  The resolution in the A and E bands is about 20\asec, 
hence the E band emission is truly more circular.   
The MSX E band emission does not show evidence of the dust ridge,
however the core region is bright.
In addition to the south--north ridge (NGC~6188) and the core region
(RCW~108--IR), 
there are three local maximum that are clear in all four bands. Table~\ref{MSX}
lists the intensity of the emission in the various regions as observed
by MSX.  The IRAS data are included for completeness, however IRAS did not
resolve RCW~108--IR.

\begin{figure}
\epsscale{1}
\plotone{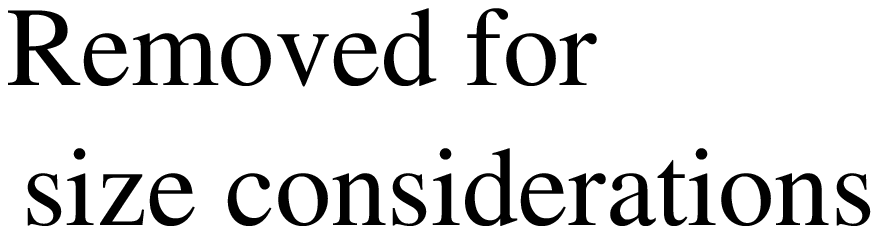}
\caption{RCW 108 with 1.2 mm contours are 3, 6, 12, 24, 48, 96, 192 $\times 
\sigma$ (235 mJy/beam). The contours are overlain on the NTT K$_s$
band image of Comer\'on \e (2005)  The beam size for the 1.2mm
observation is shown at the lower right.
\label{k108}}
\end{figure}

RCW~108--IR was observed at 1.2 mm to investigate the cold dust emission 
and morphology. 
Figure~\ref{k108} shows an NTT K$_s$--band image of RCW 108 with 1.2 mm
contours.  The stellar population coincident with the 1.2 mm emission is
extremely sparse,  probably obscured by dust. However, several bright
sources can be seen clustered near the peak of the 1.2 mm emission.  
The peak of the dust
emission coincides with the central stellar population in the K$_s$--band
image.  The extension to the south is not coincident with any visible
or X-ray source.  The $^{13}$CO data from Comer\'on \e (2005) are
peaked at the same location, slightly offset from the brightest
near-IR sources and have a similar aspect ratio. However, the
long axis of the  $^{13}$CO emission runs from the northwest to the southeast.


We compare the dust emission to mid-infrared observations 
to understand the interactions between the different
components of the regions, specifically the cold dust traced by the 1.2
mm emission and the warmer dust traced by the MSX mid-infrared
observations.
The 1.2 mm emission is concentrated in a linear
ridge running northeast to southwest, about 0.75~pc  west of NGC~6188,
approximately 1~pc in
length  with an extension to the east at
the southern edge.  From Figure~\ref{k108},  the north--south
extent of the 1.2 mm
emission is approximately 0.6 parsecs.  The cold dust emission from RCW 108 is
concentrated around one region without any secondary clumps and only
the slight extension in the south.  The peak emission is about 5 Jy at
the center of the ridge.  The bulk of the emission is within
a  ridge with the 21 $\mu$m emission extending slightly more
toward the east than the 1.2 mm emission.  The morphology of RCW~108
at 1.2 $\mu$m, 10-20 $\mu$m and 1.2 mm show the bulk of the emission
concentrated in one location near RCW~108--IR.  

Color temperatures were derived from the ratios of the fluxes of the
MSX bands assuming the regions followed a blackbody energy
distribution.  The contamination of the A band from PAH emission leads
to unexpectedly high ratios, thus the color temperature of 224~K
was calculated using the C/D ratio. This indicates that the warmer dust
is externally heated, probably by photoionization from the O stars in
NGC~6193.  Meanwhile the dense dust in the center of RCW~108--IR remains
thermally shielded 
even from the stars forming within it, with dust temperatures $\sim$ 20K.

\subsection{Dust Emission and Mass}

The dust mass and density distribution in the molecular cloud was
calculated using the 1.2 mm emission.  Dust masses were
calculated using the relation below, assuming the emission is
optically thin
($\tau \sim$ 1 at 1.3 mm when $N_H{_2} \sim  10^{26}\psqcm$; 
Hildebrand 1983, Mezger 1994) :

\begin{equation}
M_{dust}=\frac{S_{\nu}D^{2}}{\kappa_{\nu,d}B_{\nu}(T_d)}
\end{equation}

\noindent where D is the distance to the cluster, $S_{\nu}$ is the observed dust emission,
$\kappa_{\nu,d}$ is the dust opacity per unit mass column density and
$B_{\nu}(T_d)$ is the Planck function for a dust temperature $T_d$
(Chini et al. 2003). 
 The masses were calculated assuming a dust
temperature of 20 K for the cold dust the SEST observations are
sensitive to.  While there are a range of possible dust opacities, we chose 
0.37 cm$^2$~g$^{-1}$ (Chini et al. 2003)
which is a value consistent for the diffuse ISM but may increase by up
to a factor of 10 in dense regions or protostellar environments
(Ossenkopf $\&$ Henning, 1994).

The vast majority of the 1.2 mm emission from RCW 108 is concentrated
in one extended feature. 
The total mass of the region was determine from the flux within
a contour of 30$\%$ of the maximum flux in the region which is  29.5
$\pm$ 0.4 Jy.
The total dust mass was determined to be 23.1 $\pm$ 4.8 solar masses of 
cold dust. This gives a total mass of about 2300 M$_\odot$ assuming a
gas to dust ratio of about 100.

\subsection{X--Ray Morphology\label{sec_xraymorp}}

The typical method of model independent spectral analysis is to use X-ray colors in the form of hardness ratios, 
HR=(Cts$_h$-Cts$_s$)/(Cts$_h$+Cts$_s$) where ``h'' and ``s'' refer to the hard and soft bands respectively. 
In absorbed regions, hardness ratios are of inherently limited value
because of biases in the selection of bandpasses which 
lead to very non-uniform errors and limited dynamic range.  Sources
with low counts tend to be driven toward the center of the
distribution and there 
is a fundamental difficulty in breaking the degeneracy between temperature and absorption.
To compare X-ray sources in the various regions of the cluster  
we use a quartile analysis technique for model--free analysis of 
X-ray data explored by  Hong \e (2004).
  
In this form of quartile analysis,
one starts with the full ACIS band pass of E$_{lo}$= 0.3 keV   
and E$_{up}$ =8.0 keV.  These are the same bandpasses that we used in
our previous study of RCW~38 (Wolk \e 2006).   
$E_{x}$ is defined as the photon energy below which x\% of 
the photon counts are found and 
$Q_x = {E_{x}-E_{lo}\over E_{up}-E_{lo}} $ 
is defined to be  
the normalized quartile\footnote{Formally, $Q_x$ is a
  ``quantile'' however, the quantiles of 25, 50 and 75 percent are
  specifically known as ``quartiles.''}.
In essence, the median compares the hard and soft portions of the
spectrum.
The ratio of the bottom to top quartile (x= 25 and 75
respectively) is representative of a two-point slope of the spectrum. 
For the case of a single temperature, the median energy is more sensitive
to the absorption. However,
quartiles are not independent, as the absorption changes the quartile
ratio for a given temperature is different. Hong \e\ plot the data 
by normalizing the quartile ratio axis as $3\times Q_{25}/Q_{75}$ 
while the compressed  median is expressed as $\log (m/(1-m))$ where
$m$ is the median. On such a plot one
can distinguish changes in temperature from extinction and can even
distinguish thermal and non-thermal changes.

\begin{figure}[t]
\epsscale{.9}
\plotone{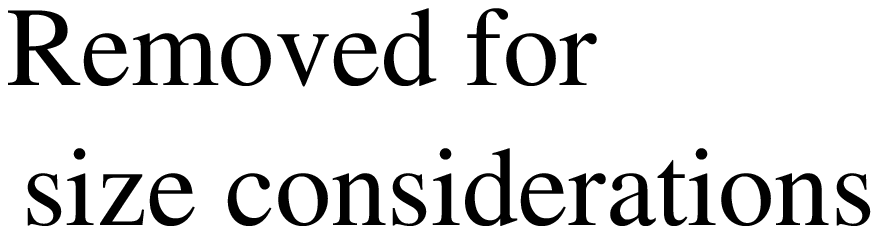}
\caption{Definitions of regions overlain on the $Spitzer$ image
  (BGR= 4.5~\micron, 5.8~\micron\ and 8.0~\micron\ respectively) 
The $Chandra$ field of view is indicated by the blue square.
The eastern region is
clearly part of NGC~6193. The northern region contains bright dust
emission.}
\label{fullregions}
\end{figure}

\begin{figure}[t]
\plotfiddle{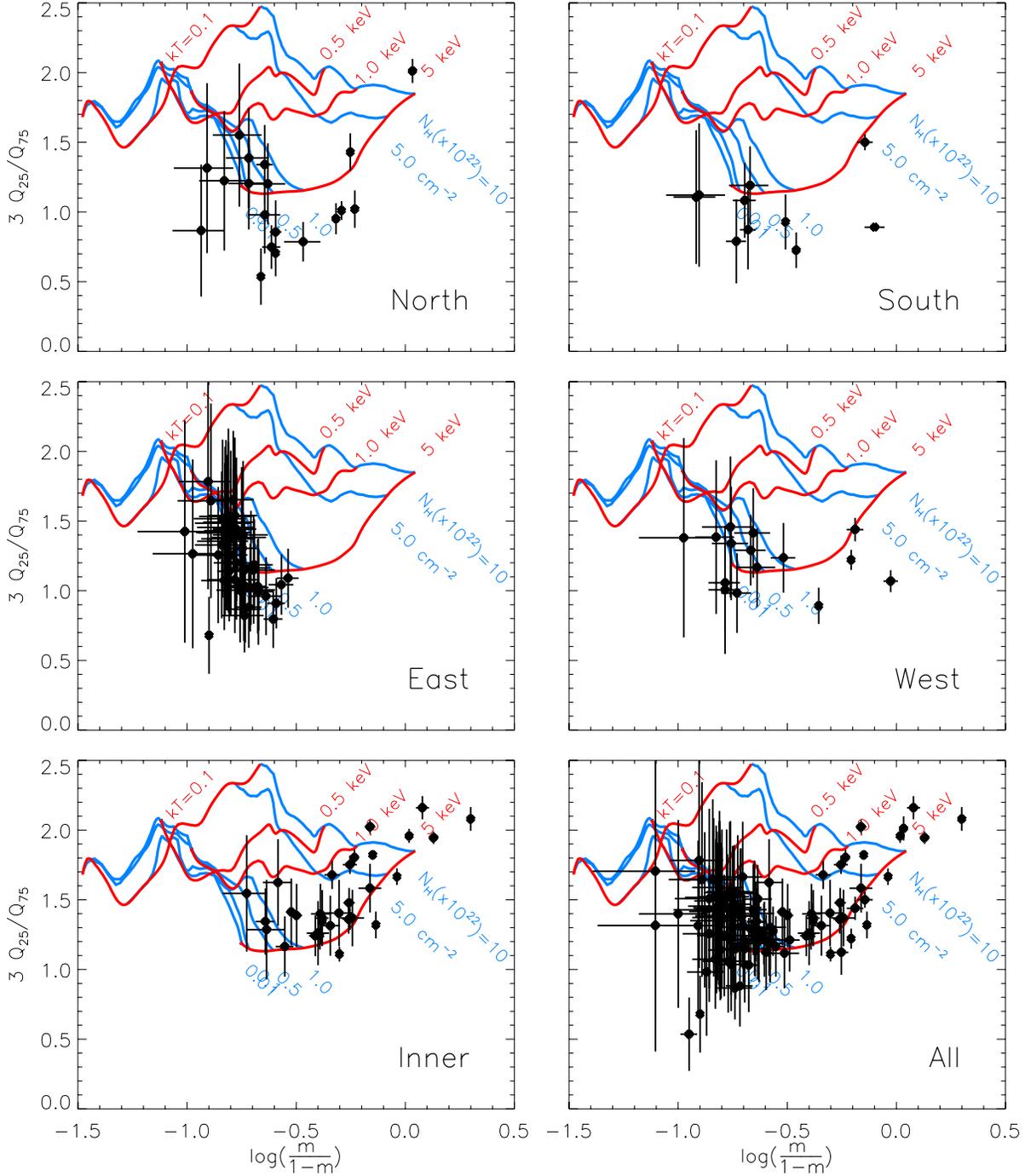}{0.0in}{00.}{450.}{520.}{0.}{-100.}
\caption{Quartile analysis of X-ray sources for the regions defined in Figure~\ref{fullregions}.
  region.  The first four frames are the extreme 3\arcmin\ to the
  west, north, east and south respectively.  The last two figures show
  the inner region and the whole field. 
In each frame, the Y axis is 3 Q$_{25}$/Q$_{75}$,  
the X-axis is the scaled median log(m/1-m), the blue lines indicate 
lines of constant Hydrogen column, the red lines indicate lines of 
constant temperature. 
The data for the first four frames are statistically similar 
but distinct from the extreme central region (frame 5).}
\label{quantiles_sides}
\end{figure}

Based on the colors of the sources as shown in 
Figure~\ref{fullfield} it is clear that
the center of the field has more embedded and/or hotter sources while the
eastern field seems cooler and less embedded.
We quantify this using the quartile values.
These regions are defined as shown in Figure~\ref{fullregions}. The bulk of
the sources are in the region east of  R.A.\
16:40:19,  A second large region is west of R.A.\
16:39:41.  The final two areas are a region to the north of declination 
-48:48:34 and a region south of declination -48:54:47.  The latter
two region are chosen to exclude sources in the other regions.  Thus
the northern region is roughly coincident with the northern extension
of the warm dust ridge.  In Figure~\ref{quantiles_sides} quartile
values of the sources in cardinal regions of the fields are compared.

The bulk of the sources to the east of the central region
are associated with NGC 6193.  Figure~\ref{fullregions} 
shows a sharp increase in 8~\micron\ emission along NGC~6188 which, at
closest approach, is $\sim 1.2$~pc 
from the central O star of RCW~108--IR (in the IRS~29 complex). 
Based on this, and the distinct change in the
quartile values for data between $\sim 1.15$~pc and  $\sim 1.5$~pc 
the core region is defined to be within about 185\arcsec of the central O star. 
The inner region is defined as a roughly square region $\sim 2.3$~pc  across 
so that all sources are included in one of the five regions. This is essentially coincident with RCW~108--IR and will henceforth be referred to as such. 

The values of  $\log (m/(1-m))$ and $3\times Q_{25}/Q_{75}$  are
tabulated in Table~\ref{quantin} and shown in
Figure~\ref{quantiles_sides}.
The errors for the quartile values are estimated using the method
of  Maritz \& Jarett  (1978) which has been found to reflect the
results of simulations, including those of Hong (2004). 
Composite quartile values for each region are shown in
Table~\ref{Region_Q}.
The difference between the central region and eastern region is very
pronounced and indicates to a large statistical significance 
that these regions contain different
kinds of sources or sources under difference circumstances. 
Comparing the sources among the other 3 cardinal directions 
indicate a similarity among them and that they are intermediate 
between those in the central region and those in the east.

These intermediate quartile values are indicative either of mixing or that
these regions contain different evolutionary stages.  In either event
it is clear that the central region is not simply a foreground object
over a background of younger sources stretching across a field.
Further, it is noted by Wolk \e (2007) that IRAS
16348-4849, which lies about 3.8~pc west of RCW108--IR, is
connected to  RCW108--IR by tendrils of warm dust, which are clear in 
the MSX A--band data (Figure~\ref{smoothzoom}). IRAS 16348-4849 has IRAS
colors consistent with its being a star forming region. Thus,
X--ray sources in the south, north and west could have IRAS 16348-4849 as
their origin.

\subsection{Diffuse X-ray Emission}

Diffuse X-ray plasma has been previously observed in several regions
of massive star formation (Townsley \e 2003). 
It is thought that OB stars excavating 
an \HII\ region 
may generate diffuse X-rays as fast winds shock
the surrounding media (Weaver et al. 1977). $Chandra$ observations have
recently revealed parsec-scale diffuse emission in 
Galactic high-mass star-forming regions RCW~38 (Wolk \e 2002), 
M17 and the Rosette Nebula 
(Townsley \e 2003).  This high energy diffuse plasma
plays a role in the evolution of the cluster by selectively clearing 
gas and dust. 

To search for diffuse X-ray emission in the vicinity of RCW~108--IR, 
we selected 6 circular regions in zones without an X-ray source, within 35 arcseconds of the cluster center (16:40:00,-48:51:40).  The diameter of
the zones  varied from 8\asec\ to 12\asec.  The total area of these 6
regions is 510.5 arcsec$^2$. The CIAO tool $dmextract$ was used to count
the number of photons in the composite region.  The total was 200
photons with a Gaussian error of photons in the composite region of
0.391$\pm$0.027 counts arcsec$^{-2}$.
An annular background region was created centered on the cluster
center with an inner radius of 148
arcsec and an outer radius of 408 arcsec for a total area of
4.36$\times 10^5$ arcsec$^2$ which was clearly not effected by diffuse emission.
Counts were extracted from the region with sources removed.  The
background region contains 133465$\pm$365 counts (0.307 $\pm 8\times
10^{-4}$ counts arcsec$^{-2}$). This suggests a 30\% excess of counts in RCW~108--IR which is about 2 $\sigma$ above the expected level.  This is
only about 20 counts and can easily be accounted for by the $>$
100 stars which are not detected in the X-ray data  but  
should be present based on the detection of 35 X-ray sources in this
highly absorbed region (see \S~\ref{secxlf}). We conclude that there is no
significant diffuse plasma within this region.  
This is consistent with expectations
discussed by Townsley \e (2003) in which they concluded winds and mass
loss consistent with an O6 star or earlier would be required to
generate such a plasma, unless there had been a recent supernova. Here
the central star has been determined to be O9.

\section{Spectral Analysis}

X-ray sources were fitted with spectroscopic
models using {\it Sherpa} (Freeman \e 2001).  
Background regions were generally selected
as annuli around the source regions. Source and background pulse height
distributions in the total band (0.3-8.0 keV)
were constructed for  each X-ray object.  The
final fits were done with CIAO version 3.3.0.1.
The CIAO script {\it psextract} was used to extract 
source spectra and to create Ancillary Response
Function (ARF) and Redistribution Matrix Function (RMF) files which
compensate for local and temporal variations in the ACIS response.
Background corrected data
from each region were fitted with an absorbed one-temperature
Raymond--Smith plasma (Raymond \& Smith 1977). Such models were found by
Wolk \e (2006) to have the lowest residual $\chi^2/{\rm (degree~of~freedom)}
-1$  in a comparison of several available models. The photoelectric absorption 
cross-sections ($\sigma(E)$) were taken from Morrison \& McCammon
(1983).  Models were fitted both unbinned, using the C-statistic 
(Cash 1979) and binned using the data variance form of the
$\chi^2$ statistic ($\chi$--DVAR).  In the case of the binned data, 
data for each source were grouped into energy bins which 
required a minimum of 8 counts per bin and background subtracted. 
  $\chi$--DVAR is a statistic with variance computed from the data and 
is appropriate if the number of counts in each bin is large 
($ >$ 5) and the shape of the Poisson distribution from which 
the counts are sampled tends asymptotically towards that of a 
Gaussian distribution. For the binned and unbinned cases, 
the optimization method was set to Levenberg-Marquardt and Powell
respectively. Absorption and temperature 
were simultaneously fitted for each
observation and the resultant 95\% emergent flux was calculated and
then scaled to 100\%.  Formal flux errors are about 4\% at 2000
counts, about 35\% at 100 counts and errors reach 100\% below
about 30 counts.  Formal errors underestimate the real uncertainty as
demonstrated by the variance in the temperatures and hydrogen columns
measured using difference techniques.  This variance can exceed the
formal errors by about 50\%.

Two advantages of the C-statistic method are that the data
are unbinned and thus sources with a small number of counts may be
fitted and the errors are symmetric and so should not
effect an ensemble analysis of the dataset. A drawback is that  
the goodness--of--fit information suffers when unbinned data are used
(Heinrich 2003).  In the end, we generally used unbinned  C-statistics
for the sources with over 3.5 ``$\sigma$'' significance. 
In a few cases, the temperatures found using C-statistics was 
above 15 keV while  $\chi$--DVAR 
temperatures were more reasonable, below 7 keV.  In other cases  C-statistic
fits were indeterminate in \nh.  In these cases, again    
the   $\chi$--DVAR  results were used. 
The result of the
single--temperature fits using C-statistics are listed in 
Table~\ref{1t}. Column 1 gives the source name from
Table~\ref{XSources}, Column 2 gives the goodness--of--fit to an
absorbed one--temperature Raymond--Smith plasma in terms of 
$\chi^2$/degree of freedom (but note the caveat above).
Columns 3--7 give the fit parameters \nh,
kT, associated errors and the log of the unabsorbed flux.  
The luminosity of each source is calculated based 
on the derived unabsorbed flux and the cluster distance of 
1.3 kpc and listed in column 8.  Fluxes and luminosities are corrected
for the 95\% encircled energy radius. The sources in 
Table~\ref{1t} are subdivided by their region as 
described in \S \ref{sec_xraymorp}. 

Two issues are present throughout the fitting process. First,  there is a
fundamental degeneracy in the fits between \nh\ and kT.  It is
possible that fits of
similar quality can be obtained by increasing \nh\ and
an appropriate increase in the temperature.  While this is especially true in
regions of high absorption such as the core region RCW~108-IR the
temperature range in question is typically about 2 keV. Temperatures
between 3 and 8 keV or less than 1 keV are very robust given enough data.
Second, since there is little
sensitivity to X-rays above 10~keV, model fits are not very reliable
much above this energy range. Because of this, if the fitted
temperatures exceed 15 keV we simply label them ``$>15$'' in the
tables.

\subsection{Two--Temperature Fits}
In many cases of stars with high flux rates, a second thermal
component is clearly needed to fit a deficiency in the temperature
model seen at low energies.
In addition to the single temperature fits,  absorbed two-temperature fits were
made for 73 sources with over 100 net counts. 
Of these, only 54 were
considered true two-temperature fits with temperatures different by
more than 10\% and each component contributing at least 10\% of the
flux. Twenty--two of these were rejected for having poor formal
constraints on the temperatures (formal errors exceeding 10~keV). 
The absorbed two--temperature fit was deemed superior to the
one--temperature fit for 22 X--ray
sources (Table~\ref{2t}). The first six columns of Table~\ref{2t}
are identical to those of  Table~\ref{1t}.  Columns 7 and 8 are the
high temperature component. 
Columns 9 -- 11 are the
log of the absorbed and modeled unabsorbed fluxes. Columns 12 and 13 
list the log of the emission measure for the cooler and warmer 
components respectively and column 14 gives log of 
the luminosity. Again, the sources are subdivided by their region as 
described in \S \ref{sec_xraymorp}.  

The results of the two
temperature fits show that given sufficient flux, about one third of the
sources are better represented as two--temperature models.  The key
datum from these fits is that the global extinction is much better 
constrained.   The eastern sample is large enough (14) to look
at the group statistically.
To the east $<$\nh$> = 0.42\pm 0.24 \times 10^{22}\psqcm$, while in the
center  \nh\ $> 2.4\times 10^{22}\psqcm$. 
Also interesting is that the cool components
are about 700 eV ($\pm$ 30 eV), while the warmer components are about
2.45~keV  ($\pm$ 1.07 keV). This indicates much more variability in the
temperature of  the hot component than the cool component. In the
case of the cool component, this result is consistent with earlier X-ray
studies of young stars which show a stable $\sim$ 800 eV corona distinct from the
high temperature corona (Sanz-Forcada \e 2003). The difference in the high
temperature component may be indicative that the sources to the west
are younger than those to the east. 
The high temperature component in the X-ray sources to
the north and south are more consistent  with those in the east.
The temperatures of the cool coronae in the core region
are harder to evaluate due to the higher absorption present there, 
the high temperature components are similar to those in the west. 
 The variability in the high temperature component may also be
indicative of different levels of flaring.  As will be shown in \S5.1,
several of these sources flared. 

\subsection{Temperature and Hydrogen Column Density Distribution} 
 
Because so few sources were well characterized by two--temperature
spectra, we use the one-temperature fits to analyze the overall
nature of the sources.  We confined the analysis to high quality data
as described below.
Since formal goodness--of--fit is inappropriate for the C-statistic,  
the quality of the fits is evaluated using three metrics.  First the 
net counts need to exceed 50 so that a meaningful fit is possible.
The second and third metrics are the ratio of the
formal errors to the measured values \nh\ and kT.  These act as  
a statement on the quality of the measurements and limits the effect
of errors in the evaluation of global properties. Overall, about 90 spectra
have errors of less than 50\% in \nh\ and kT. The mean \nh\ of these 
$\sim$ 90 sources is 5.3 $\times 10^{21}\persqcm$.  But almost 25\% of the
sources were rejected from this calculation because they were more
than 3$\sigma$ from the mean.  This indicates a very non-Gaussian
distribution of sources -- as expected from the quartile analysis.
To study the changes in the spectral characteristics of the X-rays
sources spatially across the field we divided the regions described in
\S \ref{sec_xraymorp}.

 A global comparison of the spectral fits for the sources in the 
regions is summarized in Table~\ref{Region_Spec}.  In this Table, 
restricted to bright sources with over 50 counts and formal errors in
kT and \nh\ of $<$ 50\%, column~1 indicates the source
region, column~2 indicates the number of qualifying sources in the
region. Columns 3--5
list the mean kT, median absolute deviation 
(MAD; Beers 1990)\footnote{The median absolute deviation is the
  average deviation from
the median measured as absolute values. It is an outlier resistant 
measure of the
 variability, making it useful for describing the variability of data
 with outliers.  
It is a common measure of forecast error in time series analysis.}
and the number of 3-$\sigma$ outliers in each
region. Columns 6--8  list the mean \nh, MAD and the number of
3-$\sigma$ outliers.  This spectroscopic
summary confirms the evidence given by the quartile analysis that the
5 regions contain at least 3 types of sources.  Hot and embedded in
the center, cool and low extinction to the east, while the other 
three regions are intermediate. We note that while thermally the three
intermediate regions are ordered west, south and north (hottest to
coolest) the hydrogen columns are ordered  south, north and west
(highest to lowest). The differences among the hydrogen columns in the
south and west regions are statistically
insignificant.

\begin{figure}[t]
\plotone{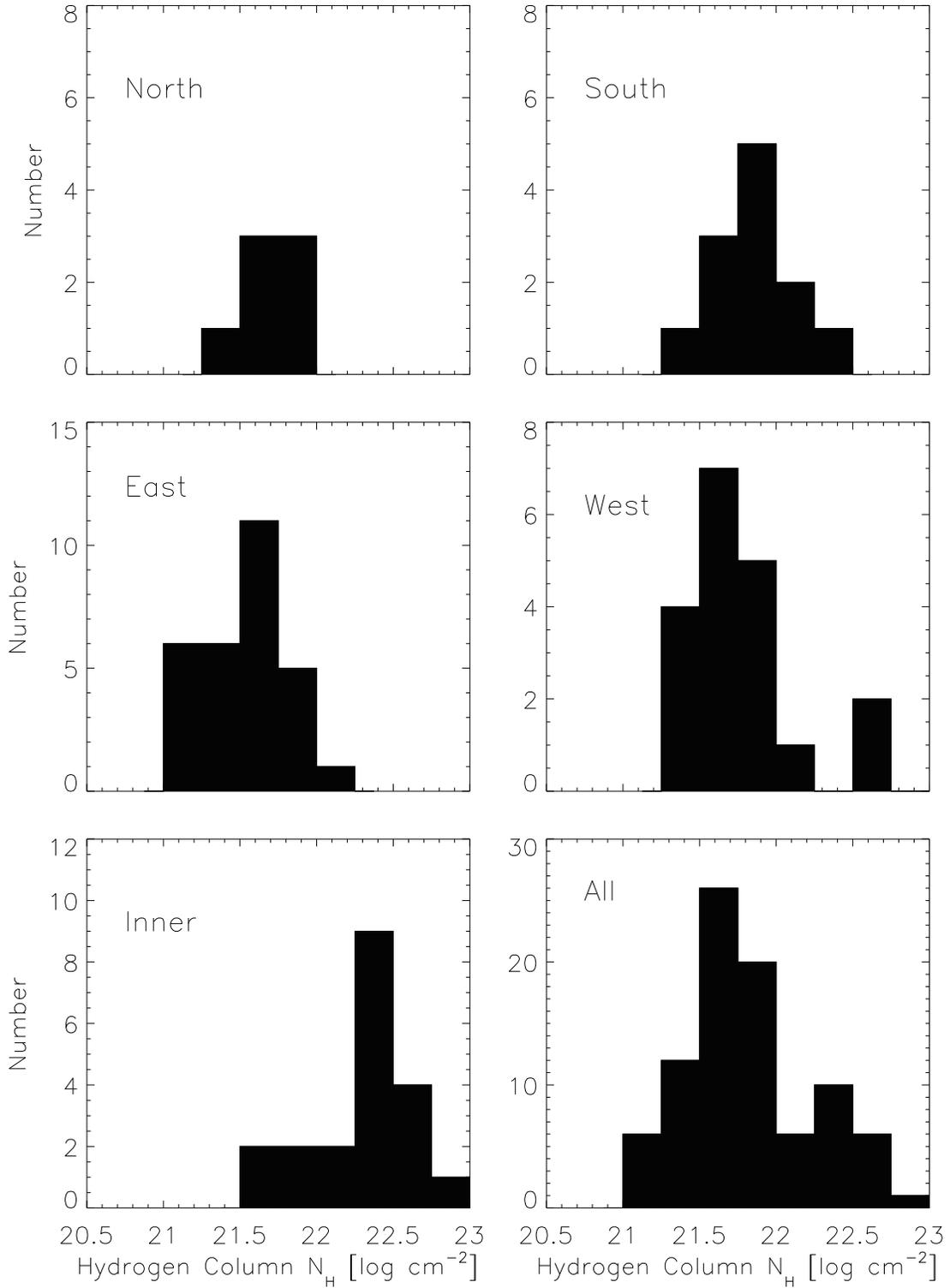}
\caption{Histograms of hydrogen absorption column, \nh, derived
through spectral fits of X-ray sources with over 50 counts and formal errors
$<$ 50\%. Bin size is 0.25 dex. 
 The six panels are marked to
indicate the location of the X-ray sources with respect to the field
of view, see Figure~\ref{quantiles_sides} for details.}
\label{nh_hist}
\end{figure}
\begin{figure}[t]
\plotone{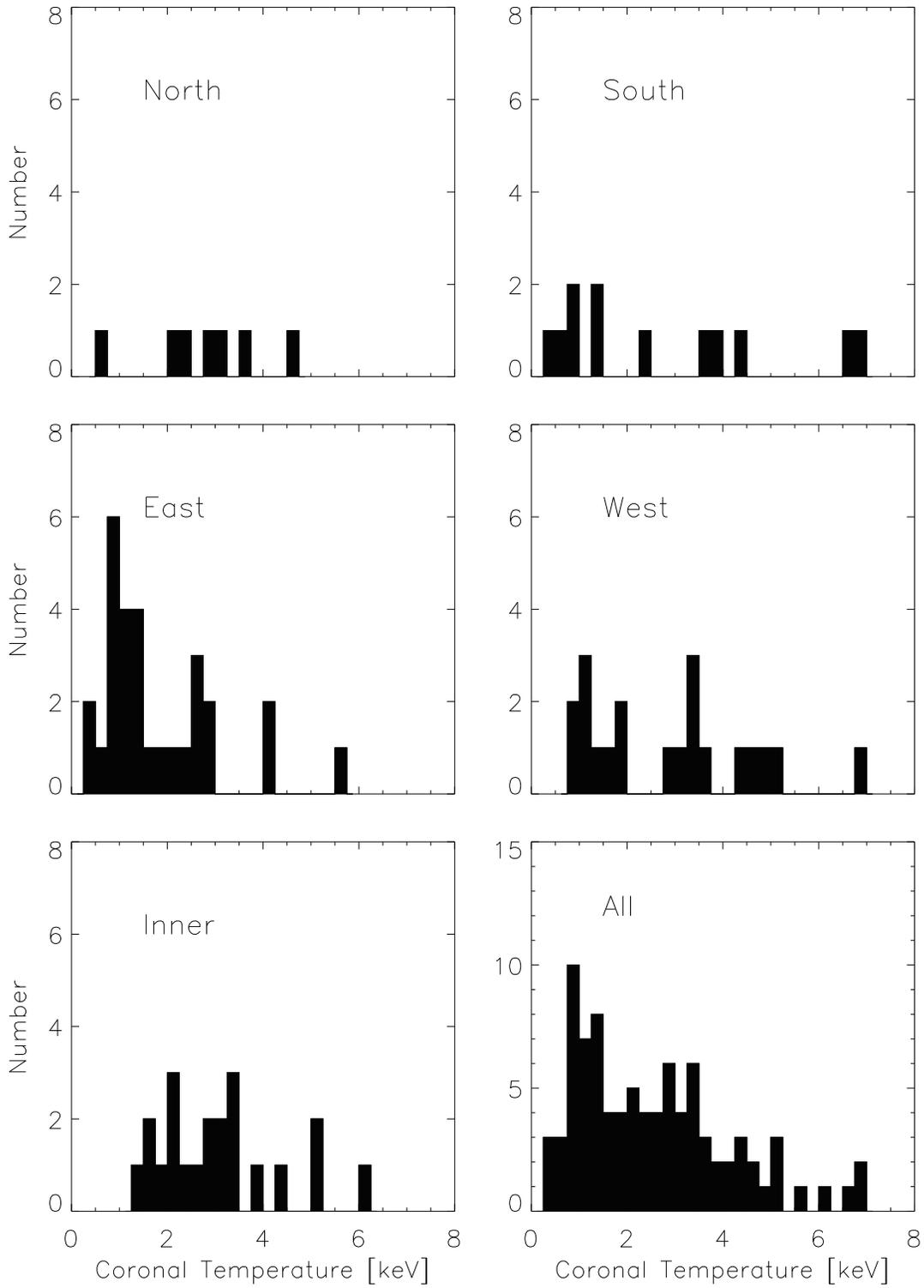}
\caption{Histograms of coronal temperature, kT, derived
through single temperature spectral fits of X-ray sources with over 50 counts and formal errors
$<$ 50\%. Bin size is 0.25 keV. 
 The six panels are marked to
indicate the location of the X-ray sources with respect to the field
of view, see Figure~\ref{quantiles_sides} for details.}
\label{kt_hist}
\end{figure}

These results are displayed schematically in Figures~\ref{nh_hist} and
\ref{kt_hist}. In these figures, the results are broken down region by
region using histograms. From Figure~\ref{nh_hist} it is clear that 
there are no bright
X-ray sources with column densities $< 10^{21} {\rm cm}^{-2}$ in the
entire field. This means that there is little foreground contamination among the
bright sources.  The distinction between RCW~108--IR and eastern
region is clear with moderate overlap in log(\nh) between 21.5 and
22 cm$^{-2}$.  Columns densities to the north are completely consistent with
those in the east while the southern and western regions possess some
more absorbed sources. Figure~\ref{kt_hist} again shows that the
sources in  RCW~108--IR and eastern regions form fairly distinct groups.
There are a few very hard sources in the eastern region. Overall,
there are 8 high quality sources with measured temperatures in excess
of 5 keV ($\sim$ 50~MK), three of these are in  RCW~108--IR , two
are to the south and one each are in the eastern and western regions.  
While the sources in the central region tend to be more absorbed and
hotter, region-by-region analysis shows 
there is no strong correlation between \nh\ and kT.

\subsection{X-ray Luminosities\label{secxlf}}

As clusters age, the luminosity of their members will drop
at different rates for stars of different masses and this will be
reflected in the cumulative distribution of X-ray luminosities
-- the  X-ray luminosity function (XLF).
Feigelson \e (2005) have attempted to understand the X-ray luminosity
function as a global property of star forming regions. For
regions at very young ages, similar to that of NGC 1333 or the Orion Nebular Cluster 
(ONC),  they
find that when high-mass stars are excluded, 
the XLF follows a somewhat flattened
log-normal distribution with a mean of log L$_X ~\approx 29.3$ and a standard
deviation of $\pm$ 1.0.  Our analysis of the XLF of RCW~38 was
consistent with this log-normal distribution (Wolk \e 2006).  

We calculate luminosities for each source using the fluxes and 
line--of--sight absorptions, determined from the C-statistic 
fits and assuming a distance of 1.3 kpc.  While we do not explicitly 
exclude high-mass stars as done by Feigelson \e (2005), we expect them
to be few in number relative to the low-mass stars and to have a
negligible impact which is uniform across all luminosity bins -- 
as seen by  Feigelson \e (2005).
Below about 50 counts, the individual estimates have errors exceeding a
factor of 2.  This is mostly due to the poor determination of \nh.
There is no evidence that such errors are markedly biased by the
C-statistic and thus the XLF derived using data with such low counts
is still valid. We do not include the 87 faintest sources in this
analysis since luminosities were not calculated for them. 

\begin{figure}[t]
\epsscale{1}
\plottwo{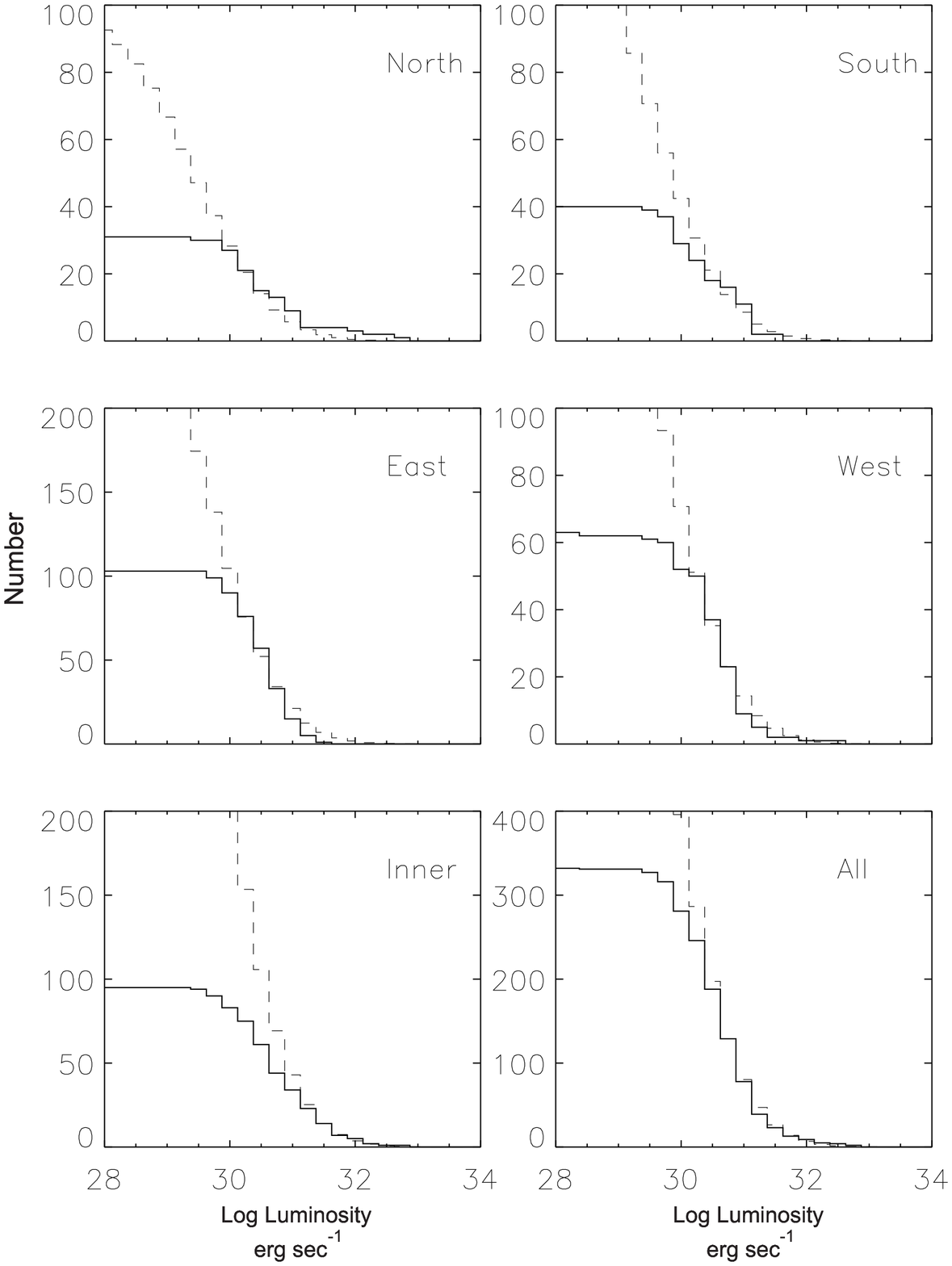}{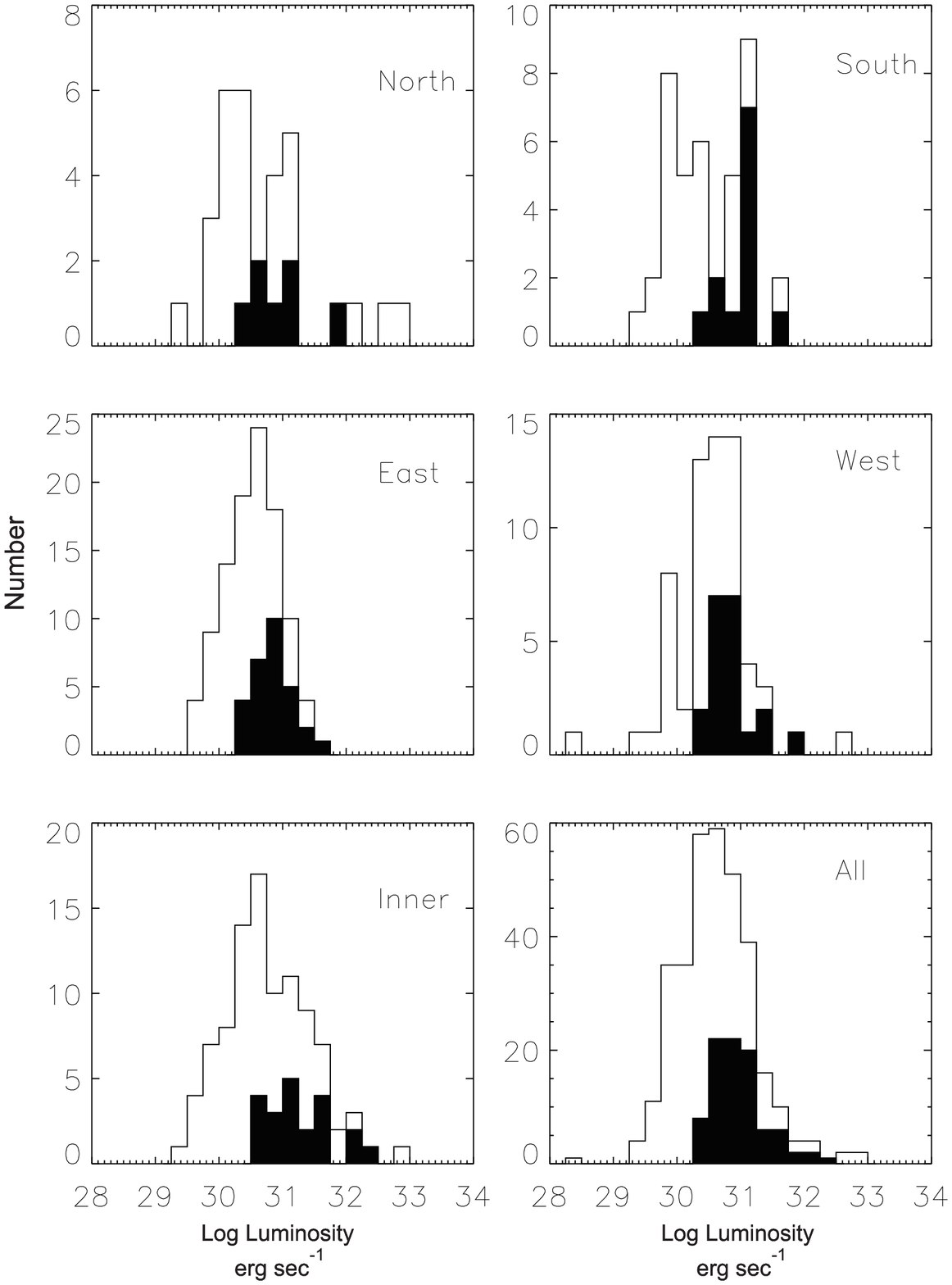}
\caption{X-ray luminosity distributions segregated by region.  Left:
The cumulative distribution for each region is shown as the solid
histogram.  The dotted histogram is from a log--normal function normalized
to equal the observed distribution at log L$_x$ $\sim 30.5$ (31 for
the inner region).  Right: A non-cumulative
distribution of the observed luminosities.  
The solid histogram indicates the luminosity
distribution of bright sources with low errors in kT and \nh. This
figure emphasizes deviations from a Gaussian profile of the XLF in the
north and the south.}
\label{xlf}
\end{figure}

In figure ~\ref{xlf} the XLFs of the 5 regions and the whole field are
shown.  On the left-hand plot the cumulative totals are shown, region
by region.  The dotted line shows the cumulative total of a log-normal
distribution 
along with observed distribution.  The cumulative total of a
log-normal distribution has been scaled to equal the observed data in
the bin  where Log(L$_X$)= 30.5.  We expect the data to complete in
this bin and the adjacent bins (see below), thus log-normal
distribution indicates what 
would be observed in an observation that
was sensitive to all low luminosity sources. 
Fitting the observed and complete portion of
the XLF with the `global' XLF offers the intriguing possibility of
estimating the total cluster size and/or distance.  Since the
distance to this cluster is not in dispute ($\pm 10\% $) we can
estimate the total number cluster members in the field at about 1600
$\pm 200$, with about 800 $\pm 100$ in the central region.

 Such extrapolations become very uncertain due to completeness issues.
 Feigelson \e (2005) provide a rough estimate of the $Chandra$ sensitivity as:
\begin{equation}
 log~L_x = 28.7 + 2 log (d/kpc) - log (t_{exp}/100. ks) +0.4 (log~
\nh -20) {\rm ~ergs~sec}^{-1}. 
\end{equation}
Assuming a distance of 1.3 kpc and estimating log \nh\=
21.5  cm$^{-2}$ in most of the field, we derive a
sensitivity limit of about 
$log L_x=29.6 {\rm ~ergs~sec}^{-1}$ in the bulk of the field and  $log L_x=29.8  {\rm ~ergs~sec}^{-1}$ or
so in the core where we assume log \nh$\sim$ 22 cm$^{-2}$.  This latter
value is clearly too small since we only estimate the absorption to the
sources that we detect.  The true mean absorption to the core region
is much higher, perhaps as high as  23.5 in log based on the IR data. 
This is reflected in 
the cutoff in the luminosity function near log (L$_x)\approx
30.5$. The luminosity is corrected for spotty absorption and we can
estimate completeness based on the location of the break in the
observed XLF from the log--normal form.  This lack of precision in
this completeness estimate leads to the error estimates given above. 

The eastern region lacks the high luminosity tail.  
 Our sample of  the XLF is taken from the southern portion 
of NGC~6193 and appear to be  ``missing'' sources as the high
  luminosity/high mass end.  This implies that the sample is depleted
  in early G stars, relative to K stars. Since we are not sampling the
 central region of NGC~6193, and only the southwestern portion, it
appears that mass segregation has kept the most luminous
members towards the center of the cluster outside of the field of view or
 moved the low luminosity sources away from the cluster center and
 into the field of view. 
Thus, we obtain an incomplete view of this cluster. 
This inference is supported by the X-ray observations of the core of NGC~6193
that were made as part of the HETG observation of the HD~150135/6 O star
 pair at the center of NGC~6193.  Skinner \e (2005) report 43 X-ray point
 sources of which,
 at least five are identified with OB counterparts. This is as many OB
 stars as we identify with over 100 X-ray sources in the southwestern
 portion of NGC~6193. Further, the earlier Chandra observation
is only about 15\% as deep as the observation we analyze here, so a smaller fraction
of B stars were detected by Skinner \e (2005).

The XLF in the west seems to also be biased away from high
luminosity sources though the limited number of stars in the region
makes the determination difficult.  
The XLFs in the north and south regions are increasingly more
poorly described by a log--normal. This is indicative that X-ray selection
is incomplete for these sparse populations.

\section{X--Ray Variability}

In this section, we examine the variability of the X-ray source
population.  Variability studies allow us to begin to assess the
plausibility of X-ray generation mechanisms.  These can be constrained
by the timescales and flux changes observed in the
variability. Variability also offers corroborating evidence that a 
source is stellar in nature and hence a possible cluster member.  
Unlike other X-ray sources almost all coronal sources vary given 
a long enough observation (see Getman \e 2005).  These variations 
can either be stochastic or impulsive deviations from a constant rate.

Various methods can be employed to investigate variability. 
This topic has been reviewed briefly by Wolk \e (2005) 
and more thoroughly by G\"udel (2004) 
and Favata \& Micela (2003). One rigorous technique commonly used is a 
one--sample Kolmogorov-Smirnov (KS) test to identify if the photon
arrival times are consistent with a constant rate.  The KS test does
not give any information on the nature of the variability in objects.
Thus we prefer the use of Bayesian Blocks (BBs; Scargle 1998).  
Bayesian Blocks not only detect variability, but
provide a method of flare detection without the biases inherent in
binning the data. 

These techniques are
discussed in some detail by Wolk \e (2005, 2006), so we do not repeat
the discussion here.  We tested each lightcurve with a ``prior
ratios''  set to approximate both 95\% and 99\% confidence that a 
flux change had occurred.  Overall 61 sources on the I array 
required more than one BB at greater than 95\% confidence.
We also tested the Gregory \& Loredo (1992) method.  
This method uses maximum-likelihood statistics and evaluates a
large number of possible break points from the prediction of
constancy. The number of sources found to vary at a given confidence 
level agreed with the Scargle method, which is still our method of choice. 
The variability results using BBs are tabulated in column~8 
of Table~\ref{XSources} which lists the number of BBs for each
source.  A number greater than one indicates that the source varied
at the 95\% confidence level during the observation.

\subsection{Flaring} 
\begin{figure}[t]
\epsscale{.9}
\plotone{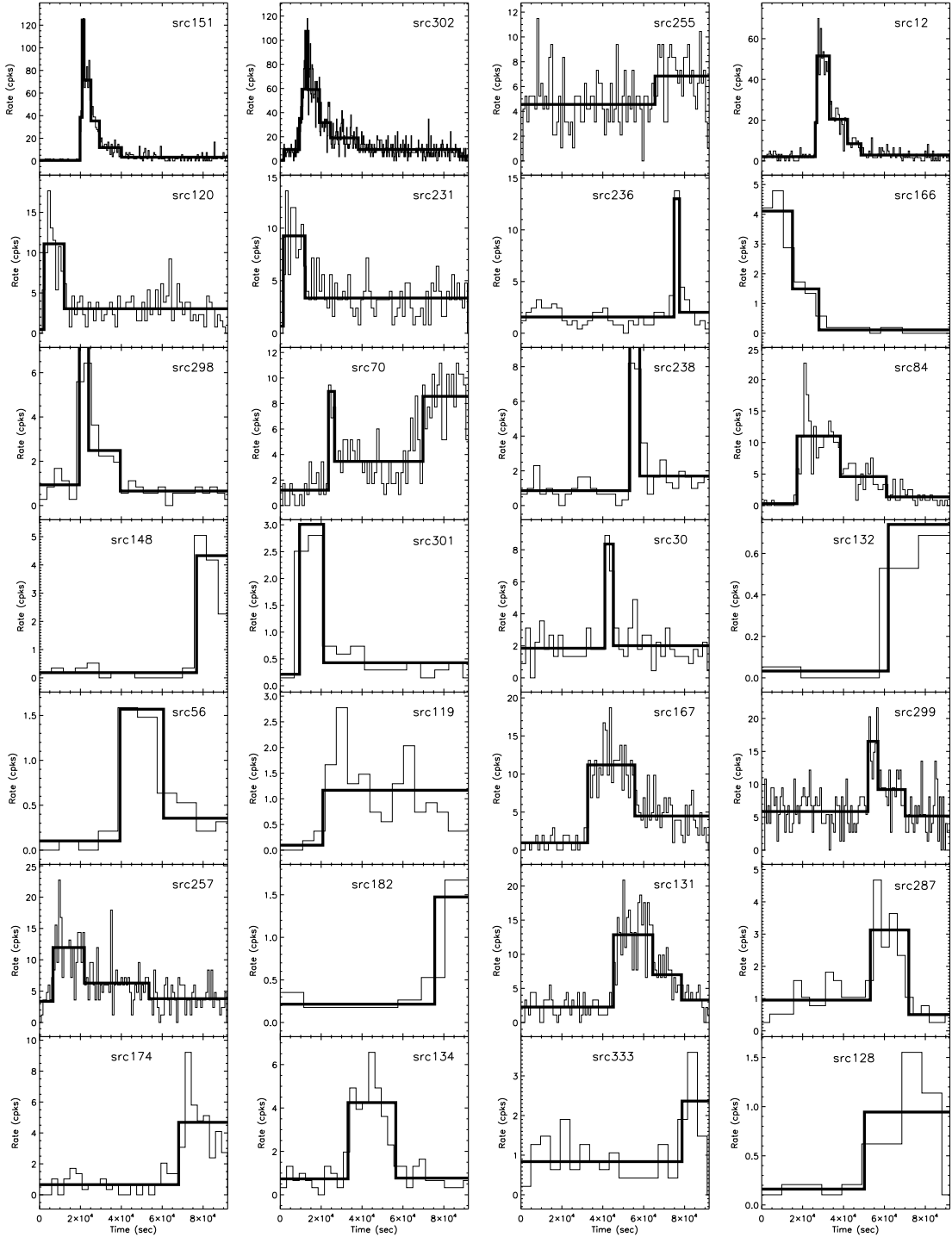}
\caption{The 28 sources seen to flare. The lightcurves are plotted
  with an average of 5 (raw) counts per bin with the Bayesian
  blocks (95\% confidence) overplotted as the thick line.  Sources are
  ordered by the metric for flare intensity, $\Delta$ (see text), 
decreasing across and then down.}
\label{flares}
\end{figure}

Amongst the brightest 30  sources (those with over 200 total counts), 
20 (66\%) were variable at $>$ 95\% confidence.   
The variability rate drops among the 56 sources with between 100 and 200  net counts, of which 
12 (21\%) were variable at $>$ 95\% confidence.  
Overall,  41\% of the bright sources are detected as variable in 88.8 ks.
The BB method converts the lightcurve to temporal periods of constant
flux, thus, one can measure the amount of rise between the blocks and
estimate the rate of change between blocks.  In their study of the
extremely deep COUP data set, Wolk \e (2005) found that most stellar
sources have a characteristic rate, R$_{char}$ and found sources were
at their characteristic levels for about  75\% of the time.  They
further found that a normalized rate of rise  
\begin{equation}
\Delta \equiv 1/R_{char}\times dR/dt > 10^{-4} \rm{s}^{-1} 
\end{equation}
was indicative of a flare.   
Following Wolk \e (2005) we define dR as the difference between the rates of
adjacent blocks and dt as the shorter of the two blocks. We choose the
value of the minimum block as the characteristic rate since we do not
have the long observation time to define a true characteristic level. 
We also require the latter period to have a higher count rate --
although the opposite may be indicative of a post flare decline.
 Twenty-eight sources have this form of impulsive variability (see 
Figure~\ref{flares}).  No star was seen to flare twice. Thirty--four
other stars were detected as variables (with 95\% confidence) but were
not seen to flare.

\subsection{Flare Rates}
These data can be used to assess flare rates.  We  
constrain the following analysis to sources with over 100 counts
to prevent biases since we are less sensitive to flares in stars with
low counts.
Thirteen of 72 (18\%) probable cluster members with over 100
counts flared in $\approx$ 90~ks. 
Assuming that all stars are the same and that there are no
stars more prone to flare than others then we conclude there are about
510~ks between flares.  This is similar the values obtained for
Solar mass stars in the ONC (one flare per 640 ks; Wolk \e 2005) 
and the very young and embedded cluster RCW 38 (one flare per 775 ks;
Wolk \e 2006).

A 100 Myr cluster, NGC~2516, was studied using a similar method and
found to have a
somewhat lower flare rate, about 1 per megasecond (Wolk \e 2004).
The flare rate might also be lower due to a sampling bias present in
the NGC~2516 study -- the lack of a low count limit.   Indeed, when we
calculate a flare rate base on 28 flares observed on 337 stars in the
full RCW108 field, the
composite flare rate seen is about 1 per 1.11 megasecond.  A similar
pattern was recently found by Caramazza \e (2007) when examining
low mass (K \& M)  stars
in the COUP field -- flare frequencies are lower in low count rate
samples. This occurs because the sensitivity to flares depends on
source statistics.  When examining low mass stars with between 100 and 200
counts, Caramazza \e find a flare frequency of about 1/680 ks -- very
close to the rate we find at similar count rates in the full RCW~108
field.  The COUP rate is about 1/800 ks below 100 counts.  
This is somewhat higher than the 1 per 1100 ks reported here.  
But the data presented here have not been filtered to exclude
high mass stars and background objects.

\section{Infrared Properties of X-ray Sources} 

Near-infrared (NIR) data were taken from three sources, the 2MASS
catalog, NTT/SOFI observation of Comer\'on \e (2005) and VLT/NACO adaptive
optics data over the very central region of RCW~108--IR (Comer\'on \&
Schneider 2007).  
The latter data indicate NIR counterparts to 5
additional X-ray sources and show that at least 3 X-ray sources are
associated with near--IR pairs. 
These data are very recent and included solely for completeness.  The NTT/SOFI data
cover a 13\arcmin $\times$ 13\arcmin\ ($\sim$5~pc) field centered on RCW 108-IR with
a somewhat deeper exposure for the central 5\arcmin ($\sim$1.9~pc).  
The observation
and reduction of these data are discussed by Comer\'on \e (2005).
Each X-ray source was matched to the nearest counterpart in each list.
Matches were accepted if the offset between the X-ray and IR sources  
fell within the following empirically
determined curve.
\begin{equation}
\ {\rm offset} < 0.0003~\theta^{1.55} + 1\arcsec \
\end{equation}
where ``offset''
 is the distance in arcseconds between the X-ray 
and NIR position and
 $\theta$ is the off--axis distance of the X-ray source, also in arcseconds. 
This curves allows offset of about 1\arcsec\ on-axis. We impose a hard
limit of 2.5\arcsec\ offset toward the edge of the field.

Once good matches were determined for each catalog the NIR catalogs were merged. 
When NIR counterparts were well observed by both primary input 
sources astrometry was
taken from the global 2MASS program and photometry was taken from the
deeper study by Comer\'on \e (2005) which lacks high precision astrometry.  NIR
counterparts were detected for 303 of the 337 brighter X-ray sources 
on the imaging array (about 90\%).  
Among the additional 87 faint sources listed on Table~2, 56 (65\%)
had matches in either to 2MASS or Comer\'on \e catalogs. This gives
confidence to the veracity of the detected faint sources.  The lower NIR
detection of the faint sources has multiple origins. First of all, the
likelihood of false X-ray detections is certainly higher. Within 4\amin,
6 counts is a clear detection at the three-sigma level. However, as 
$\theta$ becomes higher the area and the background count level
increases, hence more counts are
required for a detection. The matching radius also goes up so the
possibility of a false match increases.  If the IR sources are spread
uniformly over the field, then the area available for matching goes as
the square of the offset in eq 4. Thus, the chance of
random coincidence goes up nearly tenfold as $\theta$
increases from 60\arcsec\ to 240\arcsec.
This is
mitigated by the fact that the NIR and X-ray sources {\em are} centrally condensed.
Secondly, since the
X-ray luminosity loosely follows the bolometric luminosity, the NIR
counterparts of faint X-ray sources also become fainter. 
The mean J magnitude of the X-ray faint
group is 15.8 while the mean J magnitude of the nominal X-ray group is
14.4.  This is especially a concern in the 40\% of the X-ray field not
within the NTT/SOFI field of view since the 2MASS data have 
a brighter limiting magnitude.
Finally, as the fluxes get fainter, detection of background objects such as AGN
become more likely.

The NIR data for all X-ray sources are
tabulated in Table~\ref{nir}.  In this table,  column~1 is the
X-ray source ID, columns 2 and 3 are the RA and Dec of the NIR source,
columns 4 and 5 are the offset between the  X-ray and NIR position and
the off-axis distance respectively.  Column 6-11 are the J, H and K$_s$
band magnitudes and errors.  For photometry taken from 2MASS
the last column carries the 2MASS
quality flags, otherwise if the photometry is taken from the NTT or
VLT data
the notation ``NTT'' or ``VLT''  is used.   Table~\ref{nir} is broken up into the five
cardinal regions. 

\begin{figure}[t]
\epsscale{.8}
\plotone{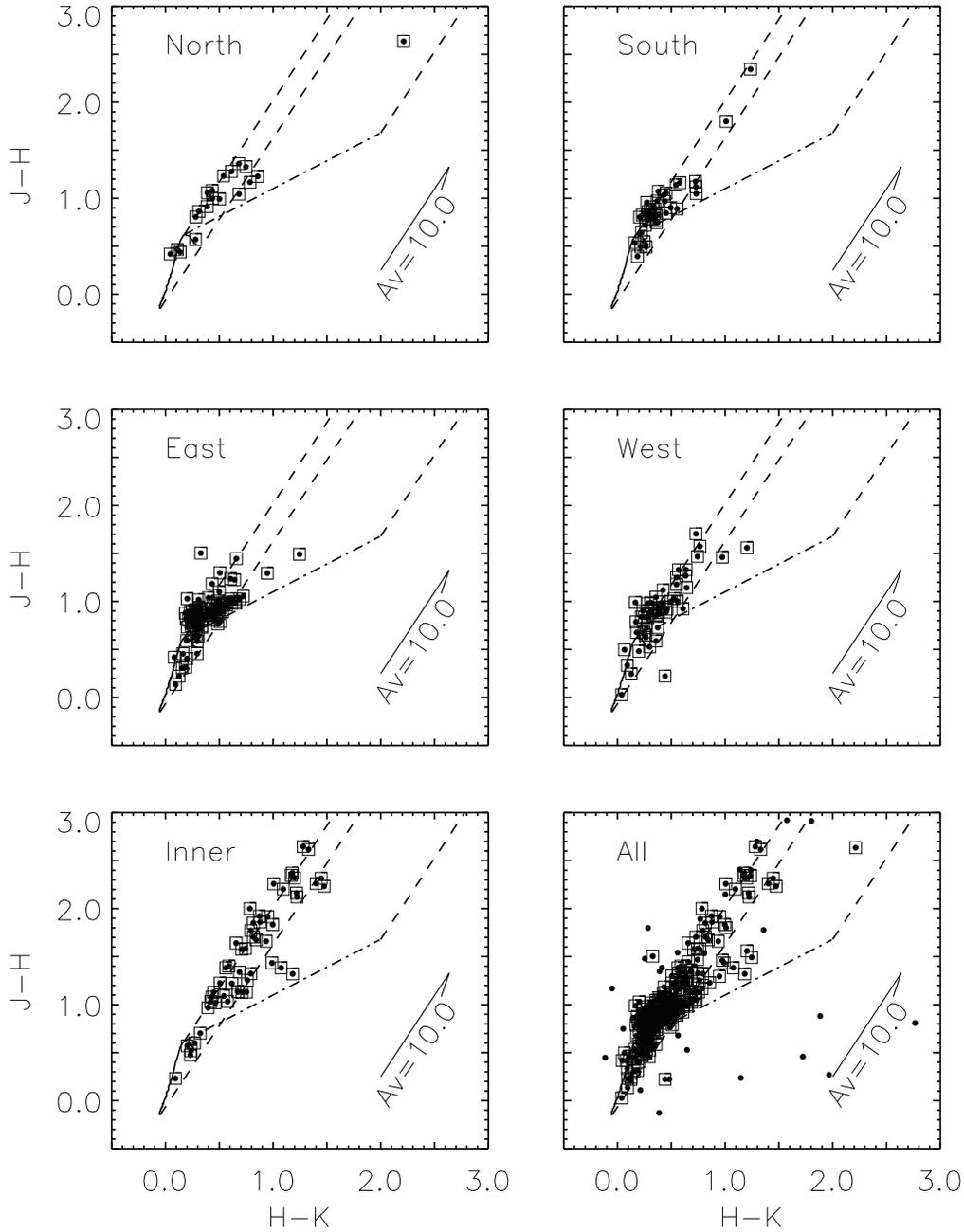}
\caption{Infrared color-color diagrams (J-H vs. H-K) for counterparts to all 
X--ray sources in the RCW~108 $Chandra$ field.  For each panel, the
solid line indicates the main sequence, the dashed line runs
parallel to the reddening vector and the dot-dashed line is the cTTs
locus (after Lada \& Adams 1990 and Meyer \e 1997). A reddening vector
indicative of A$_V = 10$ is marked.  Sources marked as points have
errors $<$ 0.25 magnitudes.  Sources marked also with squares have
errors of $<$ 0.10 magnitudes at K$_s$. The six panels are marked to
indicate the location of the X-ray sources with respect to the field
of view, see Figure~\ref{quantiles_sides} for details.}
\label{irccd}
\end{figure}

These data are used in Figure~\ref{irccd}, which shows NIR
color--color diagrams for each of the regions.
RCW~108--IR has the
highest extinction, with some sources exceeding A$_V$ of 20.  The
western region also shows significant extinction with  A$_V$
approaching 8.  The other
four regions have more mild extinction but contain a few stars with
anomalously high extinction seeming to
indicate that they may be members of  RCW~108--IR.  We also note
the lack of background giants, these are usually indicated by highly
absorbed sources that lie just above the upper track of the reddening
box -- which is set for normal dwarfs and not giant stars.   The only
region that shows such sources is  RCW~108--IR and we conclude
such sources as certainly associated with the active star formation in
RCW 108--IR.  In the remaining fields there are few background giant
candidates among the X-ray sources.

\begin{figure}[t]
\epsscale{.8}
\plotone{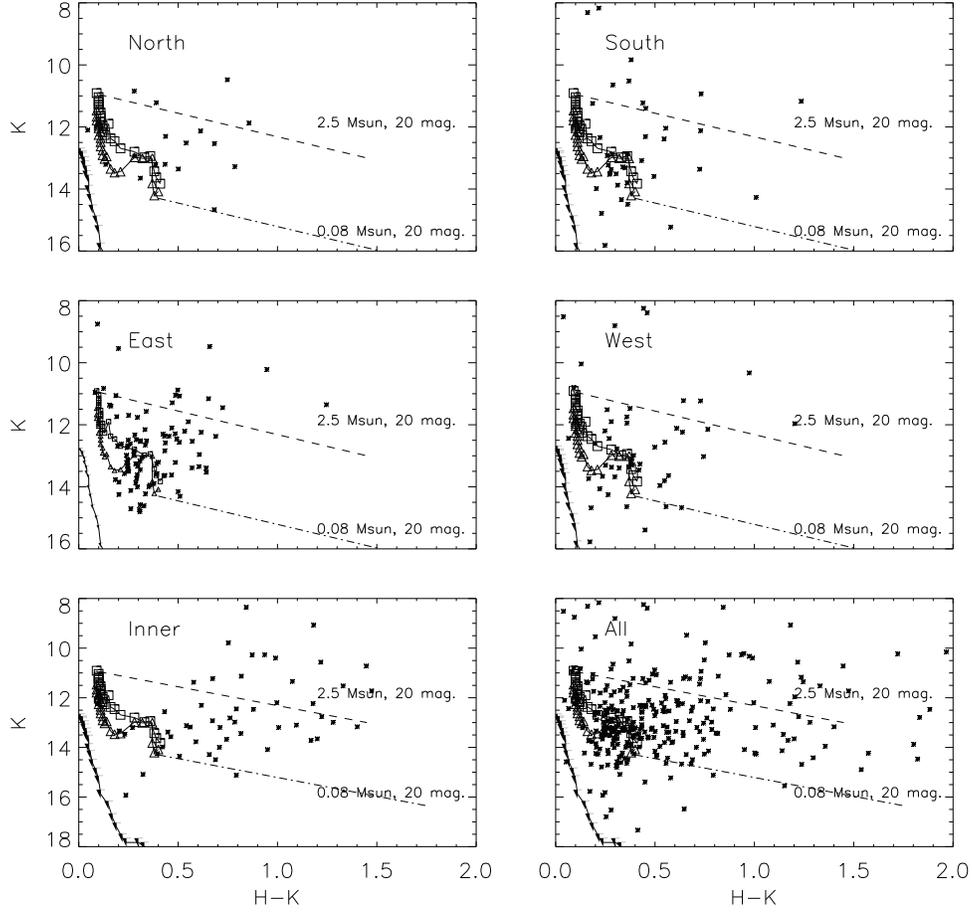}
\caption{Infrared color-magnitude diagrams 
(K$_s$ vs. H-K$_s$) for counterparts to all 
X--ray sources in the RCW~108 $Chandra$ field.  For each panel, 
the 0.5 Myr (open squares) 1.0 Myr (open triangles) and ZAMS (filled circles) isochrones for 2.5 M$_\odot$ to 0.08 M$_\odot$ are plotted (Siess \e 2000).  
Extinction of 20 visual magnitudes for a 2.5 M$_\odot$ star
indicated by the dashed lines. The dot dashed lines indicate 20 visual
magnitudes for 0.08 M$_\odot$ stars at ages of 0.5 Myr.  High mass
stars are predominately seen in the inner region.}
\label{ircmd}
\end{figure}

The IR color--color diagram provides global information about
 reddening, but does not provide much insight about the masses of the
 various sources.  To get a sense of the masses of the stars in the
 field we use the IR color--magnitude diagram shown in Figure~\ref{ircmd}.
 This figure uses isochrones from   Siess \e (SDF; 2000)
 set at  1.0 Myr and the ZAMS for reference only. 
The dashed lines indicate 20  magnitudes of visual extinction.  
The stars to the east are very
 tightly confined and most are consistent with little extinction.
 These stars range in mass but the bulk appear below the 2.5 M$_\odot$ line.
Stars in the north, south and west are more scattered.   RCW~108--IR 
shows a near perfect anti-correlation with the east.   RCW~108--IR
 also shows a large number of sources with masses above 2.5 M$_\odot$.
Either the stars in this region have a very different mass function,
 or we are only seeing the most massive sources due to extinction. 

Following
Comer\'on \e (2005) we can assess the amount of infrared excess using
the reddening free quantity: 
\begin{equation}
Q=(J-H) - 1.70 \times (H-K_s)
\end{equation}
Column 12 of Table~\ref{nir} lists $Q$ for each source.
Values of $Q <-0.10$ are indicative of an infrared excess consistent
with a disk.  Overall about 33 of the over 330 X-ray sources with
good infrared colors ($\approx$ 10\%) have  $Q <-0.10$.  There are
about 55 sources with  $Q <-0.10$ and no X-ray detection. There is some
field--to--field variation.  In  RCW~108--IR 
the fraction is somewhat higher, 13 out of 81 (16\%). The
eastern and southern regions have the lowest fraction of low $Q$ sources
at about 6\%. The fractions in the west and north are 10\% and 13\% 
respectively. From all of these assessments,
it would appear that a relatively small fraction of
sources have disks which are optically thick at K$_s$.  These results
are consistent with Comer\'on \e who found 87 such sources among over
4300 in the central field.  However $Q$
cannot distinguish among stars with less extreme optically thick disks. 
For this, mid-IR data are useful.

\subsection{Mid-Infrared Properties of X-ray Sources} 
For each source in the $Chandra$ source list, the nearest IRAC
counterpart was identified. 
 For X-ray sources with off--axis angle ($\theta$) less than
 200\arcsec, 
the maximum X-ray/IR offset allowed was 1\arcsec.
For off-axis angles between 200\arcsec\ and 600\arcsec, the maximum offset was 2.5\arcsec.
Overall, 236 X-ray sources were matched to
the IRAC source list which included about 2600 detections. 
Of the 236 counterparts, 33 of these are associated with stars with
normal photospheres while there are 58 Class~II objects and five Class~I
objects. Two are identified as probable galaxies.  
The remaining 140  sources are of unknown type due to incomplete data in
the IRAC channels. Many of these sources only have detections in 1 or
2 bands.  

\begin{figure}[t]
\epsscale{.8}
\plotone{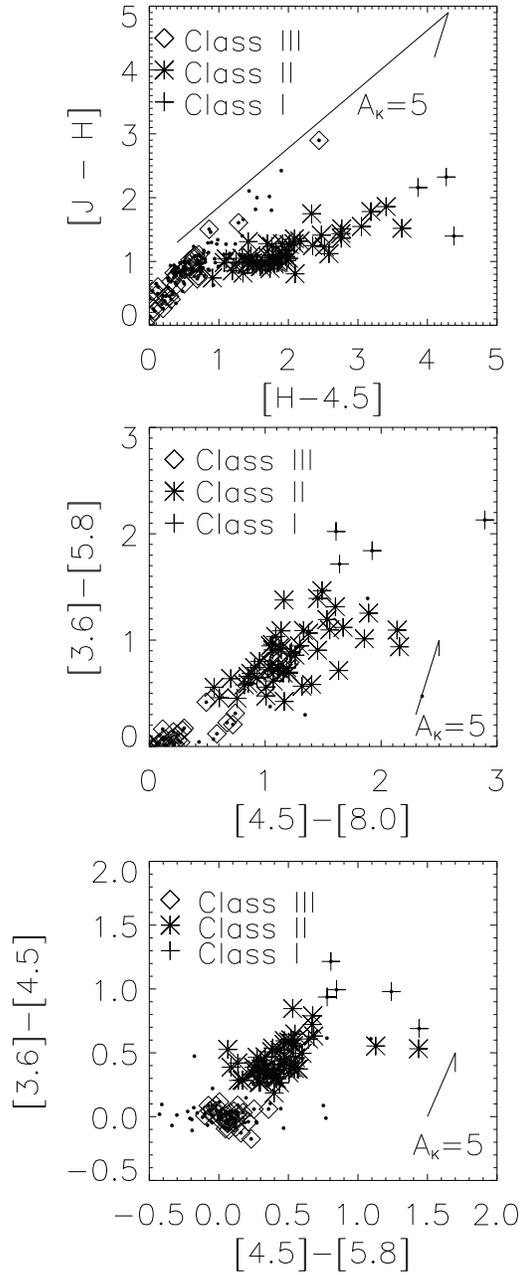}
\caption{Plot of near and mid infrared colors for 236 sources with
  detections by 2MASS and IRAC.  Class~I sources are indicated by
  ``+'', Class~II by ``*'', Class~III sources are diamonds.  Sources
  of unknown type or suspected galaxies are marked with small dots.}
\label{IRACCCD}
\end{figure}

The infrared
characteristics of the X-ray sources are listed in Table~\ref{XIRAC}. 
The IRAC Class~I sources are particularly intriguing as they should identify sites
of the most active star formation.    The key color for the
determination of Class~I sources via IRAC is the [4.5]-[5.8] color
($>$0.5) with [3.6]-[4.5] color being used as a secondary indicator.
This is shown in the bottom of Figure~\ref{IRACCCD}.
The color cutoffs are determined empirically from low mass young
stellar objects (Gutermuth \e 2007).  
The Class~I sources are clear as outliers in
H-4.5\micron\ color (top of Figure~\ref{IRACCCD}).  However, as discussed in
\S~\ref{OBCAN}, the Class~I status derived from this color index is dubious
since most of the Class~I candidates are OB star candidates and the 
empirical colors used for class determination come from low--mass stars.

We cannot detect 
transition disks since we lack 24~\micron\ data.
The IRAC data are also not very deep, complete to about magnitude 12
in channel 1 [3.6\micron]. This is about magnitude 12 in $K_s$ for a Class~III
object. These are identified as having [3.6]-[5.8] and  [4.5]-[8.0]
colors of near zero, $< 0.4$ and $< 0.75$ respectively.  Assuming a distance modulus of 10.6,
this is about abs(K)=1.4 or about 1.2 M$_\odot$ at 1 Myr (Siess, 
Dufour \& Forestini 2000).
One result of the shallow nature of the IRAC map is a bias toward detecting
stars with disks.  Hence, even in the nominally older eastern region, 68\% of 
IRAC detected sources for which disk class could be determined were
found to have disks -- meanwhile no Class~III objects were detected in
 RCW~108--IR. Because of biases inherent in the shallow IRAC data, 
we focus on the deeper JHK data for most of the further analysis.

\subsection{A$_V$ --\nh\ Relation}
If the IR data are of high enough quality (Mag$_{err} < 0.2$) 
they allow a tentative extinction estimate.  Following Wolk \e (2006),
we simply calculate the
amount of reddening required to move the object from the observed
location on the JHK color--color diagram to a location consistent with
zero reddening.  The zero reddening location is different for
a star with and without a disk and varies depending on spectral type
for pure photospheres.
To calculate the extinction, we first assume the least 
absorbed case, that all the stars have disks and deredden the
photometric colors until they intersect the cTTs locus (Meyer \e 1997). 
This is the only possible solution for stars outside the
pure reddening box (Lada \& Adams 1992).  We calculate the extinction for 
stars inside the box for each of three 
hypotheses (disk, M star, or higher mass star).
For stars which lie either below
the cTTs locus of to the left of the main-sequence or outside of the
pure reddening box by more than 2$\sigma$,  no estimate of
extinction is possible.
The error in the extinction estimate of stars under the higher mass 
hypothesis is larger because the angle between the
ZAMS and the extinction curve is very shallow.

To choose among the three hypotheses we use the measured \nh\ column.
In our previous work (Wolk \e 2006) we found 
 $N_{\rm H}/A_V = 2.0\times10^{21} {\rm ~cm}^{-2}$.
This is intermediate to results derived using $ROSAT$ data (Ryter
1996) and those of Vuong \e (2003).\footnote{$N_{\rm H}/A_V =
  1.6\times10^{21}{\rm ~cm}^{-2}$ 
was found by Vuong \e (2003). This value lead
to the same result, in terms of hypothesis choice, but has a larger
residual, in terms of the value of A$_V$ as derived from \nh\ minus the
 A$_V$ measure directly from the IR data (Wolk \e 2006).}
The expected \nh\ is calculated for each of the three hypotheses noted
above  and the one with the minimum difference with the fitted \nh\ is 
chosen as the ``correct'' extinction.  

One hundred and twenty--nine stars have both good JHK$_s$ measurements 
(with K$_s$ errors $<$7\% in about 100 cases and $<$20\% in all cases) 
and X--ray spectral fit residuals of $< 30$\%.  The 
mean deviation between the A$_V$ calculated by converting
the \nh\ column and the values calculated with the color--color diagram
technique (with outliers removed) was about 0.45 visual magnitudes.
The A$_V$ and the preferred model hypothesis (disk, M star, higher mass
star; HMS) are listed in columns 13 and 14 of Table~\ref{nir}. In cases
where the \nh\ is not constrained by at least 50\% by the X-ray spectral fits, 
no preference is given (``--'') and the minimum A$_V$ is listed in the table.

We can test the robustness of our results by comparison with
extinction measurments
published by Comer\'on \e (2005).
They publish extinctions for six sources in our ``inner'' region using the
Rieke \& Lebofsky (1985) extinction law and the (H-K$_s$) color index.
In all cases, our results show less extinction that that reported by Comer\'on \e (2005)
The mean difference was 2.8$\pm$1.3 magnitudes of visual extinction. 
An explanation for some of the deviation is that all of the
comparison stars are assumed by  Comer\'on \e to be O or B stars, while
our data found that three of the stars lacked sufficient precision to
deredden the stars in a manner consistent with their being an O star.
In terms of the K-band luminosity the differences are small,
less than 0.2 magnitudes at K$_s$. 

The extinctions and hydrogen columns 
for the stars with good X-ray spectral fits are summarized
in Table~\ref{Region_avnh}.  In this table, column~1 indicates the source
region, column~2 indicates the number of sources in the region with
\nh\ error $< 0.3$ and photometric errors below 20\%. Columns 3--5
list the mean A$_V$, MAD and the number of 2-$\sigma$ outliers in each
region. Columns 6--8  list the mean \nh, MAD and the number of
2-$\sigma$ outliers.  The \nh\ values listed here differ slightly from
those listed in Table~\ref{Region_Spec} since measurement error and not
flux rate is used as the inclusion criteria. Among the high quality
measurements with \nh\ errors $<$30\% and photometric errors $<$ 7\%,  
75 (58\%) are found to
be most consistent with having an optically thick disk at K$_s$. These
are much higher rates than indicated by the Q
parameter alone but below the IRAC determined rates. 
The disk fraction is highest in  RCW~108--IR 65\%
(24/37), but not significantly so.  This may indicate a bias in bright
X-ray sources possessing disks.  This would run counter to results from
the Orion nebula studies which found X-ray sources with disks to be 
systematically fainter than those without disks (Kastner \e 2005). It is more likely an age
effect. We note that only 37 of 105 sources in the eastern region were
bright enough for X--ray spectral fits with residuals of $< 30$\%.
Meanwhile 47 of 76 X-ray sources in  RCW~108--IR could be fitted
at that level of accuracy.   Overall we obtain a 
higher disk rate, 70\% if we additionally limit ourselves to the 42 
cases in which the net number of X-ray counts exceed 100.  The fraction
of optically thick disks as determined by the $Q$ parameter is 12\%
under these restrictions.
This indicates there is no bias in the $Q$ parameter with respect to the
X-ray data.  

The derived values of \nh\ and  A$_V$ were compared among the
regions. We fitted \nh\ as a function of A$_V$ using an 
outlier-resistant two-variable linear regression,\footnote{The IDL
procedure robust\_linefit.}  weighting the variables used in the fit
proportionally to their distance from the fit line. The results,
summarized in Table~\ref{Region_avnh_rel}, are consistent with 
 $N_{\rm H}/A_V = 2.0\times10^{21} {\rm cm}^{-2}$ 
and generally indicate no significant
change in the gas--to-dust ratio among the regions. This is
consistent with our previous result in RCW~38 (Wolk \e 2006) of 
 $N_{\rm H}/A_V = 2.0\times10^{21} {\rm cm}^{-2}$ 
with a MAD of
0.29. For the present observation the region to region MADs exceed
that of  RCW~38, but
the fit of the overall dataset is the same, with a smaller MAD.

\begin{figure}[t]
\epsscale{.8}
\plotone{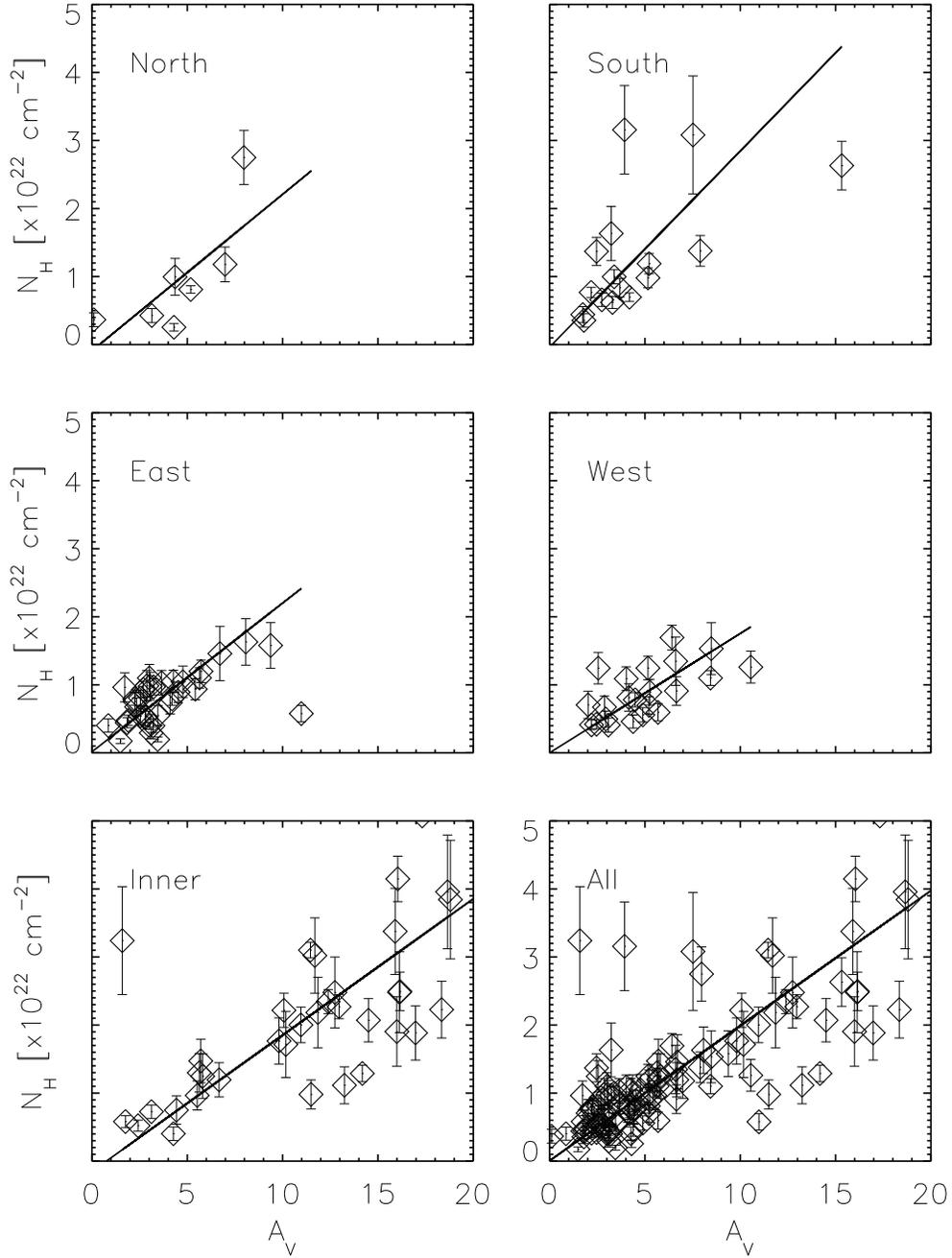}
\caption{Plot of extinction derived from JHK colors vs. \nh\
  column. Hydrogen column is measured via X-ray spectral fits, these
  have been used to choose from various possible NIR extinction
  measures.  The lines are fits to the data.}
\label{avnh}
\end{figure}

However, an anomaly is seen at high values of A$_V$ in RCW~108--IR
shown in Figure~\ref{avnh}.  
While the bi-square weighted fit to these data is
 $N_{\rm H}/A_V = 2.2\times10^{21} {\rm cm}^{-2}$,  
there is a set of
outliers with  A$_V > 11$ and \nh\ significantly below the
fit. Others sources at high  A$_V$ fall directly on the fit line. 
Two additional sources, one to the east and one to the south also fall on this
secondary line.  A similar
behavior is seen in Serpens (Winston \e 2007) and NGC 1333 (Getman \e
2002, Winston \e $in~prep.$), where at high  A$_K$
sources seem highly depleted in gas. Winston \e (2007)  point out that an alternate
possibility to gas depletion is a flattening of the reddening law at
high extinction. A third possibility, which stems from the 
fundamental degeneracy in the fits between \nh\ and kT, is that the sources are at much
higher plasma temperature than the X-ray fits indicate. This would require a
higher \nh. 
In RCW~38, Wolk \e (2006) report
five sources with high quality A$_V$ and \nh\ measurements with A$_V
>$ 12, none show significant deviation from
$N_{\rm H}/A_V = 2.0\times10^{21} {\rm cm}^{-2}$ .

\subsection{The K--band Luminosity Function}

\begin{figure}[t]
\plotone{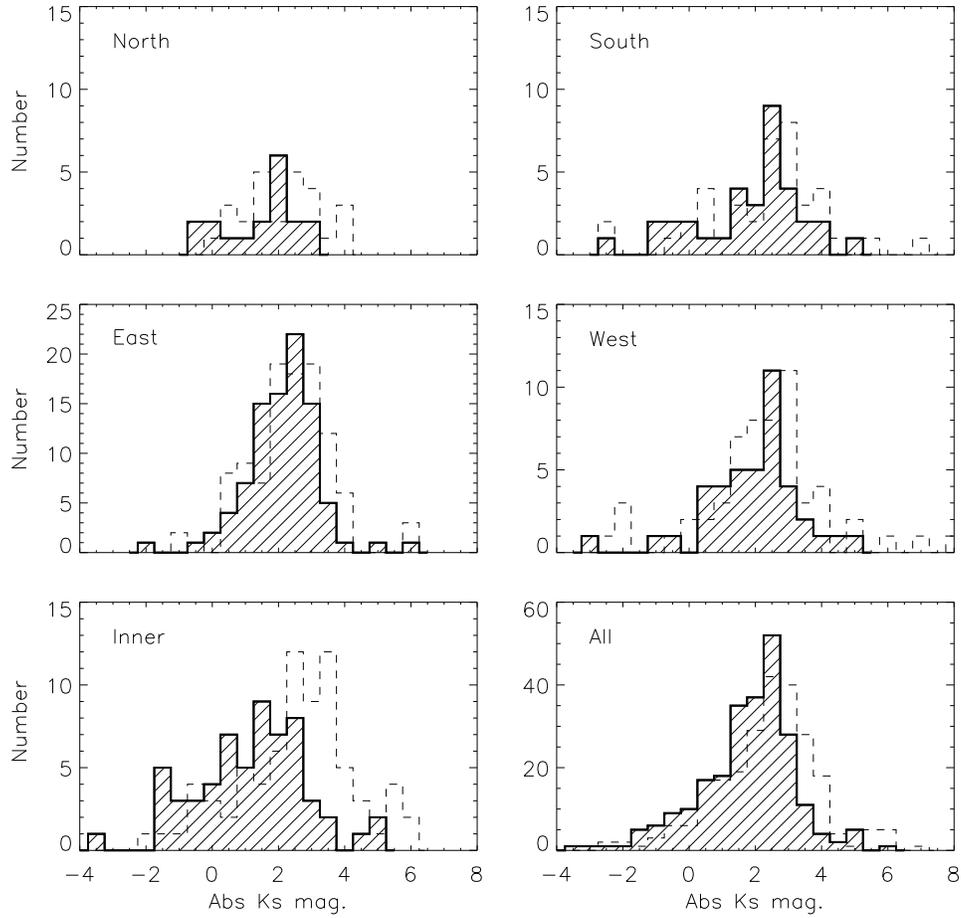}
\caption{K$_s$ band luminosity functions for the X--ray sources in the
RCW~108 $Chandra$ field. The observed magnitudes are corrected  for
the distance modulus of 10.63.  The hatched region in the histograms indicated
the data are also corrected for the 
measured extinction. While the
most luminous sources are all toward the inner region, there are
several sources with absolute K$_s$ magnitudes brighter than 0 in the
western and southern regions.}
\label{klf}
\end{figure}

We create an intrinsic K-band luminosity function (KLF)  
by correcting the observed K$_s$ band
magnitude for extinction by A$_K=0.109 \times $A$_V$ (Bessell \& Brett
1988) and the distance modulus of 10.57.  
We separately construct KLFs for each of the five regions
plus the overall field.  
We find the KLF for the full field has a mean 
absolute magnitude in the K$_s$ band of  1.66 (3 $\sigma$ outliers excluded).
However, it is clear from  Figure~\ref{klf}, that the mean of the 
KLF varies from region to region.  
In addition to comparing the mean value of the extinction
corrected K$_s$ magnitude in each field, we examine the magnitudes
which define the 90$^{th}$ and 10$^{th}$ percentiles as a way to gauge
the spread in the luminosities of a given region.
The break down is shown in Table~\ref{tab_klf}. 
Here,  columns 1 and 2 identify the region and the number of good corrected
K$_s$ band measurements, 
columns 3--5 given the mean value
of  K$_s$ the MAD and the number of outliers rejected in the
calculation of the mean.
Column 6 is the value of K$_s$ for the
90$^{th}$ percentile in the region.  This value should provide a sense
of the high mass stars in each  region. Column 7 is the value of K$_s$ for the
10$^{th}$ percentile -- this should be indicative of the completeness
of each region and have less dispersion. The last column is the spread 
in magnitudes between the 90$^{th}$ and 10$^{th}$ percentile. 

Focusing on the mean of the KLFs and the value of the 90$^{th}$ percentile,
the data indicate that the central field is the brightest and hence
most likely the youngest.  It is clear that the distribution
in RCW~108--IR is cut off at the faint end by dust extinction. Thus, it
is plausible that the distributions in the central region is the same
as for other regions.  While it is plausible, it is not likely.
If we assume the observations are complete in
the eastern region to an absolute  K$_s$ magnitude of zero 
it would require that the total number of stars
in RCW~108-IR cluster be at least 6 times the number of stars in the
eastern region to account for the 12 sources
seen in  RCW~108-IR with absolute  K$_s$ magnitude $<0$ compared with two
such sources in the eastern region.  This would be more than 3,000
total stars which is not compatible with the log-normal XLF
shown in \S~\ref{secxlf} which predicts about 800 stars in
RCW~108--IR, nor the independent line of evidence that the highest mass star in
RCW~108--IR is about O8/O9 (Comer\'on \e 2005).  A more likely
scenario is that mass segregation has deprived the sampled portion of
the east region of its high mass stars while extinction renders the
fainter population of RCW~108--IR undetected in X-rays.

The southern and western fields are 
very similar to the eastern field with the possible inclusion of a 
few brighter stars from  RCW~108--IR to the south.  The 
northern field appears also to possess a mixture of stars at
low absorption and high mass stars.  
These results, derived, primarily from the NIR data,
echo the quartile results discussed in \S3.3 as well as the regionalized
results of the one--temperature spectral fits.

For all fields except the north and the south the 10$^{th}$ percentile
is about K$_s$ of 3.2.  The  10$^{th}$ percentile to the south achieves  
K$_s \approx 3.7$, while in the north the 10$^{th}$ is a magnitude
brighter. The similarity in the depth of the KLFs is essentially a
reflection that X-ray luminosity tracks photospheric luminosity 
(cf. Feigelson \& Montmerle 1999 and references therein) and
there is an X-ray luminosity limit in the 90 ks observation.  
The fact that the brightest sources are to the center
is taken as evidence of the relative youth of this region.
The large range in values to the south is probably indicative of
some mixing, the brightest sources coming from the young, central
region, while the faintest sources are from the older
population. 

Still it is surprising that the KLF to the south goes half
a magnitude deeper than the eastern region. A point to understanding
this comes from the different evolutionary timescales of the X-ray and
IR flux from young stars.  
The {\em maximum} X-ray
  luminosity or saturation level of the corona, appears to be a
  function of the stellar diameter and hence is linearly related to the
  photospheric luminosity.  But the median X--ray luminosity of G
  stars is the  same in Orion and $\alpha$ Persei
(Favata \& Micela 2003, G\"udel 2004).   The X-ray
observations of the RCW~108/NGC~6193 field are sensitive to 
similar X-ray luminosities (modulo
extinction) in each region.  These X-ray luminosities correspond to 
similar masses (following Feigelson \e 2005).  On the other hand, the 
K magnitudes of younger sources are brighter than their photosphere due
  to the contribution of disks. Hence reach lower masses.
  As the stars age over the first few
  million years, their disks vanish and the KLF tracks the photospheric 
luminosity. However for their first 1 Myr, a substantial fraction
of the stars are over-luminous in K relative to the X-ray luminosity
  which is being used as the selection criteria.

\subsection{Possible Late-O and B Stars \label{OBCAN}}

Using the extinction corrected KLF we performed a survey of the
cluster for O and B star candidates. On the ZAMS a star of 2.7
M$_\odot$ is a late B star. 
To convert the effective temperatures resulting from the models to
colors, we used the conversion table from Kenyon and Hartmann
(1995). Both mass and luminosity were derived from the models using
the extinction corrected K$_s$ magnitude of the X-ray selected members
and the distance modulus of
10.57. On the SDF mass tracks, at 0.5 Myr a 2.7
M$_\odot$ star has an absolute K band luminosity of -0.22 as
calculated using the on-line SDF model isochrones. 

We start by assuming than an age of 0.5 Myr and solar metallicity ($Z$=0.02) is
representative for  RCW~108--IR due to the observed disk fraction
(i.e. $Q < -0.1$).  We then 
estimate the mass of the high mass candidate stars independently for their
corrected J, H and K colors.  This showed
a trend of higher mass estimates at longer wavelength,
indicative of an age underestimate. This trend was not present if we
estimated the age of  RCW~108--IR is 1 Myr. 
This age also gave consistent results
for the all regions except for the eastern region where 3 Myr was
found to give the most consistent results.  

This leads the identification of 41 X-ray
sources as candidate high mass stars in the field, excluding the eastern
region.  These are listed in Table~\ref{OB}.  Only five are candidate
O stars with masses $>7$M$_\odot$. Three of the 41 were
excluded when VLT adaptive optics data revealed that they were multiples.
For the remainder, the consistency of the mass estimates among the three colors gives us
confidence in the technique, although we are skeptical of  
the absolute age estimate due to the high measured values of A$_{V}$
in RCW~108--IR.  This table list the observed magnitudes, then the
extinction estimate, followed by the corrected absolute magnitudes
and finally the masses estimated from the absolute magnitudes.

\begin{figure}
\epsscale{1}
\plotone{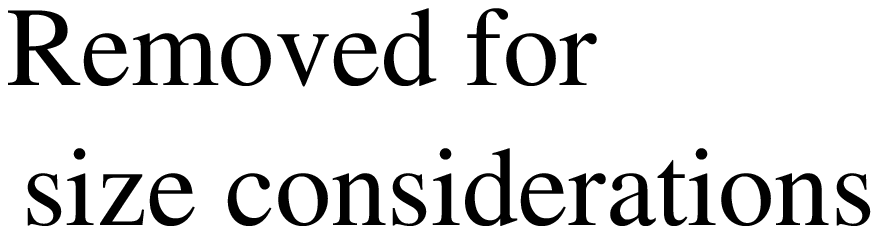}
\caption{High mass  stars in the $Chandra$ field. The large black box indicates the
ACIS-I field of view. Circles indicate the locations of high mass star candidates.
The underlying image is about 6.5~pc on a side and was adapted from
the 2MASS K$_S$ data, 
Contours are from the MSX A band data.
X-ray bright high mass stars are numbered except for the cluster in the
central field (see Figure~\ref{core}). There is
 also a clustering to the south. The eastern part of the 
field is almost free of X-ray bright high mass stars.}
\label{NTT_plus_OB_Xcont}
\end{figure}

We plot the locations of the high mass star candidates in
Figure~\ref{NTT_plus_OB_Xcont}.
Seven of the high mass star candidates are found just to the south of the 
8~\micron\ emission region plus an eighth source is located in the southern
tip of this region.
This area covers about 2.5\% of the field so the fact that about
20\% of the sources are found here indicates a causal relation
between this region and the high mass stars -- i.e. this region itself is a
sub-cluster of relatively recent star formation, independent of either
the HD stars at the center of NGC~6193 or the core of RCW~108--IR. Note
also in this region several non--X-ray detected  
possible O stars identifiable as
8\micron\ point sources in the MSX data.


About twenty of the high mass star candidates lie in the inner/core region. Four
of the  high mass star candidates are also found to be Class~I protostars on the
basis of there IRAC colors, especially 4.5\micron\ excesses. 
Since they are also identified as  high mass star candidates, their
classification as Class~I objects is dubious.  This is because the
empirical colors which are used for the stellar classification are
drawn from low mass stars.  We speculate the luminosity of the hot
stars may lead to detectable emission from cooler portions of disks
than is detectable around cool stars. This gives high mass stars
with disks IRAC colors similar to low mass protostars.

Most of the 20  high mass star candidates identified here, have been previously
identified as such by Comer\'on \e (2005).
However, Comer\'on \e confined themselves to the central 30\asec.
Of the 20  high mass star candidates, nine (sources with IDs between 112 and 131; see 
Figure~\ref{NTT_plus_OB_Xcont}) are associated with the more active
western subsection of the core and only one with the eastern
subsection of the core (Src~100). The remaining sources are somewhat further from the core.  
Among those stars a little further from the core,  
Src~52 is among the most interesting as it resides in
relative isolation, about 0.76~pc from the core region and 0.4~pc from 
the 8~\micron\ ridge which separates the active star formation from
NGC~6193, but centered on a diffuse peak of the 8~\micron\ emission (NGC~6188).  
Source~52 may be the leading part of a new sub-cluster.  
Sources 60, 61 and 62 are also associated with enhancements in the 
8~\micron\ emission. 
While these sources do not appear centered on the 8~\micron\
enhancement, Src~61 does not appear well resolved in either the X-ray
or NTT observations and might be part of a complex associated with the
8~\micron\ peak.

An age of
3.0~Myr is taken  as representative for the bright stars to the East.
There are 15 stars with an absolute magnitude in the K$_s$ band consistent
with being a B star in the eastern region.
However, due to the non-monotonic relation between mass and IR luminosity for
these photospheres, 11 of these stars are somewhat below 2.7 M$_\odot$ once data
from the other colors is compared. 
Four stars of the $\sim$ 100 in this region, which
represents a sample of the outlying region of a more evolved group of
young stars are found to be high mass stars. One of these, Source~40 is very
much associated with the NGC~6188 ridge and thus may be much younger
and lower in mass.  The bulk of the remainder of the X-ray sources are
expected to comprise a fairly complete sample of the G and K
membership of NGC~6193.
The detection of only 3-4  high mass star candidates to the east (including Source~262 in the north), 
is again indicative of mass segregation in the NGC~6193 cluster.

From the X-ray perspective the  high mass stars run the gamut ranging from
over 1000 counts to $<$16 as well as  
covering a wide range in kT.  The overall mean (2 outliers rejected)
is about 2.15 keV.  About 25\% of the sample is cool with
one--temperature component fits below 1~keV. 
This is consistent with the standard model of X-rays originating in 
shocked material in line-driven spherically symmetric winds (MacFarlane et al. 1991).
Typical temperatures in these models are about 10MK ($\sim$ 900 eV).

A similar number have temperatures
above 4 keV.  Hot X-ray emission from  high mass stars
is a somewhat recent finding.  The hard, embedded sources of
RCW~108--IR  resemble the central sources in NGC~2024 (Skinner et al. 2003).
Schulz \e (2001, 2003) reported temperatures of
about 5 keV from $\Theta^1~Orionis~C$. Schulz \e conclude
that the component of the emission at this temperature must be formed 
``near the terminal velocity of the wind,'' 
at about 7 stellar radii from the photosphere.
Gagn\'e et al. (2005) show that they can be fitted by a two-dimensional
MHD magnetospheric wind model due to  ud-Doula \& Owocki, (2002). 
Several other examples of X-ray hot O and B stars have since been found
(Lopes de Oliveira \e 2006, Rakowski \e 2006). 
Mullan \& Waldron (2006) 
have proposed a two component scenario by which the cooler X-ray
emission is generated by line-driven spherically symmetric winds, and 
a second component of the wind emerges from magnetically active
regions in  polar caps which may extend as low as 45$^{\rm o}$.
In such a scenario whether an O star is found to be a relatively hot
or soft source may simply be a function of viewing angle.  

Most of the hotter stars are associated
with  RCW~108--IR (see Figure~\ref{OBHIST}).  There is a 
certain selection effect in  RCW~108--IR 
where the mean temperature is about 2.8 keV (two outliers
rejected) and only 1 cool high mass star is found. This selection effect arises
because local absorption can exclude the detection of bright, but soft X-ray
sources when they are located deep within or on the far side the
absorbing material (i.e. star forming cloud).  
However, the inclusion of 75\% of the high mass
stars with X-ray temperature above 3~keV in the inner group indicates
that young high mass stars as observed in  RCW~108--IR have hotter
X-ray emission than their older, field counterparts.  
We note in the eastern region that
all the  high mass stars have temperatures below 3~keV, including Source~243 with
moderate \nh\ and a kT of about 300~eV.
In the southern region, Sources 269 and the embedded Source~104 are
bright with kT $>$4. Yet the remaining  high mass star candidates all have X-ray
temperatures below 1.2~keV.
Nearly a quarter of the X-ray sources detected in  RCW~108--IR are  high mass stars.

\section{Specific Interesting Sources}
While the bulk of the X-rays sources have similar characteristics,
well described in a statistical sense by Tables~\ref{Region_Q},
\ref{Region_Spec} and \ref{Region_avnh}, there are outliers.  
We examined the data set of bright X-ray sources looking specifically
for unusual sources which had enough counts for credible spectral
fits.  Of about 75 X-ray sources with 100 net counts, we found only 
six well fitted as plasma of $<$ 1 keV. Conversely,  only 
five sources were best fitted as a thermal plasma of over 10 keV,
while 13 were fitted to coronal temperature between 4 and 10 keV.
Eleven sources, including 6 of the hot sources counted above had
measured \nh\ columns of greater than 2$\times 10^{22} \psqcm$

\subsection{Cool Objects} 
The six sources with over 100 net counts and fitted temperatures of
$<$ 1 keV are Sources 158, 243, 254, 260, 267, and  294.  Most of
these have clear counterparts with moderate magnitudes in the 2MASS
survey. Two (243 and 267) were identified as candidate OB stars in the
previous section.  
 
The other four may be examples of young ($\sim$ 3 Myr) stars with no hot coronal
component. This might be indicative that a fraction (perhaps 10\%) 
of stars no longer have a hot corona by this age (Jeffries \e 2006). 
As a group they are not remarkably bright at
near-IR wavelengths so we consider them probable cluster members and 
not foreground dKe/dMe stars.

\subsection{Very Hot objects} 
At the other end of the spectrum, there are five bright sources which
were best fitted, by all techniques, as extremely hot.  The binned
data variance  spectral fits  as well as the unbinned C-statistic
returned temperatures in excess of 10 keV.
They are Sources 53, 266, 281, 284, \& 341.
Source 53 is very hard with no optical/IR counterpart making it a
very good candidate for a background AGN. It was fitted with a power
law with an index $\Gamma$= 1.23 $\pm$ 0.14 which had a very similar reduced $\chi^2$
to the thermal fits. Similarly sources 281 and 284 have no optical counterpart,
and show fairly high extinction, ($\sim 2-5 \times 10^{22}$ cm$^{-2}$)
even though they are far from the active star forming cloud, 
indicating that they are being seen through an absorber which lies
behind the region of observed star formation.
They were also fitted with a power law ($\Gamma$= 0.68 $\pm$ 0.16 and  
$\Gamma$=0.21  $\pm$ 0.42 respectively), which had a very 
similar reduced $\chi^2$ to their thermal fits. 
The optical/IR counterpart to Source~266 is offset
from the X-ray source by 1.85\asec\ which is near the limit of our
matching requirement and may be a chance superposition,  this
interpretation is enhanced by the high extinction observed to the X-ray
source, despite the fact that it is away from the main cloud. Again,
this is an indication that background absorption is being detected.   
The interpretation is much the same for Source~341,  in addition it  
shows a bright Fe line near 6.7 keV and is most likely
an AGN.

Five (or six -- see below) of the brightest 75 objects 
appear to be AGN, or other hard,
absorbed, non-stellar sources.  
Hence a little less than 10\% of the bright sample of PMS ``stars'' 
are not stars at all.  This is consistent with 
expectations derived from the Champlane survey (Hong \e
2005). Assuming the minimum detection required an average of 10
counts, then in $\approx$ 90ks,
the RCW~108 observation reached a mean flux limit of about 
$1 \times 10^{-15}~ \erg\ \ps \psqcm$ (0.5-8 keV) averaged across the
field. Preliminary results from the Champlane survey for $Chandra$ 
indicate 
that we can expect to find about 70 background AGN, cataclysmic
variables, neutron stars, black holes, and other non--PMS star 
point sources in a sample of 15 ACIS observations in the galactic
plane, but not pointed at the
galactic center, (90$^{\rm o} < l < 270^{\rm o}$), sensitive to flux levels of
10$^{-15} \erg\ \ps \psqcm$
from 0.5-2.0 keV.  In the harder band, the result is similar.
Due to the nearly opaque nature
of the dust clouds in the central $\sim$40\% of the field,
we expect that number to be significantly lower in this field,
perhaps 45 total --  a little more than 10\% of the total population of
objects.  Many of these are among the faint X-ray sources, 
X-ray sources with no
optical/IR counterparts or the sources with offset between the
X-ray position and optical/IR position exceeding 4\asec.



A second group of bright X-ray sources is less extreme. Thirteen have coronal
temperatures between 4 and 10 keV (Sources~1, 12, 70, 84, 104, 131,
151, 157, 167, 269, 278, 280 and
298).  Of these, only source 280 appears to be a clear AGN, with no
counterpart and high extinction. It is well fitted by a power law
$\Gamma$ = 1.72 $\pm$ 0.14 with a reduced $\chi^2$ equal to that of
the  one--temperature fit.  Like the other AGN candidates 
it has no observed variability.

Several of the other hot X-ray sources (104, 131
and 269), have been identified as high mass star candidates.  
Among the others, Src~131 is only
about 30\asec\ from the apparent core of RCW~108--IR (IRS~29) 
while Src~104 is well to
the south ($\sim$ 2~pc) of the cluster center but lies 
along the same line of sight
of the 8~\micron\ diffuse emission.  Source 269 is an additional
1.1~pc to the south of Src~104, but still overlaying dust absorption,

Several other hard sources  (12, 70, 84, 131, 151, 167
and 298) all show strong X-ray flares while Src~278 is an X-ray variable.  
Temperatures of between 4-10 keV
(50-110MK) are not unusual in such flares (Favata \e 2005).
Such flares are somewhat unusual among high mass stars, as such 
Source 131 does seem a bit unusual but variability has been observed in
other hot O stars, notably  $\Theta^1~Orionis~C$ (Schulz \e 2001).

\skipthis {

Hi Scott,
  Here are the powerlaw fit results.  Plots are in the individual source 
directories:
/data/ANCHORS/AExtract/YAXX/04503/data/obs4503/src*/abs_pl.ps

src, nh, nh_err, gamma, gamma_err,chi2/dof
  #, 1022/cm2, 1022/cm2,  G       G+-      X^4
 53,  1.3251,  0.2746,  1.2306,  0.1392,  0.2021=
266,  2.1270,  0.5673,  1.1990,  0.5486,  0.3896=
281,  0.8330,  0.3835,  0.6832,  0.1609,  0.3781=
284,  1.7186,  0.6884,  0.2133,  0.4150,  0.9011=
341,  0.3846,  0.2049,  1.4477,  0.2287,  0.5578=

  1,  0.4080,  0.1873,  1.3594,  0.2259,  0.2853
 12,  0.2764,  0.0502,  2.0189,  0.1074,  0.5996

 70,  1.3045,  0.1584,  2.0300,  0.1090,  0.5445
 84,  4.4056,  0.4472,  2.0110,  0.3062,  0.5233
104,  2.5550,  0.4845,  1.5607,  0.4743,  0.4150
131,  2.4370,  0.2422,  1.9072,  0.0841,  0.5852
151,  0.7111,  0.0800,  1.7004,  0.0794,  0.5141
157,  0.8217,  0.2298,  1.7959,  0.2262,  0.5159
167,  1.2214,  0.1553,  1.8654,  0.1033,  0.5578

269,  0.6374,  0.1209,  2.4289,  0.1829,  0.6245
278,  0.3673,  0.1312,  2.0943,  0.2315,  0.2424
280,  2.8716,  0.5666,  1.7156,  0.1443,  0.6477

}

\subsection{Very embedded sources}
\begin{figure}
\epsscale{1}
\plotone{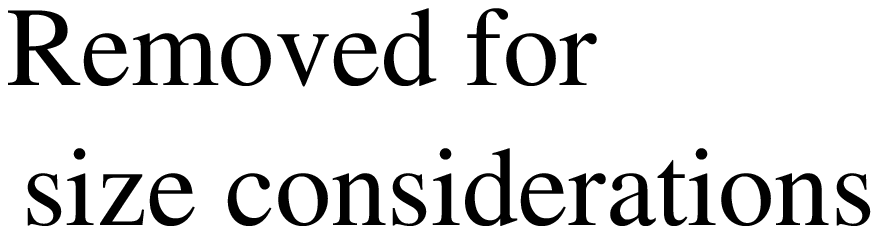}
\caption{High mass stars in the inner region.  Field of view is about
  1~pc on a side.
Boxes indicate the locations of high mass star candidates in the 
heart of RCW~108--IR. 
Contours indicate X-ray count rates of 0.5, 1, 2, 4, etc.
counts per pixel.  The underlying image is adapted from the NTT K$_s$ data
of Comer\'on \e (2005).  A dark lane splits the nebula. } 
\label{core}
\end{figure}

\begin{figure}
\epsscale{1}
\plotone{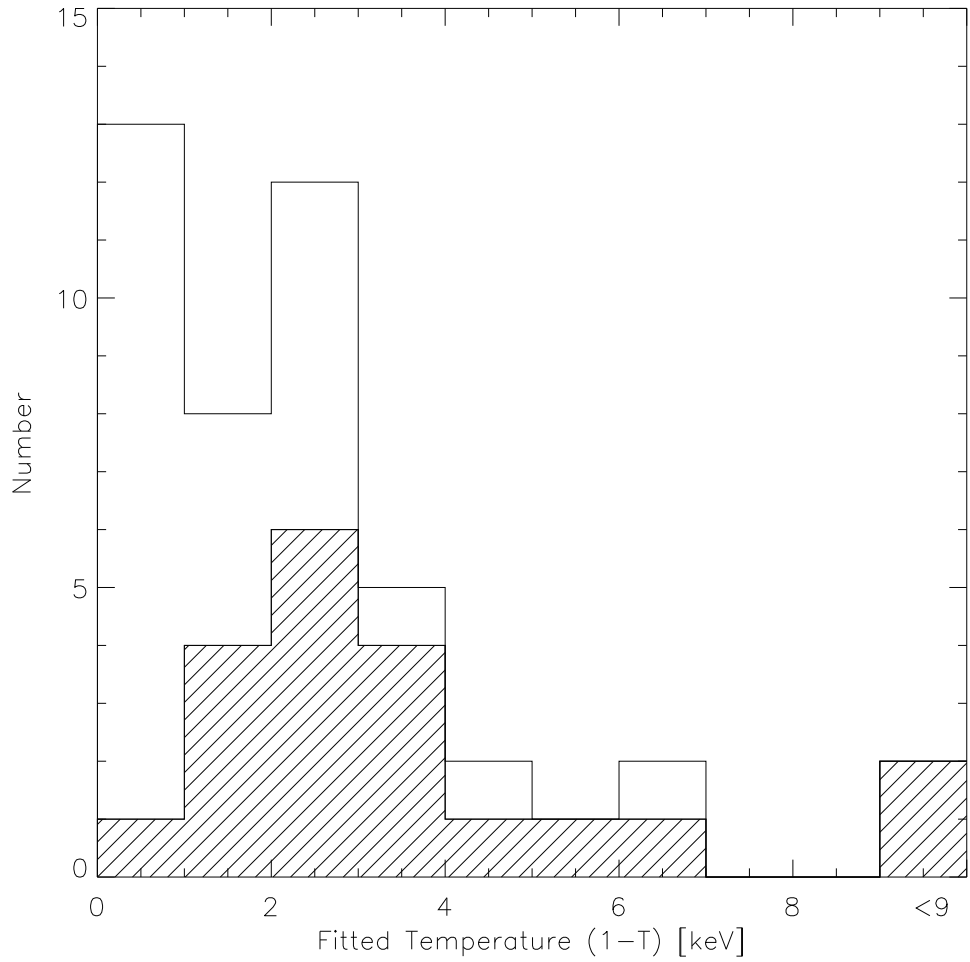}
\caption{Histogram of the temperatures found by one temperature
  spectral fits to candidate high mass  stars. Stars in the inner region (RCW~108--IR)
  are
  indicated by the hashing. These clearly represent a warmer subset
  of the group.}
\label{OBHIST}
\end{figure}

There are five additional sources that are quite absorbed but not
extremely hot.   Three of them, Src~116, Src~118 and Src~120 are
associated with IRS~29.  This has been resolved by
Comer\'on  \e (2005) into at least 6 sources and 
by the new VLT observations into more than 10 sources 
(Comer\'on and Schneider 2007). 
Three other components of IRS~29 have X-ray fluxes below the 100 count
cutoff for quality X-ray spectra used for this analysis.  
In addition, Src~116 has
been identified as high mass star candidates in \S\ref{OBCAN}. Source~120 lacked sufficient
flux at J band for classification in \S\ref{OBCAN}, however it was classified
by Comer\'on  \e  as a B0-B5 (their source 10) based on its H and K band colors. 
Our Source 124 is about 15\asec\ from IRS 29 and associated with
IRS 15 of Straw \e and source 7 of Comer\'on  \e which the latter
group classified at B5-A0.  The only outlying source (literally) in this group is
Source~92 (IRS 27)  which is about 60\asec\ away from IRS 29 (See Figure \ref{core}).  

\section{Comparison to Other Star Forming Regions}
Since 2000, the high resolution of $Chandra$ and the large collecting 
area of XMM--Newton have been used to systematically study many of
the active star forming regions within 2 kpc of the Sun (see review by
Feigelson \e 2007). $Chandra$ is especially
 adept at younger, denser and more embedded regions including
 RCW~108. 
Key among the contributions of these studies has been the 
understanding of the role of accretion and disks in flare generation
as well as the impact of flares on the developing disk system.
In addition, X-ray studies give us a independent pathway, relative to
IR observations to evaluate cluster membership and the IMF.  
In this section, we discuss the relationship of the observed results 
from RCW~108 in the context of these global questions.

Nominally, the presence or absence of a K-band emitting inner disk
 does not appear to affect X-ray emission. 
 However, the presence of accretion appears to have negative impact on X-ray
production (Flaccomio \e 2003; Stassun \e 2004; 
Preibisch \e 2005). This is manifested as a
 statistical decrease in X-rays by a factor of 2--3 in accreting vs. 
non-accreting PMS stars which are otherwise similar.  In these observations,  
we are
 sensitive to the K-band excesses, but not the accretion indicators
as a result of high optical absorption and the paucity
of near-IR spectra.
We have 17 sources
 with $Q<$-0.10 indicative of a thick disk and good X-ray spectra with
 errors in \nh\ and kT \< 33\%.  The log average flux of these sources 
is -13.26$\pm 0.41 \erg\ \ps \psqcm$.  There are 130 X-ray source with similar quality
 spectra and  Q$>$-0.10, the log average flux of these sources with no
 indication of a disk
is -13.44$\pm 0.47$. While the sources with thick disks appear
 brighter, the difference is less then 1 standard deviation. 
Overall, we only detected about one-third of the $Q<$-0.10 (disk
 possessing) stars known in
 the field. From this we can conclude that accretion is not the primary
 X-ray production mechanism since this sample should include the most
 active accretors and viewing angle effects should not absorb X-rays
 on two--thirds of the population unless disk flaring exceeds about
 60$^{\rm o}$.

Selecting sourced with $Q>$0.0 so that we are clearly looking at disk free
 stars
does not change the average flux significantly.  
Among the 17 X-ray bright sources with thick
 disks, 3 flared (17.6\%), and they were among the brighter 50\% of
 the flares.  Among the 120 X-ray bright sources with high Q, 7 flared
 (5.8\%).  The statistics flatten out a bit if we include X-ray
 sources with flares, no such flares are seen among the 20 such
 sources with low Q, but 9 are seen among the 120 fainter X-ray
 sources with high Q.  It would appear that the presence of a disk is
 conducive to strong flares, but may suppress our ability to see
 weaker flares, or suppress the mechanism of weaker flares
 altogether. We re-examined the data from RCW~38 to look for this effect
 and found flares among 11 of 172 ($\sim$6\%) of stars with high
 $Q$. This is a slightly higher rate in an observation that was about
 10\% longer.  But there were no flares among the 23 stars with 
 $Q <$-0.10. So while
 weak flares certainly seem suppressed in stars with disks in 
RCW~38 as well, there were no corresponding strong flares.

In other comparisons with RCW~38, despite the fact that RCW~108 field has
75\% the number of X-ray sources {\it and} is closer, the regions
as a whole, are comparable. We estimated
between 1400--2400 total members of RCW~38. The current full
RCW~108 $Chandra$
field contains 1400--1800 sources because it includes members of 
up to three separate star forming regions, including 
a large contribution from NGC~6193, which is probably similar in number
to RCW~38 based on the two O6 stars near its center. There is
evidence that the O4-5 star in the center of RCW~38 is triggering a
new generation of star formation.  RCW~108--IR appears to owe its
triggering to the O6 stars in NGC~6193 and
contains of order 800 stars.  However due to the highly variable
extinction the sampling of the stars near the cluster center is much
less uniform as manifest in the very choppy KLF shown in Figure~14 for
 RCW~108--IR.
 
The similarities continue when we examine the high mass star population.
In RCW~38, Wolk \e (2006) identified  31 high mass star candidates in the
field,  consistent
with the number of high mass stars expected for a cluster with about 2000
members.  We detect 15 high mass star candidates in RCW~108--IR for which we
project about 800 candidates.  This is in concurrence with the result of  Comer\'on \e
(2005) and it is very consistent with the results in RCW 38. 
The overall finding of about 42 high mass star candidates in the RCW~108
$Chandra$ field indicates that the mass function may be a bit
weighted towards high mass stars relative to RCW~38 and its
surroundings. This is especially notable in light of the evidence that the sample
from NGC~6193 is depleted in high mass stars due to mass
segregation. We estimate in \S~\ref{secxlf} that the 
south and west regions contained a total of about 750 
PMS stars. The identification of 17 high mass star candidates 
causes the regions to appear to have a top
heavy mass function.  

Two comparable star forming regions which 
have been recently reported on 
are Cepheus~B  (Getman \e 2006) and IC~1396N  (Getman \e 2007).
They were observed to a similar flux limit as the X-ray data presented
here and have some physical similarities. 
Cepheus~B is one of the most active star forming regions within a kpc.
It is an optically dark cloud subject to ionizing radiation, similar
to RCW~108-IR. 
About 390 X-ray
sources are found in Cepheus~B, including 64 associated with the embedded cluster. 
The biggest surprise of this dataset was the somewhat lower mass XLF with
an excess of low-mass (M$_\odot \sim 0.3$) stars relative to the ONC. 
Our data are not deep enough to detect such a discrepancy.  Like
RCW~108, the diffuse X-rays in this region are entirely attributable
to the integrated contribution of fainter PMS stars.  This is not
surprising since the region has only 1 known O star and it is removed
from the embedded region.   

IC~1396N is a much lower mass region. Getman \e argue that the star
formation in the embedded region has been triggered via a
radiation-driven implosion in which photoevaporation of the exterior
of the cloud by an O star, in the case of IC~1396N about 10 pc away,  
induces a shock that compresses the interior leading to
gravitational collapse (Reipurth 1983; Gorti \& Hollenbach 2002).
This is similar to the geometry seen in RCW~108 where in O6 stars 
HD~159135/6 are seen in the MSX A band data, to be 
clearly evaporating the edges of the cloud and small
concentrations of dust are forming interior to this (See Figures~3 and
5).  Also several of the smaller mid-IR peaks do seem to be
independent sites of star formation, as discussed in the previous
section and as predicted by radiation-driven implosion models. 
An alternative approach to triggering star formation, in which the photon loading
of the O stars compresses the cloud (see review by Henney 2006), 
is not as clearly supported by the data as the concentrations of warm dust interior to
NGC~6188 appear distributed in several locations in addition to the
main core of RCW~108--IR.

\section{Conclusions} 
We observed the massive, embedded young cluster RCW~108--IR and the
surrounding region using 
the $Chandra$ X-ray Observatory and in the IR using the NTT/SOFI
photometry and MSX as well as mm wave data from SEST. 
Over 420 X-ray sources were detected within the 17\amin $\times$
  17\amin\ $Chandra$ ACIS-I field of view in an 88ks exposure.  
We expect 90\% of these to be cluster members.  Near-IR
counterparts were identified for 361 of the X-ray sources. 
We performed one-temperature spectral fits to nearly 340 X-ray sources
and two-temperature spectral fits to the brighter
sources.  Since we used the C-statistics we describe high quality data
on the basis of small error instead of standard goodness-of-fit
parameters.  We looked for, but did not find, biasing of kT with respect
to \nh.  Here we summarize our results. 

\begin{itemize}

\item NGC 6193 is
seen through a uniform, foreground absorption, while the inner
region, RCW~108--IR, is embedded in a locally highly variable absorber. Further,
there is a high column density cloud  ($\sim 2-5 \times 10^{22}$
cm$^{-2}$) behind the observable  
the star forming region
through which hard AGN are seen.

\item The warm ($\sim 225K$)  dust observed along NGC~6188
is externally heated, probably by photoionization from the O stars in
NGC~6193. The dense dust in the center of RCW~108--IR remains
thermally shielded, even from the stars forming within it. There is
about 23.1 M$_\odot$ of cold dust in RCW~108--IR.  This implies a total
gas reservoir of about 2.3 $\times$ 10$^3$ M$_\odot$.

\item Intermediate temperatures among the sources in the western
  region as well as a physical connection between 
RCW~108--IR and IRAS 16348-4849 
indicate that the X-ray sources to the west of RCW~108--IR might have
  formed in association with IRAS 16348-4849.

\item  There is no significant hot (or X-ray emitting), diffuse plasma within this region.  

\item We find evidence of mass segregation in NGC 6193.  Only three
  high mass star candidates were found in the southern portion of
  this cluster while the existence of over 500 total stars is inferred.
   In addition, if one assumes a log-normal XLF, there is a deficit of 
  sources with log L$_x > 31.5$ relative to sources with 31.0 $>$log L$_x > 30.0$.
  This implies that the sample is depleted
  in early G stars, relative to K stars.  
 
\item The X-ray flare rate among brighter stars in the cluster is
  about 1 per 510 ks.  This is similar to the flare rate for G stars in
  the Orion Nebular Cluster and for lower mass stars in the ONC at a 
  similar X-ray count rate. Seven of the flares have temperatures in
  excess of 50MK.  There is a tendency for stars with disks to show
  either large flares or no flares at all.  The presence of a disk is
 conducive to strong flares, but suppresses our ability to see
 weaker flares.

\item The gas to dust ratio does not change perceptibly among the five
  regions and is consistent with  
  $N_{\rm H}/A_V = 2.0\times10^{21} {\rm cm}^{-2}$ 
as was found in
  RCW~38.  However, there seems to be a change in the relation at high
  A$_V$ for some stars.  This indicates either a gas depletion,
  a flattening of the reddening law at high A$_V$ or may be a
  manifestation of a fundamental degeneracy in the fitting of X-ray
  spectra.  The astrophysical phenomena should be
  investigated further in other embedded star forming regions.

\item Only a small fraction ($<$10\%) of the X-ray sources in the field have
  large optically thick disks as indicated by a $Q$ parameter of $>$
  -0.1. However, the evidence indicates that a much
  higher fraction 
  of stars, perhaps as high as 50\%--65\% do have optically thick
  disks.  Most these
  disks are not luminous enough to move the stars outside of the
  Lada \& Adams (1992) reddening bands.  Data from the IRAC bands indicates
  an even larger fraction of the stars may have disks.  This makes 
the region a good area for the study of transition disks.  

\item Hard X-ray emission seems more common from young embedded high mass stars
  than from the more dispersed population of OB stars.  This may
  indicate an evolutionary effect in the X-ray emission from high mass
  stars. Overall the region has more high mass star candidates than expected,
  especially in specific regions to the south and west.  This
  indicates that small episodic star formation is taking place 
independently in  several locations within the field.  This form of star formation is 
predicted by radiation-driven implosion models for star formation.

\end{itemize}

\acknowledgments
We acknowledge many helpful comments by referee M.~Gagn{\'e}. 
This publication makes use of data products from the Two Micron All
Sky Survey, which is a joint project of the University of
Massachusetts and the Infrared Processing and Analysis Center, funded
by the National Aeronautics and Space Administration and the National
Science Foundation. 
CXC guest investigator program supported this work through grant GO4-5013X.
SJW was supported by NASA contract NAS8-03060.

Facility: CXO(ACIS), Spitzer(IRAC), SEST, 2MASS


\clearpage

\clearpage

\clearpage

\clearpage

\clearpage


\begin{deluxetable}{llllrrrc}
\tablecaption{Faint X-ray sources detected in the RCW 108 Field \label{XSources}}
\tablewidth{0pt}
\tablehead{
\colhead{SRC \#} &\colhead{CXOWSBJ} &\colhead{RA}  &\colhead{DEC}   & \colhead{Offaxis}   &
 \colhead{Counts} &\colhead{Counts}  & \colhead {BB95}\\
\colhead{} &\colhead{} &\colhead{(J2000.0)}  &\colhead{(J2000.0)}   & \colhead{[arcsec.]}   &
\colhead{[Raw]} &   \colhead{[Net]} & \colhead{}}
\startdata
  1&164036.6-485217.3&16:40:36.64&-48:52:17.37&375.2&130&112.7&1   \\
  2&164035.8-485106.7&16:40:35.82&-48:51:06.76&369.5&187&169.8&1   \\
  3&164034.9-485049.5&16:40:34.90&-48:50:49.52&363.2&95&79.3&1	   \\
  4&164033.3-485023.6&16:40:33.32&-48:50:23.67&353.6&666&661.5&4   \\
  5&164032.1-485323.0&16:40:32.14&-48:53:23.02&341.8&23&17.5&1	   \\
  6&164031.9-484913.0&16:40:31.90&-48:49:13.06&365.2&59&43&1	   \\
  7&164031.8-485312.4&16:40:31.89&-48:53:12.42&336.7&138&133.2&1   \\
  8&164031.4-485311.9&16:40:31.44&-48:53:11.95&332.3&152&147.1&2   \\
  9&164031.2-485211.5&16:40:31.23&-48:52:11.56&321.6&59&51.8&1	   \\
 10&164031.0-485249.6&16:40:31.07&-48:52:49.68&324.2&61&55.7&1	   \\
 11&164030.9-484938.0&16:40:30.92&-48:49:38.04&346.1&33&27.9&1	   \\
 12&164030.6-485215.6&16:40:30.67&-48:52:15.61&316.3&738&730.8&6   \\
 13&164030.6-485026.4&16:40:30.66&-48:50:26.41&327.5&37&31.3&1	   \\
 14&164030.2-485108.3&16:40:30.20&-48:51:08.36&314.4&43&38.3&2	   \\
 15&164030.1-484957.8&16:40:30.12&-48:49:57.82&331.4&18&12.7&2	   \\
\enddata
\tablenotetext{a}{BB95 = The number of Bayesian Blocks of greater than
  95\% confidence required to describe the lightcurve.  Any number
  $> 1$ indicates variability at 95\% confidence.}
\tablenotetext{~}{REMAINDER OF DATA AVAILABLE ELECTRONICALLY}
\end{deluxetable}

\begin{deluxetable}{llllrrr}
\tablecaption{Faint X-ray sources detected in the RCW 108 Field \label{XSources_faint}}
\tablewidth{0pt}
\tablehead{
\colhead{SRC \#} &\colhead{CXOWSBJ} &\colhead{RA}  &\colhead{DEC}   & \colhead{Offaxis}   &
 \colhead{Src sig.} &\colhead{Counts} \\
\colhead{} &\colhead{} &\colhead{(J2000.0)}  &\colhead{(J2000.0)}   & \colhead{[arcsec.]}   &
\colhead{~} &  \colhead{[Net]}} 
\startdata
401&164028.1-484956.6&16:40:28.109&-48:49:56.59&313.2&3.1&8.1\\
402&164027.4-485040.1&16:40:27.481&-48:50:40.14&293.4&2.1&5.0\\
403&164027.3-485116.9&16:40:27.339&-48:51:16.89&285.0&1.9&4.2\\
404&164026.2-484749.1&16:40:26.210&-48:47:49.05&366.0&3.0&8.5\\
405&164022.5-485512.0&16:40:22.533&-48:55:11.96&307.2&2.2&5.0\\
406&164022.4-485231.4&16:40:22.425&-48:52:31.37&237.0&2.5&5.9\\
407&164022.3-485650.1&16:40:22.357&-48:56:50.06&376.8&3.0&9.4\\
408&164022.0-485025.4&16:40:22.038&-48:50:25.39&247.2&2.8&6.8\\
409&164021.0-485202.1&16:40:21.037&-48:52:02.06&220.8&2.1&5.0\\
410&164019.9-484945.7&16:40:19.991&-48:49:45.75&246.6&2.8&6.8\\
411&164019.6-485314.3&16:40:19.635&-48:53:14.30&221.4&3.1&7.1\\
412&164018.8-485317.6&16:40:18.827&-48:53:17.69&215.4&2.3&5.2\\
413&164015.1-485141.6&16:40:15.188&-48:51:41.64&163.2&3.3&8.4\\
414&164013.3-484853.1&16:40:13.316&-48:48:53.11&231.6&3.0&7.0\\
415&164010.9-484742.8&16:40:10.964&-48:47:42.86&279.0&2.2&4.6\\
\enddata
\tablenotetext{~}{REMAINDER OF DATA AVAILABLE ELECTRONICALLY}
\end{deluxetable}

\begin{deluxetable}{llllrrrrr}
\tabletypesize{\footnotesize}
\tablecaption{Cross--Reference of X-ray sources with SIMBAD counterparts
detected in the RCW 108 Field \label{XREF}}
\tablewidth{0pt}
\tablehead{
\colhead{SRC \#} &\colhead{CXOWSB~J} &\colhead{RA$^a$}  &\colhead{DEC$^a$}    
& \colhead{Alt. name} &\colhead{Ref.} 
& \colhead{Alt. name} &\colhead{Ref.} 
& \colhead{Alt. name} 
} 
\startdata
 92&164006.2-485111&16:40:06.1  & -48:51:12 & IRS 27 &SHJ87$^b$&~&~&~\\
 95&164005.9-485111&16:40:06.1  & -48:51:12 & IRS 27 &SHJ87$^b$&~&~&~\\
 96&164005.7-485213&16:40:05.6  & -48:52:12 & IRS 9 &SHJ87&~&~&~\\
 99&164004.7-485152&16:40:04.8  & -48:51:53 & IRS 20 &SHJ87&~&~&~\\
100&164004.4-485145&16:40:04.3  & -48:51:45 & IRS 22 &SHJ87&~&~&~\\
101&164003.8-485254&16:40:03.82 & -48:52:54.7& Star q & UTM04&~&~&~\\ 
107&164001.8-485137&16:40:01.8  & -48:51:39 & star 23 & CSR05&~&~&~\\
109&164001.6-485153&16:40:01.7  & -48:51:53 & star 22 & CSR05&~&~&~\\
111&164001.2-485145&16:40:01.3  & -48:51:45& star 19 & CSR05&~&~&~\\
112&164001.1-485152&16:40:01.2  & -48:51:52& star 18 & CSR05&~&~&~\\
114&164000.8-485138&16:40:00.9  & -48:51:39& star 17 & CSR05&~&~&~\\
116&164000.3-485143&16:40:00.4  & -48:51:43& star 16 & CSR05&~&~&~\\
118&164000.1-485139&16:40:00.2  & -48:51:40& IRS 29 & SHJ87 & star 12 & CSR05&~\\
120&164000.0-485142&16:40:00.1  & -48:51:42& star 10 & CSR05& star 462 & HH77&~\\
121&163959.9-485153&16:40:00.0  & -48:51:53& star 9 & CSR05&~&~&~\\
124&163959.7-485159&16:39:59.8  & -48:52:00& star 7 & CSR05&~&~&~\\
125&163959.5-485137&16:39:59.6  & -48:51:37& star 5 & CSR05&~&~&~\\
127&163959.3-485127&16:39:59.2  & -48:51:27& IRS 42 & SHJ87&~&~&~\\
130&163958.5-485236&16:39:58.5  & -48:52:36.8& star i & UTM04&~&~&~\\
131&163958.5-485204&16:39:59.6  & -48:52:05& star 2 & CSR05&~&~&~\\
136&163956.0-485148&16:39:55.98 & -48:51:48.2& star f & UTM04&~&~&~\\
138&163954.4-485201&16:39:54.5  & -48:52:01.4& star e & UTM04&~&~&~\\
140&163953.7-485145&16:39:53.7  & -48:51:46.2& star d & UTM04&~&~&~\\
145&163951.8-485141&16:39:51.83 & -48:51:41.6& star b & UTM04&~&~&~\\
174&163930.6-485102&16:39:30.67 & -48:51:02.5& HD 149834 & ~&~&~&~\\
215&163936.3-484755&16:39:36.39 & -48:47:55.2&08333-00232& GSC& star 593& HH77&~\\
217&163926.1-485147&16:39:26.10 & -48:51:47.6&08333-01365& GSC  & star 581 & HH77&CPD-48 8655\\ 
243&164043.6-484851&16:40:43.61 & -48:48:51.8&08333-00420&GSC&   star 608 & HH77 & CD-48 11055 \\
245&164041.0-485112&16:40:40.98 & -48:51:21.1& 08333-00814&GSC& star 226& HH77&~\\
259&164016.0-485816&16:40:16.06 & -48:58:15.8&08333-00074&GSC&  star 536 & HH77 & CD-48 11044 \\
267&164005.9-485908&16:40:05.89 & -48:59:08.8&08333-00240&GSC&  star 540 & HH77 & CD-48 11040 \\
306&164039.0-484843&16:40:39.08 & -48:48:44.3& 08333-01125&GSC& star 625 & HH77&~\\
418&164006.5-485214&16:40:06.6  & -48:52:15  & IRS 8         &SHJ87&~&~&~\\
427&164000.1-485214&16:40:00.1  & -48:52:16  & IRS 6         &SHJ87&~&~&~\\
\enddata
\tablenotetext{a}{SIMBAD position}
\tablenotetext{b}{Both X-ray sources $\sim$ 1.8\arcsec\ from IRS 27.}
\tablenotetext{~}{Rreferences: SHJ87=Straw \e (1987), UTM04=Urquhart \e (2004), CSR05= Comer\'on \e (2005),
HH77= Herbst \& Havlen (1977), GSC= Guide Star Catalog (Lasker \e 1990)}
\end{deluxetable}

\begin{deluxetable}{lccccccc}

\tabletypesize{\footnotesize}

\tighten \tablewidth{0pt}

\tablecolumns{8}

\tablecaption{MSX and IRAS Intensity of RCW 108 \label{MSX}}

\tablehead{
\colhead{Region} &  \colhead{Area}
 & \colhead{MSX A} & \colhead{IRAS 12\micron} & \colhead{MSX C} &
\colhead{MSX D} & \colhead{MSX E} &\colhead{IRAS 25\micron} \\
\colhead{} &  \colhead{arcsec$^2$} 
 & \colhead{Jy} & \colhead{Jy} & \colhead{Jy} & \colhead{Jy} & \colhead{Jy} & \colhead{Jy}
}

\startdata
A & 3,996         &  6.81 &  & 8.74 & 4.10 & 11.39 &  \\
B & 2,268         &  6.40 &  & 7.98 & 4.42 & 20.29 &  \\
C & 1,080         &  2.72 &  & 3.97 & 2.06 & 3.96 &  \\
Core & 2,844      &101.37 &  & 216.6 & 326.0 & 2,063.4 &  \\
Entire & 187,416 &286.31 & 208.3 & 510.2 & 489.0 & 2,999.2 & 2,978 \\
\enddata

\end{deluxetable}

\begin{deluxetable}{lcccccccc}
\tablecaption{Quartile Values of X-ray sources \label{quantin}}
\tablewidth{0pt}
\tablehead{
\colhead{SRC\#} &\colhead{Q25} &\colhead{Q25$_{err}$} &\colhead{M}&
\colhead{M$_{err}$}  & \colhead{Q75}   &\colhead{Q75$_{err}$} &
 \colhead{log(m/(1-m))} &\colhead{3*Q25/Q75} }
\startdata
East &  &  &  &  &  &  &  & \\
1 & 0.147 & 0.012 & 0.228 & 0.019 & 0.399 & 0.031 & -0.530 & 1.105    \\
2 & 0.100 & 0.005 & 0.145 & 0.011 & 0.208 & 0.017 & -0.771 & 1.442    \\
3 & 0.086 & 0.009 & 0.131 & 0.010 & 0.188 & 0.027 & -0.822 & 1.372    \\
4 & 0.108 & 0.003 & 0.153 & 0.003 & 0.224 & 0.012 & -0.743 & 1.446    \\
5 & 0.100 & 0.014 & 0.141 & 0.015 & 0.176 & 0.048 & -0.785 & 1.705    \\
6 & 0.106 & 0.006 & 0.133 & 0.012 & 0.177 & 0.019 & -0.814 & 1.797    \\
7 & 0.102 & 0.009 & 0.146 & 0.008 & 0.213 & 0.021 & -0.767 & 1.437    \\
8 & 0.124 & 0.008 & 0.180 & 0.010 & 0.254 & 0.035 & -0.659 & 1.465    \\
9 & 0.095 & 0.007 & 0.130 & 0.014 & 0.199 & 0.025 & -0.826 & 1.432    \\
10 & 0.095 & 0.015 & 0.146 & 0.011 & 0.249 & 0.060 & -0.767 & 1.145   \\
11 & 0.080 & 0.007 & 0.101 & 0.008 & 0.137 & 0.071 & -0.949 & 1.752   \\
12 & 0.111 & 0.003 & 0.168 & 0.005 & 0.294 & 0.015 & -0.695 & 1.133   \\
13 & 0.111 & 0.012 & 0.144 & 0.023 & 0.273 & 0.118 & -0.774 & 1.220   \\
14 & 0.089 & 0.014 & 0.154 & 0.019 & 0.226 & 0.109 & -0.740 & 1.181   \\
15 & 0.029 & 0.012 & 0.068 & 0.048 & 0.240 & 0.134 & -1.137 & 0.363   \\
\enddata
\tablenotetext{~}{REMAINDER OF DATA AVAILABLE ELECTRONICALLY}
\end{deluxetable}

\begin{deluxetable}{lrcrcr}
\tablecaption{Median Quartile Value -- by Region \label{Region_Q}}
\tablewidth{0pt}
\tablehead{
\colhead{Region}  &\colhead{\#}&\colhead{ $\log (m/(1-m))$}&\colhead{$\pm$}& \colhead{$3\times Q_{25}/Q_{75}$}
&\colhead{$\pm$}}
\startdata
North  &31& -0.58&  0.09&   1.02 &  0.07 \\
South  &41& -0.59&  0.07&   1.22 &   0.05 \\ 
East   &106 & -0.77&  0.01&   1.23 & 0.04  \\
West   &63& -0.60&  0.05&   1.11 &  0.08 \\
Inner 1\arcmin\ &31& -0.30 &0.06&   1.56& 0.06 \\
annulus from 1\arcmin\ to 3.5\arcmin\ &65&  -0.30&  0.04& 1.34 &   0.05\\ 
Field  &337& -0.67  &0.27   & 1.32  & 0.33  \\
\enddata
\end{deluxetable}

\begin{deluxetable}{lcccccccc}
\tabletypesize{\small}
\tablecaption{One Temperature Spectral Fits \label{1t}}
\tablewidth{0pt}
\tablehead{ 
\colhead{SRC\#} &\colhead{$\chi^2$ $^a$} & \colhead{\nh}
 & \colhead{\nh err }&\colhead{kT} &  \colhead{kT err} & 
\colhead{    Abs. Flux}& \colhead{    Unabs. Flux} & \colhead{Log L$_{x}$} \\ 
\colhead{}  & \colhead{per d.o.f.} & 
\colhead{10$^{22}~cm^{-2}$} & \colhead{10$^{22}~cm^{-2}$} &
\colhead{keV} & \colhead{keV} &\colhead{                         } &
\colhead{                         } &\colhead{ergs sec$^{-1}$}}
\startdata
East &     &      &      &      &      &        &        & \\
  1 & 0.62 & 0.55 & 0.15 & 5.51 & 1.87 & -13.73 & -13.53 & 30.77   \\
  2 & 0.58 & 0.44 & 0.08 & 1.10 & 0.08 & -13.90 & -13.40 & 30.91   \\
  3 & 0.53 & 0.79 & 0.09 & 0.61 & 0.07 & -14.29 & -13.28 & 31.03   \\
  4 & 0.47 & 0.17 & 0.03 & 2.65 & 0.31 & -13.14 & -12.95 & 31.36   \\
  5 & 0.29 & 1.03 & 0.18 & 0.56 & 0.11 & -14.87 & -13.63 & 30.67   \\
  6 & 0.51 & 0.34 & 0.12 & 1.14 & 0.13 & -14.42 & -13.99 & 30.31   \\
  7 & 0.37 & 0.09 & 0.06 & 2.34 & 0.54 & -13.87 & -13.72 & 30.58   \\
  8 & 0.38 & 0.40 & 0.10 & 2.52 & 0.54 & -13.80 & -13.52 & 30.79   \\
  9 & 0.39 & 0.58 & 0.12 & 0.95 & 0.15 & -14.42 & -13.78 & 30.53   \\
 10 & 0.35 & 0.04 & -- & 2.93 & 1.13 & -14.20 & -14.12 & 30.19 \\
 11 & 0.25 & 0.01 & -- & 1.02 & 0.18 & -14.57 & -14.48 & 29.83 \\
 12 & 0.56 & 0.20 & 0.04 & 4.01 & 0.54 & -13.05 & -12.89 & 31.41   \\
 13 & 0.39 & 0.73 & 0.16 & 0.82 & 0.15 & -14.67 & -13.86 & 30.45   \\
 14 & 0.37 & 0.01 & -- & 2.04 & 0.84 & -14.39 & -14.34 & 29.97 \\
 15 & 0.29 & 0.02 & -- & 0.09 & 0.01 & -13.80 & -13.11 & 31.20 \\
\enddata
\tablenotetext{a}{$\chi^2$/d.o.f is included for completeness.  These
  models were calculated with C statistics,  it is
unclear how $\chi^2$/d.o.f  relates to  goodness of fit.  The formal
errors are unbiased however. --
see text.}
\tablenotetext{~}{REMAINDER OF DATA AVAILABLE ELECTRONICALLY}
\end{deluxetable}

\begin{deluxetable}{lccccccccccccc}
\rotate
\tabletypesize{\scriptsize}
\tablecaption{Sources with Two--Temperature Spectra \label{2t}}
\tablewidth{0pt}
\tablehead{ \colhead{SRC\#}  &\colhead{$\chi^2$ $^a$} & 
\colhead{\nh} & \colhead{\nh\ err} &   \colhead{kT$_1$}   & \colhead{kT$_1$ err} & \colhead{kT$_2$} &
\colhead{kT$_2$ err} & \colhead{Abs.} & \colhead{Unabs.} &
\colhead{Unabs.} & \colhead{EM$_1$} &  \colhead{EM$_2$}& \colhead{L$_{x}$} \\
\colhead{} & \colhead{per d.o.f.} & 
\colhead{10$^{22}~cm^{-2}$} & \colhead{} & \colhead{keV}  & \colhead{}&\colhead{keV} &
\colhead{} &\colhead{Flux} & \colhead{Flux$_1$} &\colhead{Flux$_2$}
&\colhead{log} & \colhead{log} &  \colhead{log}\\
\colhead{} & \colhead{} & 
\colhead{} &   \colhead{}   & \colhead{} & \colhead{} &\colhead{} & \colhead{} &
\colhead{ergs cm$^{-2}$} & \colhead{ergs cm$^{-2}$} &
\colhead{ergs cm$^{-2}$} & \colhead{ cm$^{-6}$} &   \colhead{ cm$^{-6}$} & \colhead{ergs$\ps$}
}
\startdata
East &  &  &  &  &  &  &  &  &  &  &  && \\
2   & 1.07 & 0.47 & 0.09 & 0.86 & 0.13 & 2.09 & 0.77 & -13.86 & -13.77 & -13.69 & 53.02 & 53.51 & 30.88\\
4   & 0.95 & 0.25 & 0.04 & 0.75 & 0.14 & 2.67 & 0.41 & -13.18 & -13.90 & -12.99 & 52.85 & 54.18 & 31.37\\
7   & 0.88 & 0.51 & 0.10 & 0.54 & 0.27 & 2.33 & 0.70 & -13.92 & -13.84 & -13.69 & 52.78 & 53.47 & 30.85\\
12  & 0.95 & 0.35 & 0.04 & 0.86 & 0.10 & 5.30 & 1.83 & -13.06 & -13.53 & -12.95 & 53.29 & 54.11 & 31.46\\
30  & 0.38 & 0.48 & 0.07 & 0.70 & 0.11 & 2.72 & 0.96 & -13.77 & -13.55 & -13.68 & 53.67 & 53.48 & 31.00\\
230 & 0.96 & 0.59 & 0.05 & 0.14 & 0.01 & 0.98 & 0.08 & -13.55 & -12.05 & -13.02 & 54.83 & 54.18 & 32.30\\
233 & 1.80 & 0.44 & 0.08 & 0.21 & 0.04 & 1.07 & 0.12 & -13.94 & -13.31 & -13.56 & 53.91 & 53.39 & 31.19\\
245 & 0.99 & 0.21 & 0.03 & 1.00 & 0.07 & 2.96 & 0.33 & -12.78 & -13.33 & -12.65 & 53.57 & 54.51 & 31.74\\
248 & 1.16 & 0.07 & 0.07 & 0.92 & 0.26 & 1.96 & 0.70 & -13.77 & -14.37 & -13.70 & 52.45 & 53.51 & 30.69\\
255 & 0.73 & 0.10 & 0.05 & 1.05 & 0.09 & 3.55 & 0.93 & -13.33 & -13.87 & -13.29 & 53.06 & 53.83 & 31.12\\
257 & 0.64 & 0.28 & 0.04 & 0.86 & 0.09 & 2.41 & 0.42 & -13.32 & -13.61 & -13.17 & 53.21 & 54.01 & 31.27\\
North &  &  &  &  &  &  &  &  &  &  &  && \\
134 & 0.39 & 0.95 & 0.13 & 0.44 & 0.11 & 2.67 & 0.73 & -13.79 & -13.39 & -13.47 & 53.48 & 53.70 & 31.18\\
262 & 1.09 & 0.24 & 0.07 & 1.21 & 0.43 & 2.94 & 1.00 & -13.50 & -13.94 & -13.38 & 53.08 & 53.77 & 31.03\\
RCW~108--IR &  &  &  &  &  &  &  &  &  &&  &  & \\
116 & 1.06 & 2.45 & 0.21 & 1.21 & 0.43 & 2.80 & 0.47 & -13.36 & -13.09 & -12.96 & 53.91 & 54.20 & 31.58\\
118 & 0.90 & 3.48 & 0.12 & 0.57 & 0.08 & 2.61 & 0.13 & -12.63 & -11.88 & -12.06 & 54.82 & 55.11 & 32.65\\
136 & 0.84 & 1.15 & 0.10 & 0.71 & 0.15 & 2.42 & 0.55 & -13.15 & -12.68 & -12.82 & 54.41 & 54.32 & 31.86\\
South &  &  &  &  &  &  &  &  &  &  &  && \\
64  & 0.74 & 1.58 & 0.11 & 0.69 & 0.09 & 2.72 & 0.82 & -13.60 & -12.66 & -13.37 & 54.06 & 53.78 & 31.73\\
267 & 0.78 & 0.01 & 0.06 & 0.44 & 0.11 & 1.16 & 0.32 & -13.97 & -14.15 & -14.23 & 53.01 & 52.77 & 30.42\\
West &  &  &  &  &  &  &  &  &  &  &  && \\
158 & 1.01 & 0.37 & 0.08 & 0.23 & 0.06 & 0.90 & 0.12 & -14.18 & -13.81 & -13.76 & 53.12 & 53.43 & 30.82\\
164 & 1.22 & 0.19 & 0.08 & 0.89 & 0.14 & 2.15 & 0.96 & -13.90 & -14.07 & -13.84 & 52.77 & 53.37 & 30.67\\
170 & 1.02 & 0.80 & 0.14 & 0.76 & 0.33 & 3.60 & 1.16 & -13.67 & -13.93 & -13.42 & 52.84 & 53.70 & 31.01\\
\enddata
\tablenotetext{a}{Data Variance $\chi^2$/d.o.f. statistic has mean of
  1 for a good fit.} 
\end{deluxetable}

\begin{deluxetable}{lrrrrrrr}
\tablecaption{X-ray Spectral Properties by Region  \label{Region_Spec}}
\tablewidth{0pt}
\tablehead{
\colhead{Region}  &\colhead{\#}&\colhead{kT} &\colhead{MAD}
&\colhead{Rejected} &\colhead{\nh}   & \colhead{MAD}   &
\colhead{Rejected}\\
\colhead{ }  &\colhead{ }&\colhead{[keV]} &\colhead{ }
&\colhead{ } &\colhead{[$\times 10^{22} $cm$^{-2}$]}   & \colhead{}   &
\colhead{ }
}
\startdata
North     &7    &2.78&  0.49 &   0  &  0.54 &   0.09  &  0\\ 
South     &12   &2.79&  0.66 &   0  &  0.69 &   0.10  &  1\\ 
East      &29   &1.44&  0.15 &   3  &  0.41 &   0.04  &  1\\ 
West      &20   &2.87&  0.39 &   0  &  0.53 &   0.06  &  3\\ 
RCW~108--IR     &21   &3.07&  0.29 &   0  &  2.29 &   0.28  &  2\\ 
Field     &89   &2.54&  0.17 &   0  &  0.53 &   0.04  & 20\\
\enddata
\end{deluxetable}

\begin{deluxetable}{lrrrrrrrrrrrrrr}
\rotate
\tabletypesize{\footnotesize}
\tablecolumns{15}
\tablewidth{0pc}
\tablecaption{Near--IR Observations of X-Ray Sources
\label{nir} }
\tablewidth{0pt}
\tablehead{ \colhead{SRC\#} & \colhead{IR R.A.} & \colhead{IR Dec.} & 
\colhead{offset} & \colhead{off-axis} & \colhead{J} &  \colhead{Jerr}
& \colhead{H} & \colhead{Herr} & \colhead{K} &  \colhead{Kerr}  & \colhead{$Q$} &
\colhead{A$_V$} &  \colhead{model}  & \colhead{flags}}
\startdata
East &  &  &  &  &  &  &  &  &  &  &  &  &  & \\
  1 & 16:40:36.70 & -48:52:16.0 & 1.504 & 375.2 & 12.52 & 0.01 & 11.58 & 0.006 & 11.08 & 0.007 & 0.07 & 3.46 & M star & NTT       \\
  2 & 16:40:35.80 & -48:51:06.0 & 0.777 & 369.5 & 13.43 & 0.016 & 12.45 & 0.01 & 11.90 & 0.010 & 0.04 & 1.85 & Disk & NTT 	 \\
  3 & 16:40:34.90 & -48:50:48.0 & 1.524 & 363.2 & 13.64 & 0.018 & 12.84 & 0.012 & 12.62 & 0.014 & 0.44 & 2.24 & Disk & NTT	 \\
  4 & 16:40:33.40 & -48:50:23.0 & 1.015 & 353.6 & 13.52 & 0.017 & 12.74 & 0.011 & 12.50 & 0.014 & 0.35 & 1.49 & Disk & NTT	 \\
  5 & 16:40:32.20 & -48:53:23.0 & 0.614 & 341.8 & 15.38 & 0.04 & 14.56 & 0.029 & 14.23 & 0.032 & 0.25 & 3.65 & HMS & NTT	 \\
  6 & 16:40:32.00 & -48:49:12.0 & 1.439 & 365.2 & 13.58 & 0.017 & 12.73 & 0.011 & 12.30 & 0.012 & 0.13 & 1.2 & Disk & NTT	 \\
  7 & 16:40:31.90 & -48:53:12.0 & 0.426 & 336.7 & 14.20 & 0.023 & 13.37 & 0.016 & 13.12 & 0.019 & 0.41 & 2.31 & -- & NTT	 \\
  8 & 16:40:31.50 & -48:53:12.0 & 0.616 & 332.3 & 13.63 & 0.018 & 12.74 & 0.011 & 12.21 & 0.012 & -0.02 & 0.86 & Disk & NTT	 \\
  9 & 16:40:31.40 & -48:52:12.0 & 1.719 & 321.6 & 16.45 & 0.095 & 14.94 & 0.048 & 14.62 & 0.059 & 0.95 & 10.98 & Disk & NTT	 \\
 10 & 16:40:31.10 & -48:52:49.0 & 0.745 & 324.2 & 14.29 & 0.024 & 13.45 & 0.016 & 13.21 & 0.019 & 0.43 & 2.43 & -- & NTT	 \\
 11 & 16:40:31.00 & -48:49:38.0 & 0.797 & 346.1 & 14.92 & 0.033 & 13.96 & 0.021 & 13.39 & 0.022 & -0.01 & 1.55 & -- & NTT	 \\
 12 & 16:40:30.70 & -48:52:15.0 & 0.682 & 316.3 & 14.09 & 0.022 & 12.60 & 0.011 & 11.36 & 0.008 & -0.63 & 3.45 & Disk & NTT	 \\
 13 & 16:40:30.70 & -48:50:26.0 & 0.573 & 327.5 & 15.17 & 0.037 & 14.35 & 0.026 & 14.00 & 0.030 & 0.22 & 4.1 & HMS & NTT	 \\
 14 & 16:40:30.20 & -48:51:09.0 & 0.632 & 314.4 & 14.65 & 0.029 & 13.61 & 0.018 & 13.21 & 0.019 & 0.36 & 3.97 & -- & NTT	 \\
 15 & 16:40:30.20 & -48:49:59.0 & 1.424 & 331.4 & 17.51 & 0.117 & 16.81 & 0.106 & 16.53 & 0.129 & 0.23 & 0.28 & -- & NTT	 \\
\enddata
\tablenotetext{a}{VLT adaptive optics data indicate this source is a double.}
\tablenotetext{~}{FLAGS:
VLT = data are from the Very Large Telescope, reduction
discussed in Comer{\'o}n \& Schneider (2007).
 NTT = data are from New Technology Telescope, reduction
discussed in Comer{\'o}n \e 2005.
  AAA, UAB, UAU, AAU, etc. = data are taken from 2MASS with these
  listed photometry codes for the JHK channels respectively: 
A-- error $<$10.9\%, B -- error $<$15.6\%, 
E -- This category includes detections where the goodness-of-fit
quality of the profile-fit photometry was very poor,  
U -- upper limit.}
\tablenotetext{~}{REMAINDER OF DATA AVAILABLE ELECTRONICALLY}
\end{deluxetable}

\begin{deluxetable}{lccrrrrrrrrc}
\tablecaption{IRAC detections of X-ray sources in the RCW 108 Field \label{XIRAC}}
\tablewidth{0pt}
\tablehead{
\colhead{SRC \#} &\colhead{Offset} &\colhead{Offaxis}  &\colhead{Ch1}   & \colhead{Ch2}   &
\colhead{Ch3} &  \colhead{Ch4} &\colhead{Ch1e}   & \colhead{Ch2e}   &
\colhead{Ch3e} &  \colhead{Ch4e} & \colhead{IR Class}\\
\colhead{} &\colhead{[arcsec.]} &\colhead{[arcsec.]}}
\startdata
East &  &  &  &  &  &  &  &  &  &  & \\ 
1 & 0.481 & 375.2 & 10.51 & 10.13 & 9.88 & 9.42 & 0.01 & 0.01 & 0.02 & 0.05 &II   \\ 
2 & 0.447 & 369.5 & 11.06 & 10.75 & 10.40 & 9.90 & 0.02 & 0.02 & 0.03 & 0.03 &II  \\
3 & 0.421 & 363.2 & 12.50 & 12.53 & 12.72 & -- & 0.03 & 0.03 & 0.12 & -- & --      \\
4 & 0.373 & 353.6 & 12.14 & 12.10 & -- & -- & 0.04 & 0.05 & -- & -- & --           \\  
6 & 0.441 & 365.2 & 11.46 & 11.05 & 10.60 & 9.79 & 0.01 & 0.01 & 0.03 & 0.03 &II  \\
7 & 0.364 & 336.7 & 12.88 & -- & -- & -- & 0.02 & -- & -- & -- & --                \\
8 & 0.233 & 332.3 & 11.41 & 11.05 & 10.72 & 9.84 & 0.01 & 0.01 & 0.03 & 0.05 &II  \\
10 & 0.063 & 324.2 & 13.20 & 13.18 & 13.61 & -- & 0.04 & 0.04 & 0.27 & -- & --     \\
11 & 0.261 & 346.1 & -- & 12.26 & -- & -- & -- & 0.03 & -- & -- & --               \\
12 & 0.382 & 316.3 & 10.99 & 10.62 & 10.06 & 8.46 & 0.01 & 0.01 & 0.02 & 0.03 &II \\
16 & 0.476 & 307.2 & 11.32 & 11.31 & 11.26 & 11.35 & 0.01 & 0.01 & 0.05 & 0.10 & III \\
17 & 0.322 & 298.2 & 13.30 & 13.17 & -- & -- & 0.04 & 0.06 & -- & -- & --          \\
18 & 0.425 & 313.4 & 12.85 & 12.84 & -- & -- & 0.03 & 0.04 & -- & -- & --	   \\
19 & 0.524 & 298.2 & 12.39 & 12.39 & 12.51 & -- & 0.02 & 0.03 & 0.11 & -- & --	   \\
\enddata
\tablenotetext{~}{REMAINDER OF DATA AVAILABLE ELECTRONICALLY}
\end{deluxetable}

\begin{deluxetable}{lcrrrcrr}
\tablecaption{Bulk Near-IR Extinction and Hydrogen Column -- by Region \label{Region_avnh}}
\tablewidth{0pt}
\tablehead{
\colhead{Region}  &\colhead{Number of}&\colhead{Mean A$_V$}& \colhead{MAD}
&\colhead{Rejected}  &\colhead{Mean \nh}& \colhead{MAD}
&\colhead{Rejected}\\
\colhead{ }  &\colhead{good fits}&\colhead{ }& \colhead{ }
&\colhead{ }  &\colhead{[$\times 10^{22} $cm$^{-2}$]}&
\colhead{ } &\colhead{ }
}
\startdata
North     &8     &4.58&  1.11 & 2 & 0.67&  0.17 &   2\\ 
South     &16    &3.31&  0.36 & 3 & 0.92&  0.12 &   3\\ 
East      &35    &3.01&  0.22 &6  & 0.75& 0.06 &   2\\ 
West      &22    &4.66&  0.44 &2  &0.84&  0.09 &   1\\ 
RCW~108--IR     &36    &10.99&  1.01&1  & 1.98 & 0.20 &   2\\ 
Field     &117   &4.07&  0.23 &29 &0.92&  0.05 &   22\\ 
\enddata
\end{deluxetable}

\begin{deluxetable}{lccc}
\tablecaption{\nh -- A$_V$ relation -- by region  \label{Region_avnh_rel}}
\tablewidth{0pt}
\tablehead{
\colhead{Region}  & \colhead{Number of}&\colhead{slope}&\colhead{variance$^2$}\\
\colhead{ }        & \colhead{good fits}&\colhead{[$\times
    10^{21}$]}& \colhead{ }
}
\startdata
North     &8      &2.3&  0.39 \\ 
South     &16     &2.9&  0.43 \\ 
East      &35     &2.1&  0.24 \\ 
West      &22     &1.8&  0.30 \\ 
RCW~108-IR     &36     &2.2&  0.80   \\ 
Field     &117    &2.0 & 0.24  \\ 
\enddata
\end{deluxetable}

\begin{deluxetable}{lccccccc}
\tablecaption{Details on the KLF -- by Region \label{tab_klf}}
\tablewidth{0pt}
\tablehead{
\colhead{Region}  &\colhead{\# -in region}&
\colhead{Mean K$_s^a$}& \colhead{StDev$^b$}
&\colhead{Rejected} &\colhead{90$^{th}$percentile}& 
\colhead{10$^{th}$Percentile}&{Spread}}
\startdata
North     &18    &1.48&  0.26 & 0&-0.27  & 2.83& 3.10  \\
South     &34    &1.96&  0.26 & 1&-0.41 & 3.68& 4.10   \\
East      &91    &2.14&  0.09 &4 &0.94  &3.21 & 2.27   \\
West      &41    &2.07&  0.20 &1 & 0.64 &3.34 & 2.70  \\
RCW~108--IR     &60    &1.24&  0.21 &1 &-1.06 & 3.17& 4.43   \\
Field     &244   &1.85&  0.09 &6 &-0.15 & 3.21&3.36   \\
\enddata
\tablenotetext{a}{Absolute K$_s$ magnitude.}
\tablenotetext{b}{Three sigma outliers excluded.}
\end{deluxetable}

\begin{deluxetable}{lrrrrcrrrrrr}
\tabletypesize{\small}
\tablecolumns{12}
\tablewidth{0pc}
\tablecaption{Probable High Mass Stars \label{OB} }
\tablewidth{0pt}
\tablehead{ \colhead{SRC\#} & \colhead{J} & \colhead{H} & 
\colhead{K} & \colhead{A$_V$} & \colhead{\nh}&  \colhead{Abs. J}
& \colhead{Abs. H} & \colhead{Abs. K} &  \multicolumn{3}{c}{[M$_\odot$] Est.$^a$}\\
           \colhead{ } & \multicolumn{3}{c}{Observed} &
&\colhead{[$\times 10^{22}$cm$^{-2}$]} 
& \multicolumn{3}{c}{Ext. Corrected} 
& \colhead{J} &  \colhead{H}  &\colhead{K} }
\startdata
East &      &       &      &       &       &       &       &       & & & \\
  29 & 11.59&  10.14 &  9.48 &  7.53 &  0.16 & -1.11 & -1.75 & -1.93&  $>7$ &  $ >7$ &  $>7$   \\
  40 & 12.46&  11.17 & 10.22 &  3.11 &  0.29 &  1.01 &  0.05 & -0.70&   2.2&   2.5&  2.7\\
 243 &  8.98&   8.85 &  8.76 &  0.00 &  0.34 & -1.59 & -1.72 & -1.81&   $>7$ &  $ >7$ &  $>7$\\
 245 & 10.15&   9.74 &  9.54 &  0.00 &  0.16 & -0.42 & -0.83 & -1.03&   6.0&   7.0&  $>7$\\
\tableline
North &      &       &      &       &       &       &       &       & & & \\
  60  &16.42 & 13.78 & 11.57&  11.51&   8.31&   2.60&   1.20&  -0.27&  $<$2& $<$2&   3.25\\
  61  &12.56 & 11.23 & 10.48&   5.18&   0.81&   0.52&  -0.25&  -0.66&  3.0 & 3.25& 3.5\\
 146  &11.70 & 11.13 & 10.85&   0.31&   0.37&   1.04&   0.50&   0.24&  2.5 & 2.5 & 2.7\\
 262  &12.67 & 11.61 & 11.22&   4.30&   0.25&   0.88&   0.29&   0.18&  2.7 & 2.7 & 2.7\\
\tableline
RCW~108--IR &      &       &      &       &       &       &       &       & & & \\
  52  &15.43 & 13.20 & 11.73&  11.87&   2.77&   1.52&   0.55&  -0.15& 2.0  & 2.5  & 3.25\\
  62  &14.49 & 12.17 & 10.72&  13.24&   1.11&   0.18&  -0.72&  -1.30& 3.25 & 3.5  & 4.0 \\
  64  &11.86 & 10.89 & 10.52&   3.30&   0.62&   0.36&  -0.26&  -0.41&  3.25&  3.25&  3.5\\
  69  &16.11 & 13.98 & 12.76&  18.80&   3.84&   0.24&   0.12&   0.12& 3.25 & 3.0  & 3.0\\
  76  &18.56 & 14.95 & 12.88&  26.01&   2.24&   0.65&  -0.17&  -0.55& 3.0  & 3.0  & 3.5\\
  92  &12.86 & 11.20 & 10.27&  10.31&   2.22&  -0.61&  -1.17&  -1.44& 3.75 & 4.0  & 4.0\\
 100  &13.65 & 11.95 & 11.13&   9.81&   1.77&   0.32&  -0.34&  -0.52& 3.25 & 3.25 & 3.5\\
 111  &15.47 & 12.85 & 11.52&  18.34&   2.23&  -0.27&  -0.92&  -1.06& 3.5  & 3.75 & 4.0\\
 112  &11.58 & 10.26 &  9.07&   1.59&   3.24&   0.56&  -0.59&  -1.67& 3.0  & 3.5  & 4.0\\
 114  &12.41 & 11.95 & 10.23&   0.00&   3.95&   1.84&   1.38&  -0.34& $<$2 & $<$2 &   3.25\\
 116  &13.07 & 11.15 & 10.28&  12.39&   2.34&  -0.99&  -1.59&  -1.66& 4.00 & 3.75 & 4.0\\
 124  &15.79 & 13.41 & 12.24&  16.15&   2.49&   0.66&   0.02&  -0.11& 3.0  & 3.0  & 3.0\\
 131  &13.95 & 11.79 & 10.57&  12.96&   2.27&  -0.28&  -1.05&  -1.43& 3.5  & 4.0  & 4.0\\
 136  &11.67 & 10.54 &  9.79&   2.42&   0.52&   0.42&  -0.45&  -1.05& 3.5  & 3.5  & 3.75\\
 140  &12.83 & 11.39 & 10.40&   4.70&   0.94&   0.93&   0.00&  -0.68& 2.5  & 3.0  & 3.5\\
 143  &12.99 & 11.95 & 11.38&   4.69&   0.75&   1.09&   0.56&   0.29& 2.5  & 2.5  & 2.7\\
 144  &17.88 & 14.62 & 12.79&  27.35&   5.78&  -0.40&  -0.73&  -0.79& 3.5  & 3.5  & 3.5\\
\tableline
South &	     &	     &	    &	    &	    &	    &	    &	    &	   &	  &\\
  54  &11.71 & 10.94 & 10.65&   1.39&   0.39&   0.75&   0.12&  -0.08&  2.7 &  3.0 &  3.0\\
 104  &14.75 & 12.41 & 11.17&  15.32&   2.63&  -0.14&  -0.84&  -1.08&  3.5 & 3.75 & 3.75\\
 222  &12.91 & 11.85 & 11.40&   5.16&   0.98&   0.88&   0.38&   0.26&  2.7 &  2.7 & 2.7\\
 259  & 8.88 &  8.39 &  8.17&   2.74&   0.66&  -2.46&  -2.66&  -2.70&  $>$7& $>$7 &   $>$7\\
 263  &12.72 & 11.67 & 10.94&   5.84&   1.19&   0.50&   0.07&  -0.28&  3.0 &  3.0 &  3.25\\
 267  & 9.01 &  8.48 &  8.31&   0.00&   0.40&  -1.56&  -2.09&  -2.26&  $>$7& $>$7 &  $>$7\\
 269  &11.29 & 10.22 &  9.84&   4.59&   0.21&  -0.58&  -1.16&  -1.24&  3.75&  4.0 & 4.0 \\
 318  &12.63 & 11.66 & 11.23&   5.42&   0.94&   0.53&   0.15&   0.06&  3.0 &  3.0 &  3.0\\
\tableline
West  &	     &	     &	    &	    &	    &	    &	    &	    &	   &	  &\\
 162  & 9.77 &  8.85 &  8.39&   6.42&   1.69&  -2.61&  -2.84&  -2.88& $>$7 & $>$7 &   $>$7\\
 215  &11.23 & 10.89 & 10.80&   0.00&   0.22&   0.66&   0.32&   0.23& 3.0  & 2.7  & 2.7\\
 171  & 9.63 &  9.11 &  8.81&   0.00&   0.29&  -0.94&  -1.46&  -1.76& 4.0  & 4.0  & $>$4\\
 172  &12.77 & 11.30 & 10.33&   5.21&   0.70&   0.73&  -0.18&  -0.81&  2.7 & 3.25 & 3.5\\
 174  & 8.59 &  8.56 &  8.52&   0.00&   0.06&  -1.98&  -2.01&  -2.05& 5.0&   5.0& 5.0$^b$\\
 217  &10.41 & 10.17 & 10.04&   0.00&   0.75&  -0.16&  -0.40&  -0.53&  3.5 & 3.5  &3.5\\
 286  & 8.91 &  8.69 &  8.25&   0.00&   0.01&  -1.66&  -1.88&  -2.32&  $>$7&  $>$7 &   $>$7\\
 289  &13.01 & 11.87 & 11.22&   3.46&   0.38&   1.46&   0.69&   0.27&  2.0 & 2.2  &2.7\\
 296  &13.67 & 11.96 & 11.23&  10.55&   1.26&   0.12&  -0.46&  -0.50&  3.25& 3.5  &3.5\\
\enddata
\tablenotetext{a}{Age estimated to be 1 Myr for all regions except
  east, 3 Myr used for east.}
\tablenotetext{b}{Consistent mass estimate only found for 0.5 Myr.}
\end{deluxetable}


\begin{thebibliography}{}

 
\bibitem[Niemela et al.(2004)]{2004ApJS..154..363A} Niemela, V.S. \& Gamen, R.C.,\ 2005, \mnras, 356, 974

\bibitem[Walborn (1972)]{2004ApJS..154..363A}Walborn N.R.,\ 1972, \aj, 77, 312







\bibitem[Allen et al.(2004)]{2004ApJS..154..363A} Allen, L.~E., et  al.\ 2004, \apjs, 154, 363 

\bibitem[Beers et
    al.(1990)]{1990AJ....100...32B} Beers, T.~C., Flynn, K., \& Gebhardt, K.\ 1990, \aj, 100, 32 
\bibitem[Bessell \& Brett (1988)]{1988PASP..100.1134B} Bessell M.~S.,
  Brett J.~M., 1988, PASP, 100, 1134 
\bibitem[Caramazza et al.(2007)]{2007A&A...471..645C} Caramazza, M., 
Flaccomio, E., Micela, G., Reale, F., Wolk, S.~J., \& Feigelson, E.~D.\ 
2007, \aap, 471, 645 
\bibitem[Cash(1979)]{1979ApJ...228..939C} Cash, W.\ 1979, \apj, 228,
  939
\bibitem[Chini et al.(2003)]{2003A&A...409..235C} Chini, R., et al.\ 2003, 
\aap, 409, 235  
\bibitem[Comer{\'o}n et al.(2007)]{2007A&A...433..955C} Comer{\'o}n, F., \& Schneider, N., 2007, \aap ~$Accepted$
\bibitem[Comer{\'o}n et al.(2005)]{2005A&A...433..955C} Comer{\'o}n, F., Schneider, N., \& Russeil, D.\ 2005, \aap, 433, 955 
\bibitem[Favata \& Micela (2003)]{2003SSRv..108..577F} Favata F., Micela G., 2003, SSRv, 108, 577 
\bibitem[Favata et al. (2005)]{2005ApJS..160..469F} Favata F.,  Flaccomio E., Reale F., Micela G., Sciortino S., Shang H., Stassun  K.~G., Feigelson E.~D., 2005, ApJS, 160, 469 
\bibitem[Fazio et al.(2005)]{2005Fazio..313F}Fazio, G.G., et al., 2004, ApJS, 154, 10
\bibitem[Feigelson et al.(2007)]{2007prpl.conf..313F} Feigelson, E., 
Townsley, L., G{\"u}del, M., \& Stassun, K.\ 2007, Protostars and Planets 
V, 313 
\bibitem[Feigelson et al. (2005)]{2005ApJS..160..379F} Feigelson E.~D., et al., 2005, ApJS, 160, 379 
\bibitem[Fitzgerald(1987)]{1987MNRAS.229..227F} Fitzgerald, M.~P.\ 1987, \mnras, 229, 227 
\bibitem[Flaccomio et al.(2003)]{2003ApJ...582..398F} Flaccomio, E., Damiani, F., Micela, G., Sciortino, S., Harnden, F.~R., Jr., Murray, S.~S., \& Wolk, S.~J.\ 2003, \apj, 582, 398 
\bibitem[Freeman (2001)]{2001SPIE...4477..76} Freeman, P.E.,  Doe, S. \& Siemiginowska A.,SPIE Proceedings vol.4477, 76
\bibitem[Gagn{\'e} et al.(2005)]{2005ApJ...628..986G} Gagn{\'e}, M., Oksala, M.~E., Cohen, D.~H., Tonnesen, S.~K., ud-Doula, A., Owocki, S.~P. Townsend, R.~H.~D., \& MacFarlane, J.~J.\ 2005, \apj, 628, 986 
\bibitem[Getman et al. (2005)]{2005ApJS..160..319G} Getman K.~V., et al., 2005, ApJS, 160, 319 
\bibitem[Getman et al.(2006)]{2006ApJS..163..306G} Getman, K.~V., Feigelson, E.~D., Townsley, L., Broos, P., Garmire, G., \& Tsujimoto, M.\ 2006, \apjs, 163, 306 
\bibitem[Getman et al.(2007)]{2007ApJ...654..316G} Getman, K.~V., Feigelson, E.~D., Garmire, G., Broos, P., \& Wang, J.\ 2007, \apj, 654, 316 
\bibitem[Gorti \& Hollenbach(2002)]{2002ApJ...573..215G} Gorti, U., \& Hollenbach, D.\ 2002, \apj, 573, 215 
\bibitem[G{\" u}del (2004)]{2004A&ARv..12...71G} G{\" u}del M., 2004, A\&ARv, 12, 71 
\bibitem[Gutermuth \e (2007)]{guter07} Gutermuth R.A., Myers, P.C., Megeath, S.T., Allen, L.E., Pipher, J.L., Muzerolle, J. Porras, A. Winston, E., Fazio, G., 2007,
\apj ~$Submitted$
\bibitem[Heinrich(2003)]{2003sppp.conf...52H} Heinrich, J.\ 2003, Statistical Problems in Particle Physics, Astrophysics, and Cosmology, 52 
\bibitem[Henney(2006)]{2006astro.ph..2626H} Henney, W.~J.\ 2006, ArXiv Astrophysics e-prints, arXiv:astro-ph/0602626 
\bibitem[Herbst \& Havlen(1977)]{1977A&AS...30..279H} Herbst, W., \& Havlen, R.~J.\ 1977, \aaps, 30, 279 
\bibitem[Hildebrand(1983)]{1983QJRAS..24..267H} Hildebrand, R.~H.\ 1983, \qjras, 24, 267
\bibitem[Hillenbrand \& White (2004)]{2004ApJ...604..741H} Hillenbrand L.~A., White R.~J., 2004, ApJ, 604, 741 
\bibitem[Hong et al.(2005)]{2005ApJ...635..907H} Hong, J., van den Berg, M., Schlegel, E.~M., Grindlay, J.~E., Koenig, X., Laycock, S., \& Zhao, P.\ 2005, \apj, 635, 907 
\bibitem[Hong, Schlegel, \& Grindlay (2004)]{2004ApJ...614..508H} Hong J., Schlegel E.~M., Grindlay J.~E., 2004, ApJ, 614, 508 
\bibitem[Kaltcheva \& Georgiev(1992)]{1992MNRAS.259..166K} Kaltcheva, N.~T., \& Georgiev, L.~N.\ 1992, \mnras, 259, 166 
\bibitem[Kastner et al.(2005)]{2005ApJS..160..511K} Kastner, J.~H., Franz, G., Grosso, N., Bally, J., McCaughrean, M.~J., Getman, K., Feigelson, E.~D., \& Schulz, N.~S.\ 2005, \apjs, 160, 511 
\bibitem[Lada \& Adams (1992)]{1992ApJ...393..278L} Lada C.~J., Adams F.~C., 1992, ApJ, 393, 278 
\bibitem[Lopes de Oliveira et al.(2006)]{2006A&A...454..265L} Lopes de Oliveira, R., Motch, C., Haberl, F., Negueruela, I., \& Janot-Pacheco, E.\ 2006, \aap, 454, 265 
\bibitem[Maritz \& Jarett (1978)]{2004ApJS..73..194}Maritz, J. S. \&  Jarett. R. G. 1978, J. of the American Statistical Association, 73, 194
\bibitem[Megeath et al.(2004)]{2004ApJS..154..367M} Megeath, S.~T., et al.\ 2004, \apjs, 154, 367 
\bibitem[Meyer, Calvet, \& Hillenbrand (1997)]{1997AJ....114..288M} Meyer M.~R., Calvet N., Hillenbrand L.~A., 1997, AJ, 114, 288 
\bibitem[Mezger(1994)]{1994Ap&SS.212..197M} Mezger, P.~G.\ 1994, \apss, 212, 197 \bibitem[Moffat \& Vogt(1973)]{1973A&AS...10..135M} Moffat, A.~F.~J., \& Vogt, N.\ 1973, \aaps, 10, 135 
\bibitem[Morrison and McCammon(1983)]{1983ApJ...270..119M} Morrison, R., McCammon, D.\ 1983.\ ApJ 270, 119-122. 
\bibitem[Mullan \& Waldron(2006)]{2006ApJ...637..506M} Mullan, D.~J., \& Waldron, W.~L.\ 2006, \apj, 637, 506 
\bibitem[Preibisch et al.(2005)]{2005ApJS..160..401P} Preibisch, T., et al.\ 2005, \apjs, 160, 401 
\bibitem[Price et al.(2001)]{2001AJ....121.2819P} Price, S.~D., Egan, M.~P., Carey, S.~J., Mizuno, D.~R., \& Kuchar, T.~A.\ 2001, \aj, 121, 2819 
\bibitem[Rakowski et al.(2006)]{2006ApJ...649L.111R} Rakowski, C.~E., Schulz, N.~S., Wolk, S.~J., \& Testa, P.\ 2006, \apjl, 649, L111 
\bibitem[Raymond \& Smith (1977)]{1977  ApJS..248...564} Raymond, J.C. and Smith B.W., 1977, ApJS 35, 419
\bibitem[Reipurth(1983)]{1983A&A...117..183R} Reipurth, B.\ 1983, \aap, 117, 183 
\bibitem[Rieke \& Lebofsky(1985)]{1985ApJ...288..618R} Rieke, G.~H., \& Lebofsky, M.~J.\ 1985, \apj, 288, 618 
\bibitem[Rodgers et al.(1960)]{1960MNRAS.121..103R} Rodgers, A.~W., Campbell, C.~T., \& Whiteoak, J.~B.\ 1960, \mnras, 121, 103 
\bibitem[Sanz-Forcada et al.(2003)]{2003ApJS..145..147S} Sanz-Forcada, J.,  Brickhouse, N.~S., \& Dupree, A.~K.\ 2003, \apjs, 145, 147 
\bibitem[Scargle (1998)]{1998ApJ...504..405S} Scargle J.~D., 1998, ApJ, 504, 405 
\bibitem[Schulz et al.(2001)]{2001ApJ...549..441S} Schulz, N.~S., Canizares, C., Huenemoerder, D., Kastner, J.~H., Taylor, S.~C., \& Bergstrom, E.~J.\ 2001, \apj, 549, 441 
\bibitem[Schulz et al.(2003)]{2003ApJ...595..365S} Schulz, N.~S., Canizares, C., Huenemoerder, D., \& Tibbets, K.\ 2003, \apj, 595, 365 
\bibitem[Siess, Dufour, \& Forestini (2000)]{2000A&A...358..593S} Siess L., Dufour E., Forestini M., 2000, A\&A, 358, 593 
\bibitem[Skinner et al.(2005)]{2005MNRAS.361..191S} Skinner, S.~L., Zhekov, S.~A., Palla, F., \& Barbosa, C.~L.~D.~R.\ 2005, \mnras, 361, 191 
\bibitem[Stassun et al.(2004)]{2004AJ....127.3537S} Stassun, K.~G., Ardila, D.~R., Barsony, M., Basri, G., \& Mathieu, R.~D.\ 2004, \aj, 127, 3537 
\bibitem[Straw et al.(1987)]{1987ApJ...314..283S} Straw, S., Hyland, A.~R., Jones, T.~J., Harvey, P.~M., Wilking, B.~A., \& Joy, M.\ 1987, \apj, 314, 283
\bibitem[Townsley et al. (2003)]{2003ApJ...593..874T} Townsley L.~K., Feigelson E.~D., Montmerle T., Broos P.~S., Chu Y.-H., Garmire G.~P., 2003, ApJ, 593, 874 
\bibitem[ud-Doula \& Owocki(2002)]{2002ApJ...576..413U} ud-Doula, A., \& Owocki, S.~P.\ 2002, \apj, 576, 413 
\bibitem[Urquhart et al.(2004)]{2004A&A...428..723U} Urquhart, J.~S., Thompson, M.~A., Morgan, L.~K., \& White, G.~J.\ 2004, \aap, 428, 723 
\bibitem[Vazquez \& Feinstein(1992)]{1992A&AS...92..863V} Vazquez, R.~A., \& Feinstein, A.\ 1992, \aaps, 92, 863 
\bibitem[Vuong et al. (2003)]{2003A&A...408..581V} Vuong M.~H., Montmerle T., Grosso N., Feigelson E.~D., Verstraete L., Ozawa H., 2003, A\&A, 408, 581 
\bibitem[Weaver et al.(1977)]{1977ApJ...218..377W} Weaver, R., McCray, R., Castor, J., Shapiro, P., \& Moore, R.\ 1977, \apj, 218, 377 
\bibitem[Werner et al.(2005)]{2005Werner..313F}Werner, M.W., et al., 2004, ApJS, 154, 1
\bibitem[Winston \e (2007)]{winston07} Winston, E.,  Megeath, S.T., Wolk, S.J.,  Muzerolle, J, Gutermuth, R., Hora, J.L., Allen, L.E.,  Spitzbart, B., Myers, P., Fazio, G., 2007, \apj ~$Accepted$
\bibitem[Whiteoak(1963)]{1963MNRAS.125..105W} Whiteoak, J.~B.\ 1963, \mnras, 125, 105 
\bibitem[Wolk et al. (2002)]{2002ApJ...580L.161W} Wolk S.~J., Bourke T.~L., Smith R.~K., Spitzbart B., Alves J., 2002, ApJ, 580, L161 
\bibitem[Wolk et al. (2004)]{2004ApJ...606..466W} Wolk S.~J., et al., 2004, ApJ, 606, 466 
\bibitem[Wolk et al. (2005)]{2005ApJS..160..423W} Wolk S.~J., Harnden F.~R., Flaccomio E., Micela G., Favata F., Shang H., Feigelson E.~D., 2005, ApJS, 160, 423 
\bibitem[Wolk et al. (2006)]{2006AJ....132.1100W} Wolk, S.~J., Spitzbart, B.~D., Bourke, T.~L., \& Alves, J.\ 2006, \aj, 132, 1100 
\bibitem[Wolk et al. (2007)]{2006ASP} Wolk, S.~J., Comer\'on, F., \&
  Bourke, T.~L., Handbook of Star Forming Regions Vol. II. The Southern Sky -B. Reipurth ed. {\it submitted}.



\end{thebibliography}
\end{document}